\documentclass{aanolinenumber}

\usepackage{graphicx}
\usepackage{natbib}
\usepackage{txfonts}
\usepackage{lipsum}
\usepackage{subcaption}      
\usepackage{lscape}            
\usepackage{placeins}

\chardef\us=`\_

\begin{document}
   \title{{Les Trente Glorieuses}: 29 years of helioseismic observations with the Luminosity Oscillations Imager}

   \subtitle{}

   \author{T. Appourchaux
          \inst{1}
          }

   \offprints{T. Appourchaux}

   \institute{Institut d'Astrophysique Spatiale, UMR8617, Universit\'e Paris-Saclay, 91405 Orsay Cedex, France\\
              \email{Thierry.Appourchaux@cnrs.fr}}
    \date{Received 25 September 2025; accepted with minor revision 12 November 2025}

   \abstract  
  {The Luminosity Oscillations Imager (LOI) of the Variability of Solar Irradiance and Gravity Oscillations (VIRGO) instrument aboard the SoHO mission has been operating for almost the past 30 years.}
   {The goal of this article is to report on the observation of p-mode parameters throughout these 30 years that covers two solar cycles using the LOI scientific and guiding pixels as well as the guiding high-voltage housekeeping.  The article also provides a definitive description of the data for the final SoHO archive.  The LOI time series analyzed here starts on 1 April 1996 0:00 TAI and ends on 31 March 2025 23:59 TAI with a 60-s cadence.}
   {I use the level-0 data of the LOI for computing level-2 corrected time series for each pixel taking into account engineering calibration, presence of attractors (locked values), the SoHO-Sun distance, the spacecraft roll I use the scientific pixels for deriving Sun-as-a-star signals, individual $(l,m)$ signals, East-West and North-South difference signals, guiding pixel signals and East-West and North-South high voltage signals to compute 29 one-year power spectra.  A 29-year long time series of the Sun-as-a-star signal is also used for pushing the detection of p modes to low frequency.  The power spectra are globally fitted (fully over the p-mode envelope) using Maximum Likelihood Estimators using a C++ code with two versions of the code depending on the presence of $l=0$ modes (\texttt{progFIT} and \texttt{guiFIT}).  The Fourier spectra for the individual $(l,m)$ signals are locally fitted (around each p-mode) using Maximum Likelihood Estimators using an IDL code for $l \le 3$.}
   { I report on the effect of solar activity upon mode frequencies, linewidths, height and energy rate.  I report on the variation as a function of frequency for frequency, $a_2$ coefficient and linewidth changes, as well as the average over the degree and the frequency of these changes.  Using the 29-year time series, I report on the frequencies, linewidths and mode height fitted with \texttt{progFIT}.  Using the collapsogram technique, I also report on the detection of modes below 1600 $\mu$Hz, making the lowest frequencies detected with an instrument observing the Sun in intensity.  I also report on the detection of p mode in the high voltage and guiding pixel signals with a mode height about 5 to 10 times larger than what is observed in the Sun-as-a-star signal for $l=1$.  The ratios of the observed mode visibilities for the different signals are provided following a calibration of the size of the guiding pixels.  While the visibility ratios for the signals excluding the limb are in good agreement with theory, those covering the solar limb are in strong disagreement.}
   {} 

 \titlerunning{29 years of observations with the Luminosity Oscillations Imager}
   \maketitle

\section{Introduction}
Helioseismology started in the late 1970's with the first detection of 
global solar p modes by \cite{AC79}, confirmed by \cite{GG80}, using 
radial velocity measurements; and by \cite{MWHH83} using irradiance 
measurements with ACRIM  
Monitor \citep[Active Cavity Radiometer 
Irradiance,][]{RCW79} onboard SMM (Solar Maximum 
Mission).  These breakthroughs were followed by the first inversion 
of the solar structure performed by \cite{DOG84} and \cite{TDWD84} 
using the so-called Duvall's law \citep{TDJH84}.  Helioseismology 
really became a useful tool when it demonstrated that the equation of 
state could be in error, and was indeed in error \citep{AK92}.

The diagnostic potential of helioseismology for our physic had 
already been perceived in the mid 80's when the space mission 
SoHO (Solar and Heliospheric 
Observatory, an ESA/NASA mission, Special volumes of {\it Solar 
Physics}, 1995, {vol. 162};1997, {vol. 170};  and 1997,{vol. 175})) was 
approved.  The need for uninterrupted data can be fulfilled by a space 
mission, but also by ground-based networks of instrument, networks on 
which the Sun never sets: GONG (Global Oscillation Network 
Group, dedicated volume of {\it Science}, 1996, {vol. 272}), 
IRIS  \citep[Investigation on the Rotation and Interior of the 
Sun,][]{SE91} and BiSON \citep[Birmingham Solar-Oscillations Network,][]{YE91}.  

The detection of solar oscillations using solar radial velocities is 
feasible from the ground.  It is not
so obvious how to perform such measurements using irradiance or radiance 
fluctuations because of transmission variations through the Earth atmosphere.  Following the detection of p modes with the ACRIM 
instrument, \cite{CF84} used photometers aboard balloons to reduce the 
impact of the Earth atmosphere on the measurement.  A similar photometer \citep[IPHIR: InterPlanetary Helioseismology by IRradiance 
measurements,][]{CF88} was flown aboard PHOBOS II \citep[Soviet Mars 
exploration mission, see][]{Siddiqi2016}.  This instrument returned excellent 
helioseismic measurements as shown by \citet{TTCF92}.  

In 1987, the SoHO mission was selected by the European Space Agency 
as part of it Horizon 2 programme.  The mission consisted of 
in-situ, coronal, particle, and helioseismic instruments.  Following the at-the-times recent 
developments in helioseismology, three helioseismic instruments were 
included for obtaining solar radial-velocity and irradiance 
measurements using either imagers or looking at the Sun as a star.  
GOLF \citep[Global Oscillations at Low Frequency,][]{AG95}
uses a Sodium cell to measure solar radial velocities look at the Sun 
as a star, while MDI \citep[Michelson Doppler Imager,][]{PS95} uses a 
one-million pixel detector for obtaining spatially resolved data.  The 
VIRGO  \citep[Variability of IRradiance and Gravity Oscillations,][]{CFJR95} instrument measures the irradiance of the Sun as a star, the 
spectral radiance with photometers,  and the spectral radiance spatially resolved over the 
solar disk.  The latter is performed by the Luminosity Oscillations 
Imager (LOI) whose concept was described by \cite{BAAJ88}, and later 
built as a space instrument \citep{TABA97}.  

The LOI is one of a kind, that is the only instrument that detected
in intensity the low degree p modes from the ground \citep{TA95a}, 
and that has been measuring the low degree p modes from space since 1996.  
Its design was specifically aiming at detecting modes with a different sensitivity for degree greater than 1; thereby optimizing the chance of detecting the solar g modes using different signals. Instruments that detects 
p modes in velocity, such as GOLF and BiSON \citep{AMB2007} or HMI \citep[Helioseismic and Magnetic Imager,][]{PS2012} and GONG \citep{TA2024}, benefit from a better signal-to-noise ratio 
allowing them to detect modes below 1 $\mu$Hz.  In addition, 
instruments such as GONG, MDI and HMI make images of the Sun permitting 
them to derive by inversion the internal structure from the surface 
below the convection zone.  Nevertheless, it must be emphasized that 
scientific progress should not always be ranked on the ability of 
making interesting discoveries, but also on the ability of 
confirming new discoveries.  The failure to confirm g-mode 
detection can serve as an example \citep{TA2000, Appourchaux2010}.

The LOI is not just merely another helioseismic instrument, it can 
contribute to the understanding of several sources of systematic 
errors.   For instance, it was demonstrated by \cite{TT97} that 
observations of solar p modes in intensity look indeed different 
from those performed in velocity.  This resulted in the discovery of 
p-mode profile asymmetries \citep{TT98}. It was also shown by \citet{Schou2002} that rotation inversion are very sensitive to different
frequency inferences made with different instrument.  Last but not least, the LOI 
has a different mode sensitivities as it observes in intensity; the 
mode leakage is therefore significantly different from that of 
velocity instruments, as are the systematic errors associated with this 
leakage.

The goals of this article are: 1) to explain how the time series of the LOI pixels are constructed, to be provided to the solar community; 2) to show that an instrument detecting 
solar p modes in intensity confirms most of the discovery performed 
in velocity; 3) to report on the limb amplification of the p modes.    The LOI time series analyzed here starts on 1 April 1996 0:00 TAI and ends on 31 March 2025 23:59 TAI with a 60-s cadence.  

 The first section of the article describes the data reduction, and production of level-1 data.  The second section explains which times series were computed.  The third section explains how the resulting power spectra were  fitted for p-mode parameter extraction.  The following sections provide results of the variations of the p-mode parameters over two solar cycles, the impact of gaps, the frequency error bars, the mode amplitude and height.  The last section gives results from the various guiding signals before concluding.

\section{From level 0 to level 1: calibration}
The design of the LOI instrument is described in \cite{TABA97}.  The 
data reduction has been heavily modified since then.  Especially, the occurrence
of locked value on the pixel data, known as {attractors}, requires a special treatment
for identifying them and flagging them.  The attractors are values that are locked at some binary values, locking that is most likely occurring in the Voltage-to-Frequency converters used for the Analog to Digital conversion in the VIRGO instrument.
The Sunphotometers (SPM) of VIRGO are also affected by this locking.  Other corrections related to
the loss of the pointing of the High Gain Antenna (HGA) are also required.  All these
steps are described hereafter.

   \begin{figure*}
   \centering
   \includegraphics[angle=180,width=9.cm]{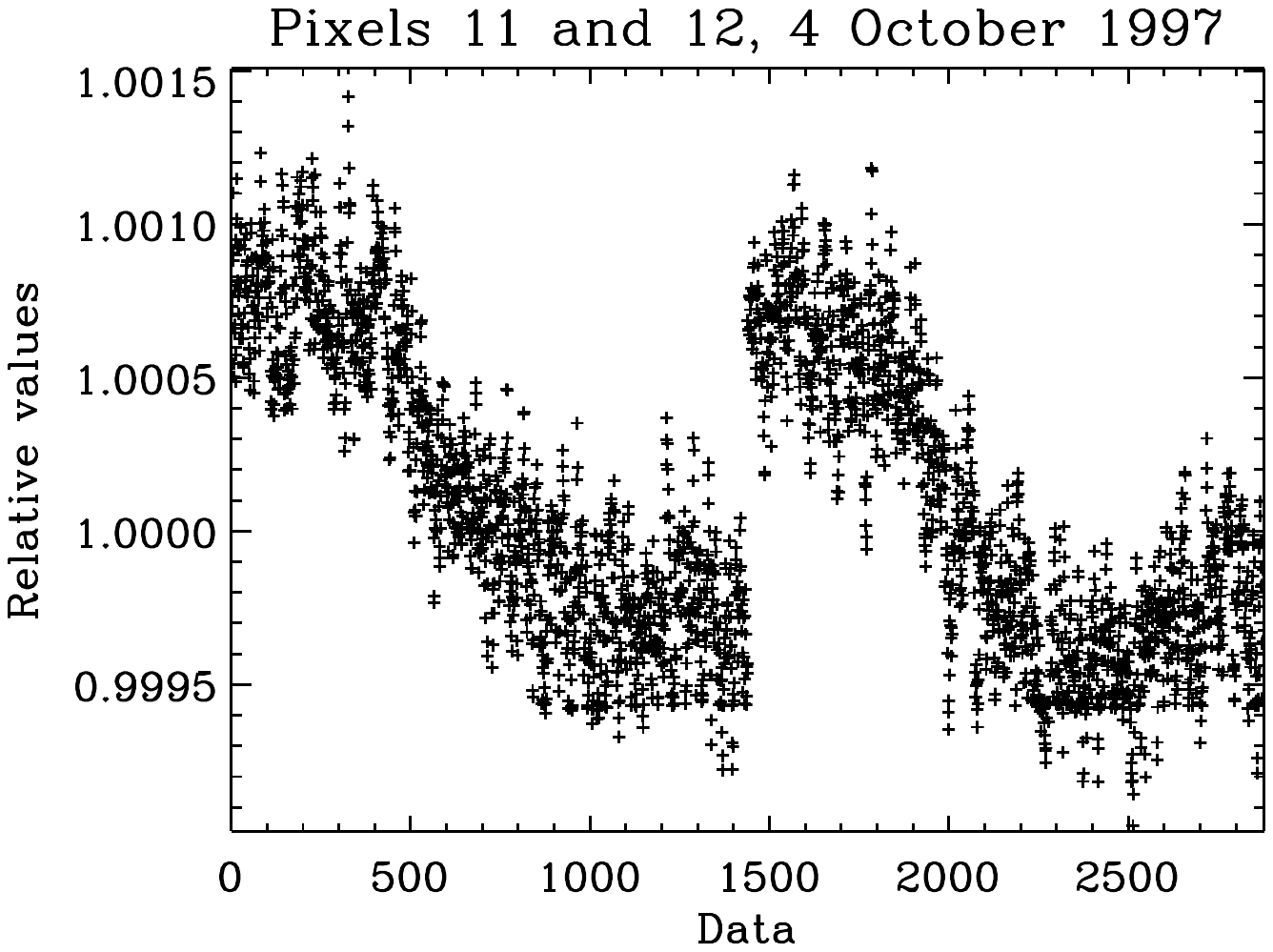}
   \includegraphics[angle=180,width=9.cm]{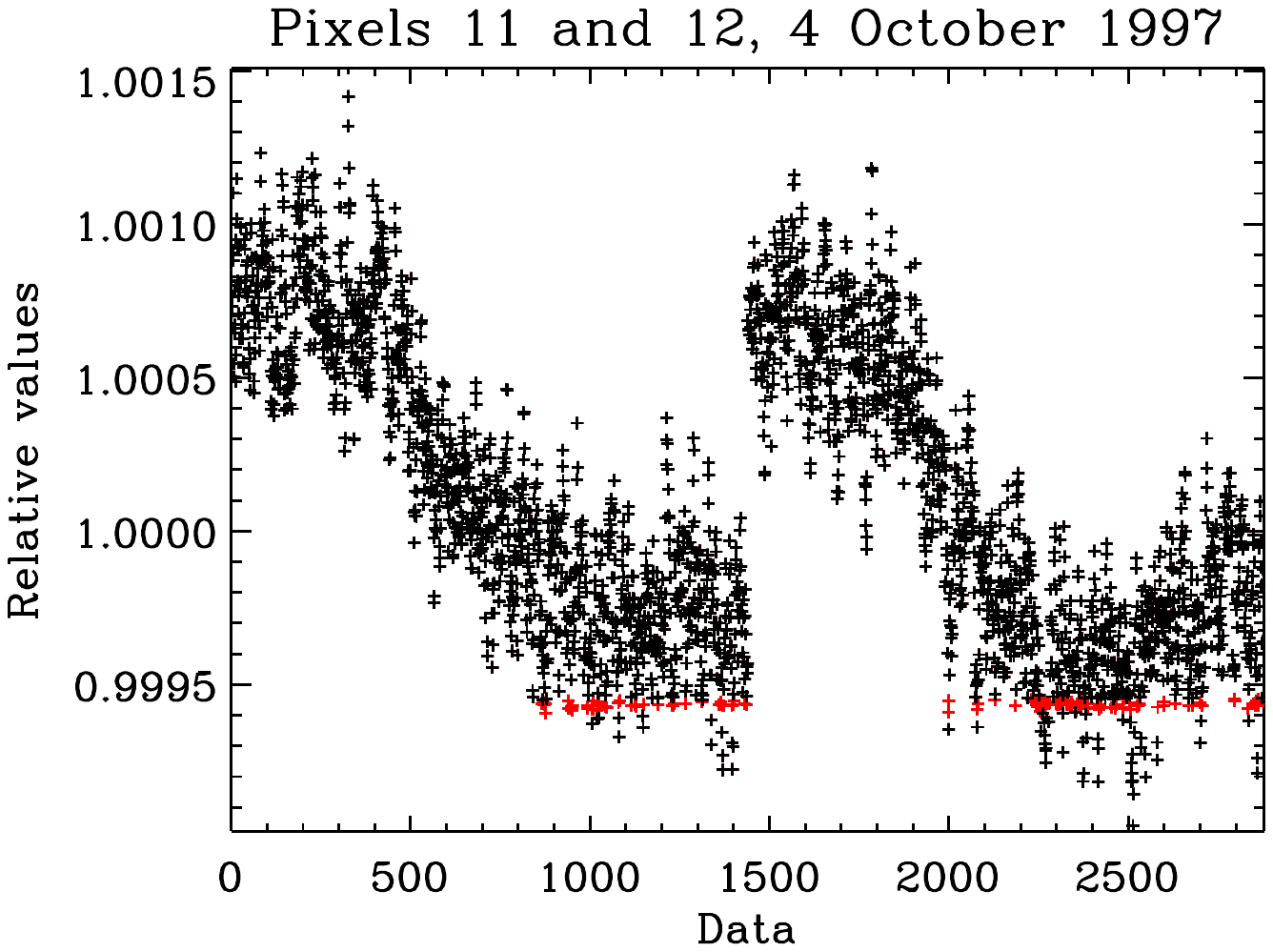}
      \caption{Relative pixel intensity as a function of time in minutes for two pixels: raw (Left), identified attractors in red (Right)}
         \label{attractor_example}
   \end{figure*}
   
         \begin{figure*}
   \centering
   \includegraphics[angle=180,width=9.cm]{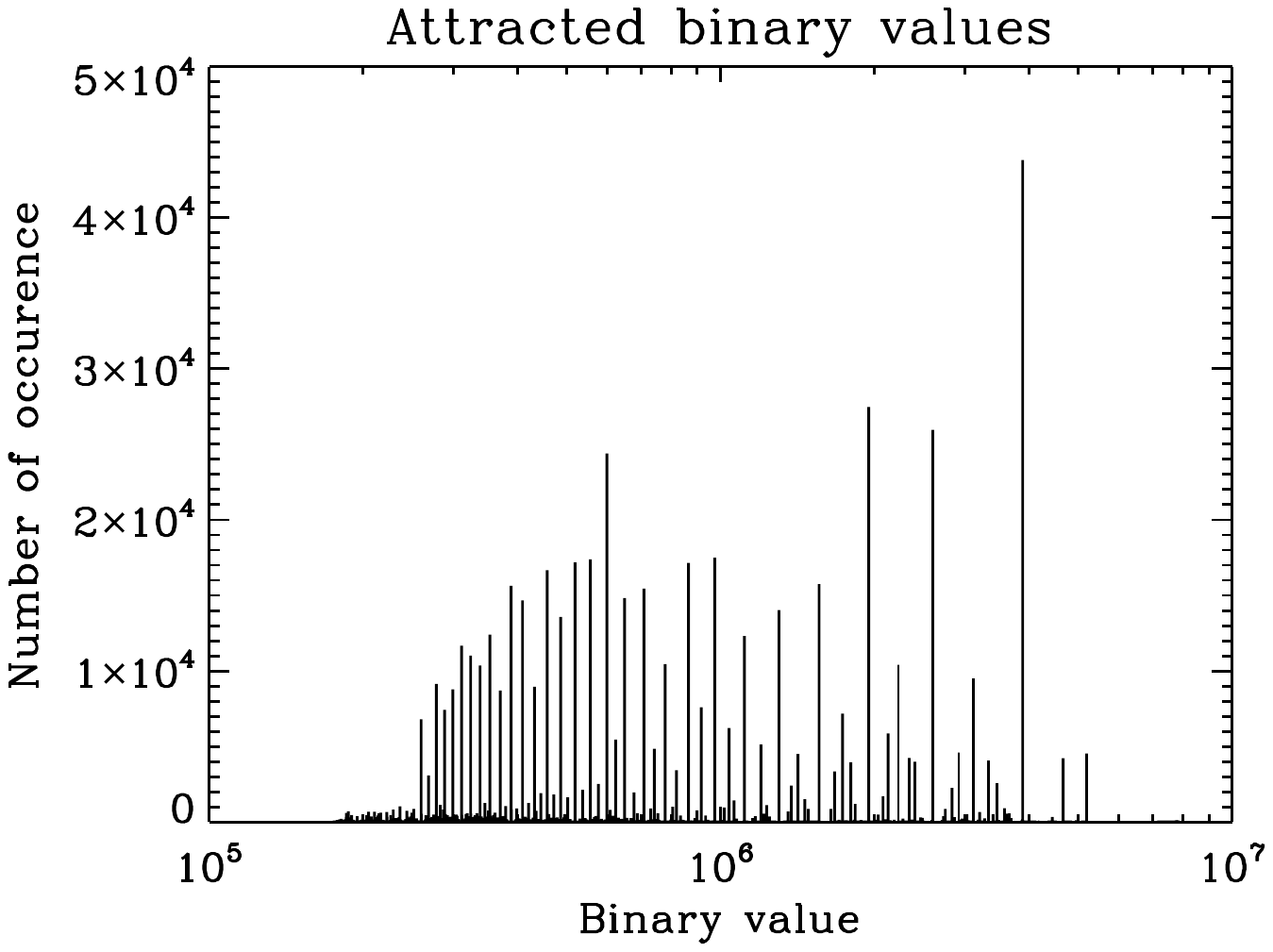}
   \includegraphics[angle=180,width=9.cm]{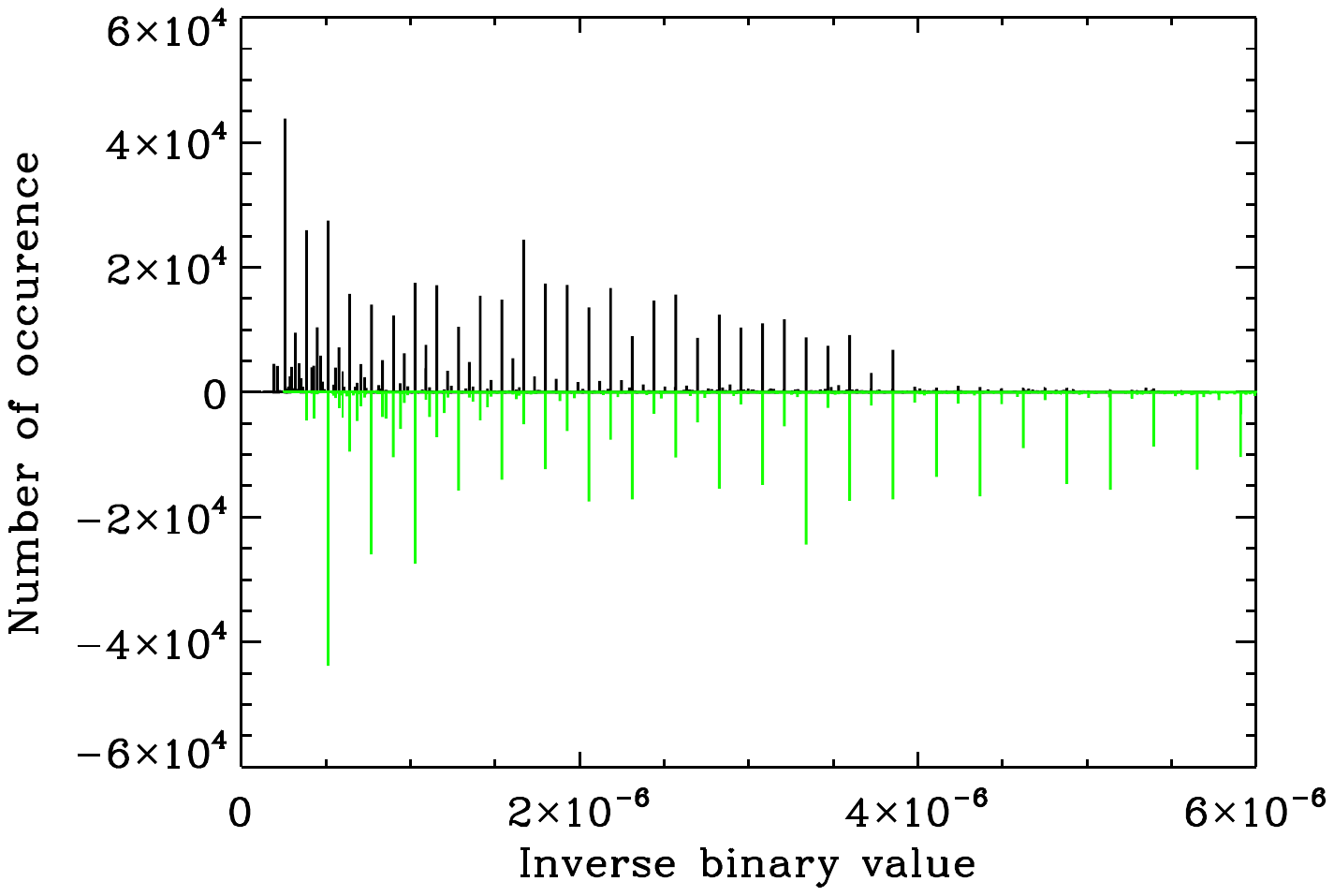}

      \caption{Compilation of all attracted bits over the 12 scientific pixels over 29 years as a function of the binary value (left) and as a function of the inverse binary value (right).  For the latter in addition, shown in green and mirrored the same compilation but as a function the inverse binary value multiplied by a factor 2.}
      \label{attracted_values}
   \end{figure*}

\subsection{Level 0: depacketization}
After depacketizing of the transmitted data, the level-0 data 
are built from the three transmissions of the VIRGO data; this is 
done at the VIRGO data center (VDC) located at the Instituto de Astrofisica de Canarias, Tenerife, Spain.  Preliminary detection of the attractors
is done at the VDC but I use my own algorithm described hereafter for flagging them.

\subsection{Level 0: attractors detection and flagging}
The attractors are best detected in the raw binary data before conversion to engineering values.  The detection is performed day by day, pixel by pixel.  The procedure is as follows:
\begin{itemize}
\item First, detect if there are any jumps greater than 0.4\% of the median in consecutive data points
\item Second, if no jumps are detected proceed with the next step, otherwise no attractor detection performed.
\item Third, I compute the mean and the rms of the data.  We arbitrarily limit the detection of the attractors to values within $\pm 4\,\sigma$ of the mean. 
\item Third, I compute the histogram of the data using a bin size of 1\% of the data variation (maximum-mininimum)
\item Fourth, the histogram is binned with a box car over 30 bins (with edge mirrored)
\item Fifth, I compute the ratio of the histogram to that of the binned histogram.
\item Sixth,  the attractors are then detected when the ratio is greater than 5.6$\sigma_s$, where the $\sigma_s$ is the estimated value of the rms ratio for the associated bin value (shot noise).  The attractor values are then replaced by a value of -10.
\end{itemize}
The procedure works in nearly all circumstances.  Given the fact, that this procedure is applied daily more than 120, times in total in the course of 29 years, there are still circumstances that would make this procedure fails.  For instance, when there is a sunspot appearing on a pixel there is a large variation of several \% that would distort the histogram.  In order to reduce, the over detection or under detection, we decided to conduct a visual check of the detected attractors, especially when the daily variations are larger than some predetermined value (>0.3\%) typical of a small sunspot crossing a pixel.

Figure~\ref{attractor_example} gives an example of attractor detection.  It shows the first detected attractor.  The occurrence of these attractors does not depend upon the pixels but on the digital value itself.  Since the instrument has a decreasing throughput of about a factor 16 in 29 years, when the pixels with the smallest  flux (Pixels 9 to 12) have an attractor, the other pixels with larger surface will also have the attractor at the same value at a later time.
Figure~\ref{attracted_values} gives an histogram of all of the digital values corresponding to attractors.  There is a clear periodicity in the inverse of the binary value shown by Fig~\ref{attracted_values}.  The periodicity corresponds to a ratio between the inverse binary value of about a factor 1.0625, which exactly corresponds to a multiplication by 1+1/16.  In binary calculation,  this multiplication corresponds to adding the original binary value (1) to the binary value shifted down by four bits (1/16).   The fact that there is also a correspondence with a factor two which is a binary value shifted down one bit may give clues to the source of the attractors.  There are also counters behind the VFC that can count up to 24 bits.  It is possible that the locking occurs not in the VFC but more likely in the counters themselves.  

The flag of the attractors are kept until the outliers are cleaned (See Section 3.2).  At most, there are no more than six attractors at any one time, thereby affecting half of the LOI pixels.  The impact of the attractors on the data is studied in Section 3.2.
   
                 \begin{figure}
   \centering
   \includegraphics[angle=180,width=9.cm]{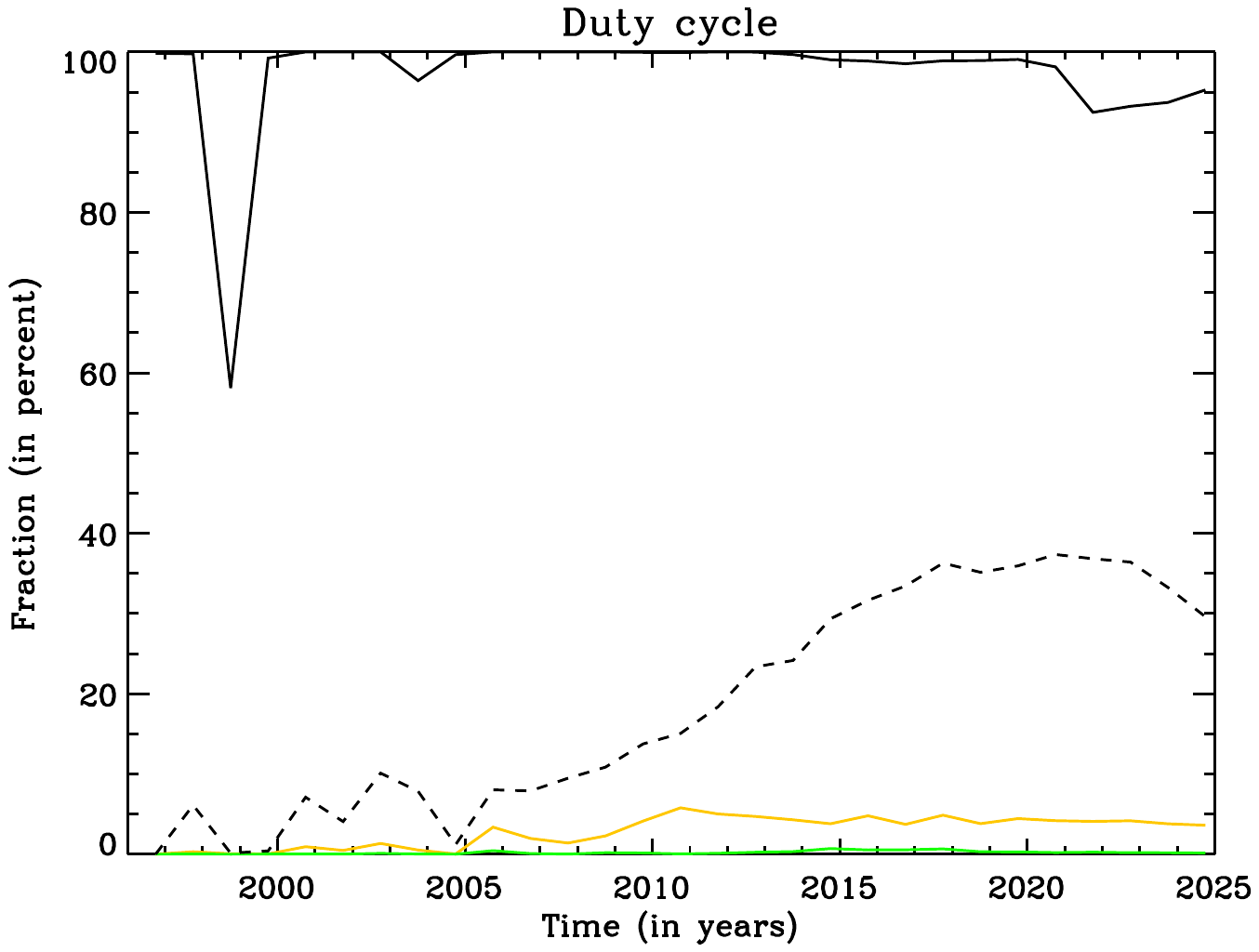}
      \caption{Duty cycle as a function of time for the 12-pixel data (continuous line), for the fraction of time any pixel having at least one attractor (dashed line), for the fraction of time at least one pixel having an attractor lasting longer than 5 minutes (orange line), for the fraction of time at least two pixels having an attractor lasting longer than five minutes (green lines).}
      \label{duty_cycle}
   \end{figure}
   
           \begin{figure}
   \centering
   \includegraphics[angle=180,width=9.cm]{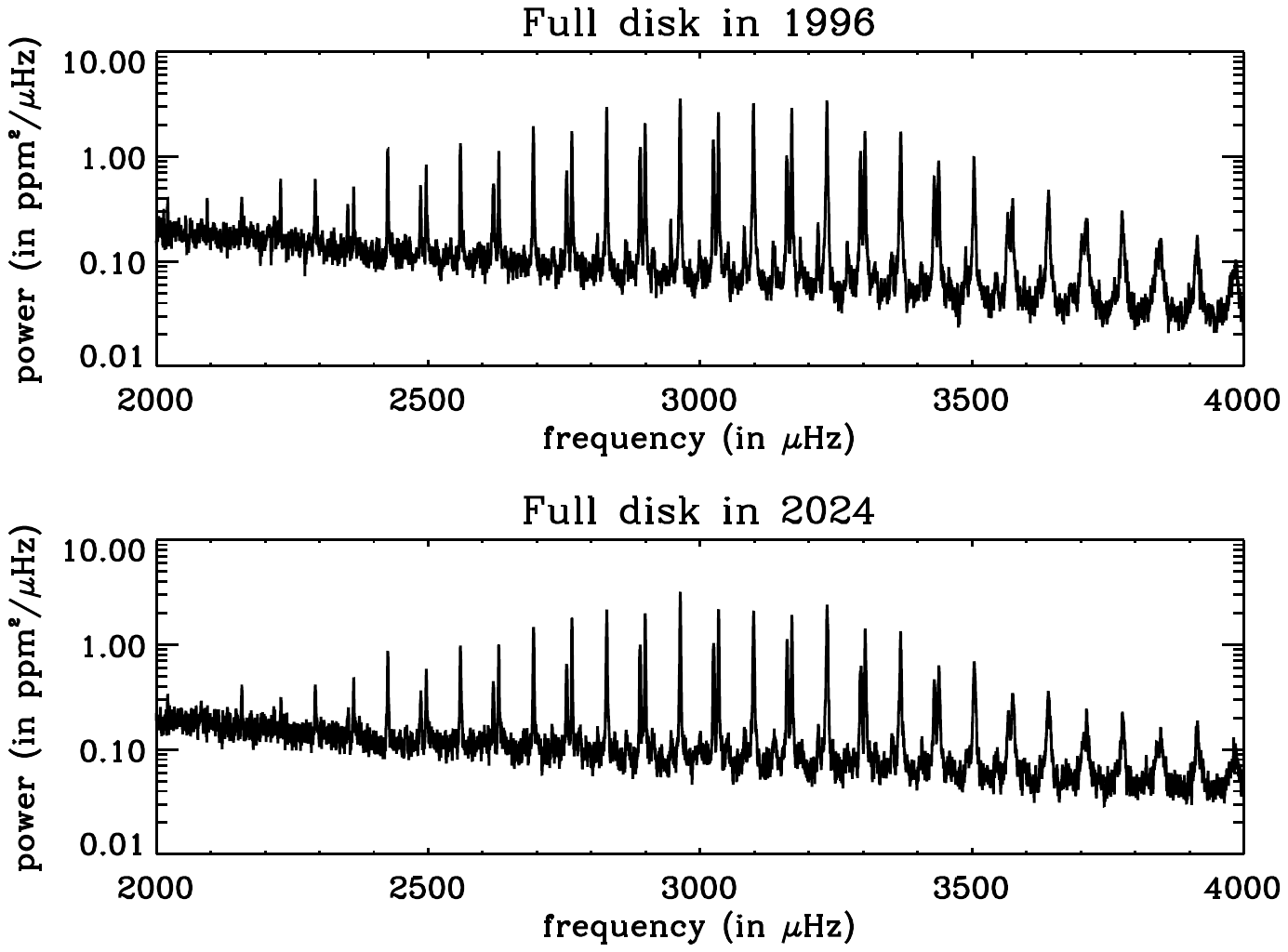}
      \caption{power spectrum as a function of frequency for the full disk signal for 1996 (top) and 2024 (bottom) for one year of data.}
      \label{power-FD}
   \end{figure}

        \begin{figure*}
   \centering
   \includegraphics[angle=180,width=9.cm]{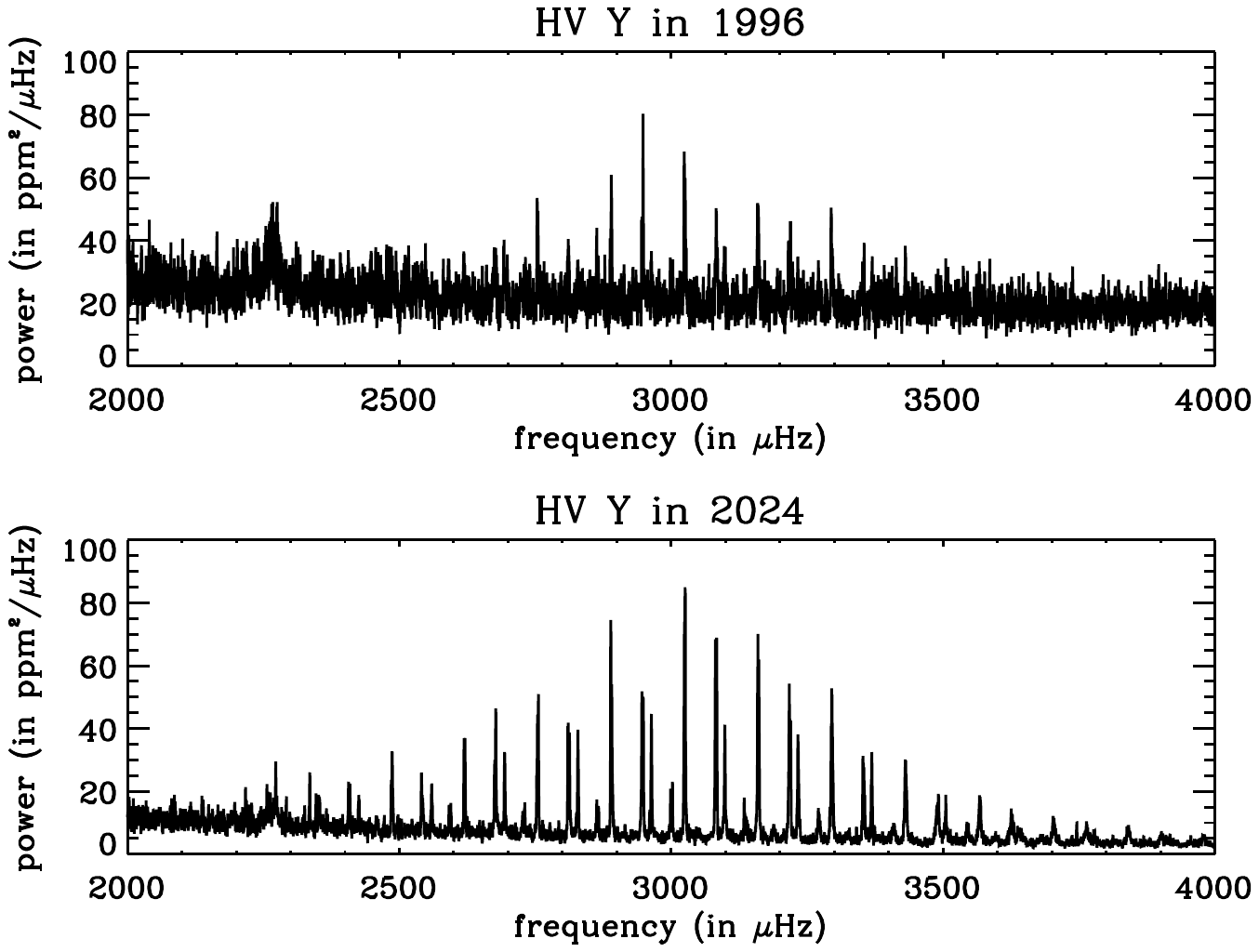}
   \includegraphics[angle=180,width=9cm]{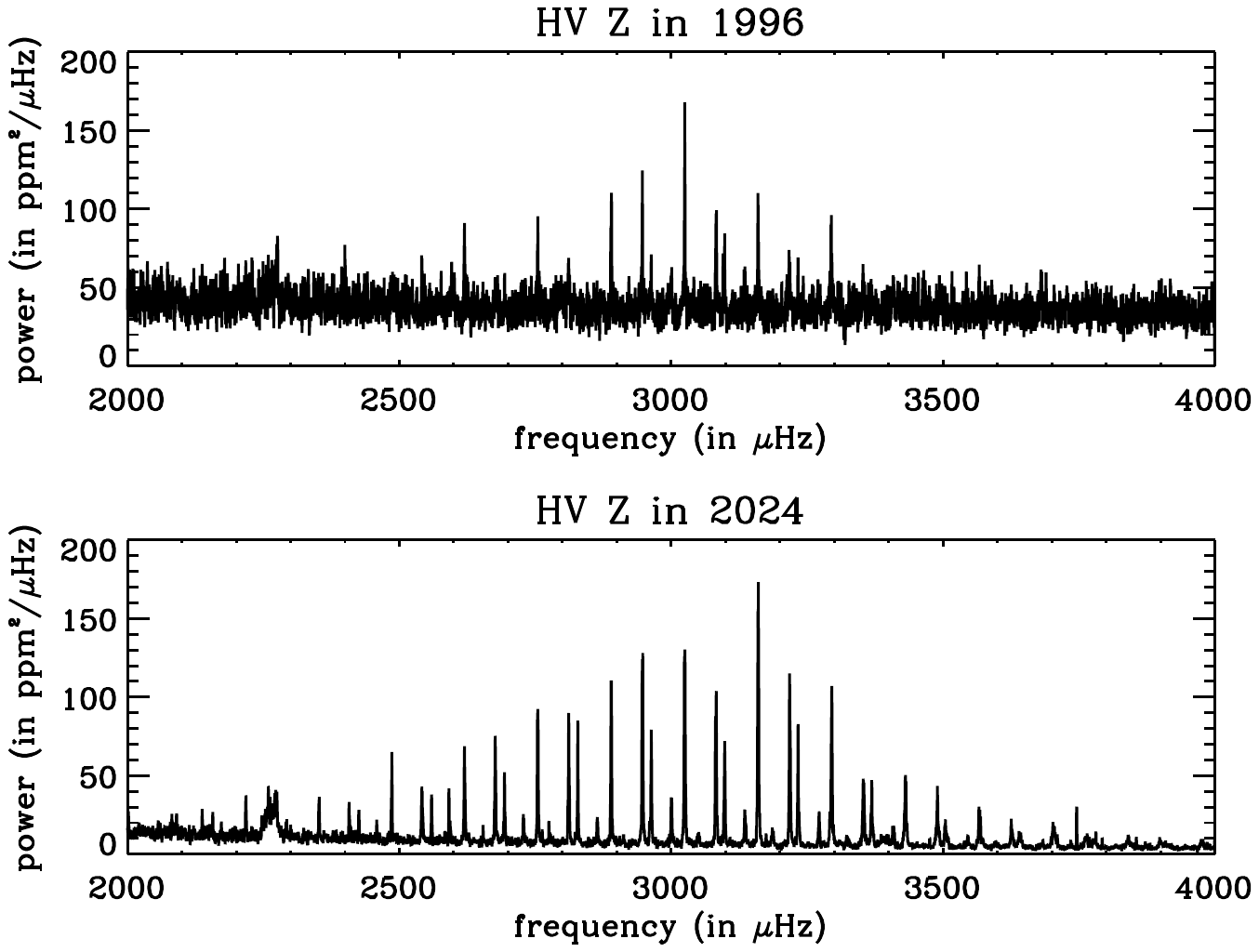}
      \caption{(Left) power spectrum as a function of frequency for the Y high voltage for 1996 (top) and 2024 (bottom) for one year of data.  (Right) power spectrum as a function of frequency for the Z high voltage for 1996 (top) and 2024 (bottom) for one year of data.}
      \label{power-guiding-HV}
   \end{figure*}
   
           \begin{figure}
   \centering
   \includegraphics[angle=180,width=9.cm]{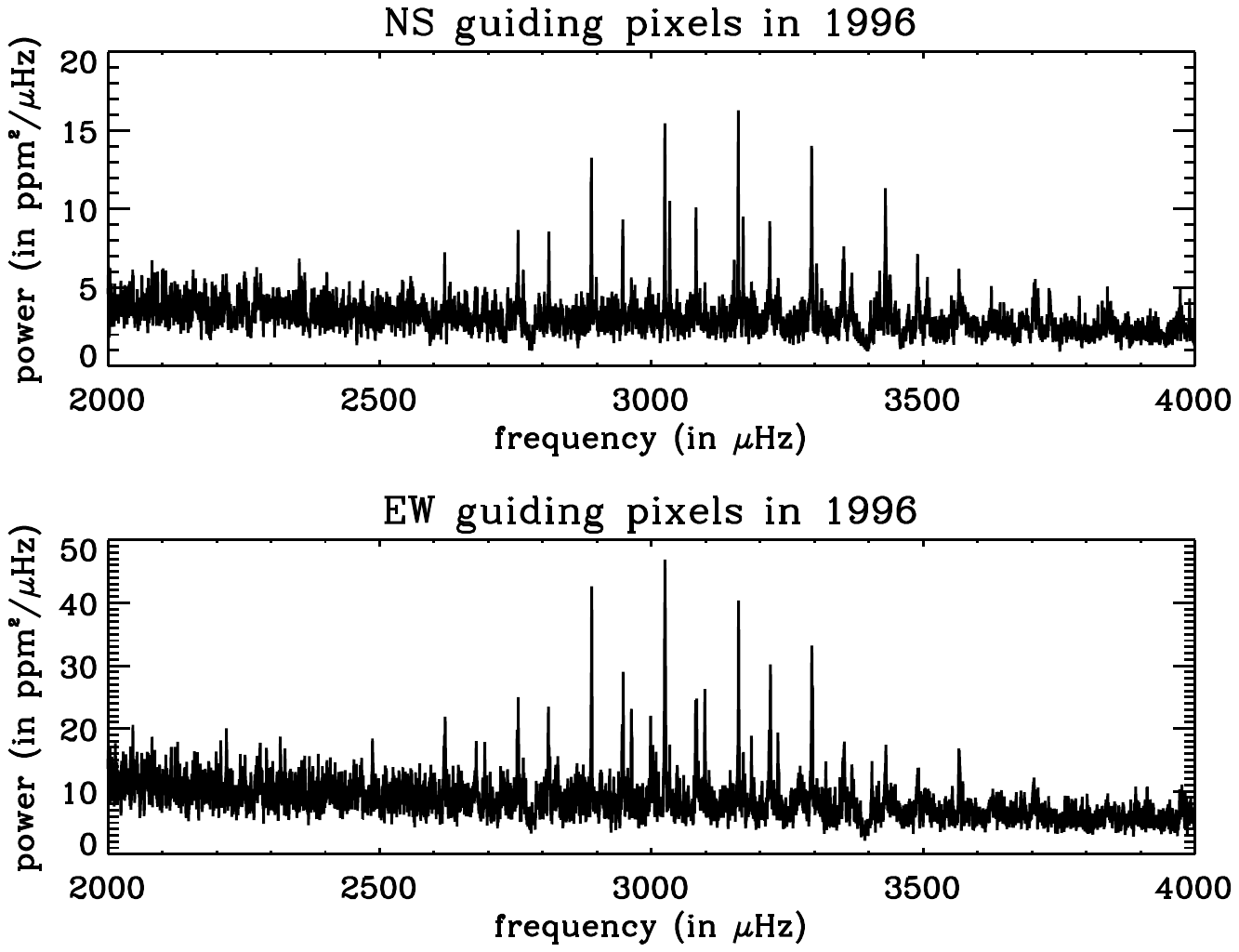}
      \caption{Power spectrum as a function of frequency for 1996, for the North-South (NS) guiding pixels (Top) and for the East-West (EW) guiding pixels (Bottom) for one year of data.}
      \label{power-guiding}
   \end{figure}

\subsection{Level 1: Engineering data conversion}
The level-0 data are then calibrated to engineering units using the calibration performed on the ground before launch.  The calibration of the offset for each pixel was refined in flight with the calibration-mode switch off.
The data are then corrected for temperature effects and pixel-sensitivity variations.

\subsection{Level 1: Sun distance and spacecraft attitude corrections} 
The 16~pixels are then converted to flux measured
at 1~AU using the SoHO-Sun distance derived from the 
orbital data \citep[See][ for the configuration of the pixels on the detector]{TABA97}.  The correction takes into account the real shape of 
the pixels, the limb darkening at 500 nm, and the in-flight size of 
the solar image at 1~AU.  The precision of the correction is  better than 
0.1 
\% for all of the scientific pixels. In addition a relativistic 
correction is applied taking into 
account the spacecraft velocity with respect to the Sun; this is minute.  All of these corrections are described in the LOI VDC document available on 
the VIRGO home page under www.ias.u-psud.fr/virgo/html/software.html.
In addition, for taking into account the pixel response that depends on the the spacecraft (S/C) attitude, I used
the sensitivities derived by \citet{TABA97} for correcting for the impact of the varying angle of incidence of the sunlight on the LOI instrument.

\section{From level 1 to level 2}
\subsection{Roll maneuver correction}
After May 25, 2003, the HGA failed to move for sending data towards the Earth.  It was then decided to leave the HGA in a fixed position and rotate the spacecraft by 180 degrees every 3 months for transmitting to Earth\footnote{See https://soho.nascom.nasa.gov/soc/soho\_events/SOHO-Spacecraft-Events.csv}.  When the S/C is rolled by 180 degrees, the pixel numbering is redefined in the process, i.e. pixels 1, 2, 5, 6, 7, 8 becoming pixels 4, 3, 8, 7,  6, 5 and vice versa; and the guiding pixels 13, 14 becoming 16, 15 and vice versa.

During the roll maneuvers, the rotation is not done exactly around the line of sight of the instrument requiring an adjustment the pointing of the instrument.  In the process, this slight depointing produces a slight change in the flux measured by each pixel.  
The impact of the roll is to slightly change the pixel response.  The correction of these changes are done for each pixel, as follows:
\begin{itemize}
\item First, get date and time of the roll by using the S/C attitude data
\item Second, the pixel flux is fitted over one day ending 200 minutes before the roll, and starting 200 minutes after the roll.
\item Third, the flux before and after the roll are then extrapolated for each fit at the time of the roll.  
\item Fourth,  the correction is done by normalizing the flux after the roll to that of the flux before the roll. 
\end{itemize}
This normalization makes sure that the time series is homogenous since the start of the mission.

\subsection{Outliers cleaning}
The various interruptions due either to the extended loss of contact (the so-called SoHO vacations, mid 1998), the roll maneuvers, the presence of attractors or other unforeseen events produce large jumps in the pixel data that need to be corrected for.  We use an automatic outlier detection based upon the Peirce criterion \citep{Peirce1852,Gould1855}.  The correction for each pixel is done as follows:
\begin{itemize}
\item First, detect outliers with the Peirce criterion over a duration of 10 days.  The detection and correction are done automatically.  The detected outliers are filled by interpolation.
\item Second, detect outliers with the Peirce criterion over a duration of 360 days.  The detection can be done either automatically (default) or interactively (choice of the user).  If the correction is done interactively, the user chooses two locations over which to perform the interpolation.  The latter is used when very large jumps cannot be corrected automatically.  This step is done recursively until no correction is necessary.
\item Third, detect outliers with the Peirce criterion over a duration of 40 days with the same procedure as for the previous step
\end{itemize}
During this process, the attractors are detected as outliers and corrected by interpolation, thereby filling the gaps created by these attractors.  Figure \ref{duty_cycle} shows the impact of the attractors on the duty cycle of the data.  At the end of the mission, the data have attractors about 40\% of the time.  { The increase of the number of attractors is mainly due to the fact that the photon flux on the pixels decreased by factor 16 over the 29 years, hence the change in digital count is smaller at the end of the 29 years compared to the beginning, making the attractors lasting longer at the end than at the beginning}.  But since the attractors are not always continuous, the interpolation can easily fill them in when they are shorter than five minutes.  The number of pixels having an attractor lasting more than five minutes is at most 5\%, while it is less than 0.5\% for two attractors.

\subsection{Detrending and filtering of pixel data}
For each pixel, the data are detrended using a triangular two-day smoothing window.  The relative values are then computed and all spikes greater than 30 $\sigma$ are removed.  { Since the 12 pixels may have some missing value (set to zero), a further filtering is done by setting all pixels to zero if at least one pixel is missing in all the three quadruplets: Pixel 1 to 4,  Pixel 4 to 8, and Pixel 9 to 12.  The 12 pixels are then not set to zero if for any datum at most two pixel quadruplets have missing data.}  The time series analyzed of all 16 pixels are available from soho.nascom.nasa.gov/data/archive.html.

\section{Generation of time series}
\subsection{Sun-as-a-star time series}
The data set for the Sun-as-a-star time series or Full-disk (FD) time series is derived from the average of the 12 level-2 scientific pixels.  The time series are made
of two sets: one time series of 29 years duration, 29 time series of one-year duration.  Examples for the power spectra for the first year and the last year are shown on Figure ~\ref{power-FD}.
The power spectra are corrected { by dividing by the duty cycle} (See Figure \ref{duty_cycle}).

\subsection{Spherical harmonics extraction}
For combining the pixel data, I used the spherical harmonic decomposition computed as in \cite{TABA90},
using the limb darkening from \cite{Allen1975}.  The filters are normalized according to Eq. 
(17) of \cite{TALG98}.  This 
normalization allows a symmetrical leakage matrix 
\citep{TALG98}.  Example for
such leakage matrix can be found in \citet{TARS98}.  Since the 
apparent solar diameter changes along the orbit, the spherical 
harmonic filters vary slowly with time.  To alleviate this problem, 
time-independent filters are applied to the pixels for extracting the 
$(l,m)$ modes.  The filters are computed as an average over one year 
of filters computed 
weekly.  Over this period the $B$ angle has a mean close to zero.  
\cite{LG98} showed that it 
was possible to detect the influence of a non-zero $B$ angle on the 
p-mode data.  Therefore, this average minimizes the effect of the $B$ 
angle 
on the data.  After combining the pixels with the filters, the time 
series are Fourier transformed; the positive frequencies provide the 
signal for $+m$, and the negative frequencies provide the conjugate 
of the signal for $-m$ \citep{TALG98}.

For each $(l,m)$ harmonic, I extracted two data sets: one time series of 29 years duration and 29 time series of one-year duration.

\subsection{East-West and North-South time series}
I also extracted the difference between the North-South (NS) pixels and the East-West (EW) pixels.  For the NS pixels, this is simply the same data set
as for $l=1, m=0$.  For the EW pixels, this is the difference between the $l=1, m=1$ complex time series and its conjugate.
For each, I extracted 29 time series of one year duration.  These data sets were used to compare the result obtained with the high voltage (HV) signals, which are sensitive to the difference between the North and South guiding pixels and between the East and West guiding pixels.

\subsection{EW / NS HV and guiding pixel time series}
I also extracted from the engineering data, the high voltage time series for the Y direction (North-South) and the Z direction (East-West).  The voltages were converted to arcsec then to ppm.  The value of 8.8 ppm/marcsec is derived from the outer radius of the scientific pixels given above and from the calibration performed in April 1996 \citep{TABA97}.  Figure~\ref{power-guiding-HV} shows two power spectra at the beginning of the mission and near the end of the mission.  The improvement in signal-to-noise ratio is striking but is not understood.  On the other hand, the guiding pixels also show the p modes but only for the first year (1996) (Figure \ref{power-guiding}) since the noise increases by such a factor that no modes can be detected after 1997, which is not understood either.  Similar power spectra were derived with six months of data by \cite{TA98d}.

\section{p-mode parameter extraction}
They are 30 Sun-as-a-star power spectra, 29 North-South difference power spectra, 29 East-West difference power spectra, two guiding pixel power spectra, 29 North-South high voltage power spectra,  29 East-West high voltage power spectra, and 29 $\times$ 15 $(l,m)$ power spectra.

\subsection{Power spectra fitting}
The power spectra are fitted using Maximum Likelihood Estimation (MLE) \citep{JS92, TALG98}.  The power spectra are fitted globally (fully over the p-mode envelope), an approach first used for fitting stellar power spectra as in \citet{TA2008}.  In that latter article, the MLE for HD49933 was made difficult due to the large linewidth of the modes making the identification of the pair $l=0-2$ with respect to the $l=1$ difficult.  It was already suggested by \citet{TA2008} that a Bayesian approach would be better providing a more conservative identification of the modes; such an approach was implemented by \citet{Benomar2009}.  This Bayesian approach that I suggested is now widely used in asteroseismology \citep{Handberg2011, Corsaro2014, Lund2017}.  The Bayesian fit of the power spectra is remarkably efficient when the S/N ratio in the power spectrum is low, less than about 3, but when the signal ratio is higher the MLE fit and the Bayesian fit provide to the same estimators with very similar error bars.  In addition, the Bayesian fit requires the creation of Monte Carlo Markov Chains which can be very slow in any interpreted language (Python, IDL).  

The code used by \citet{TA2008} for the CoRoT star HD49933 was indeed an IDL code.  There were already attempts to write a C++ code (\texttt{progFIT}) based on this IDL code \citep{Neiner2004} which was not yet available in 2008.  The development of that code continued until 2010, but the code was not still fully functioning (Neiner, private communication).  The code was then picked up from its ashes in 2021 when its use for the Planetary transits and oscillations of stars (PLATO) mission was considered an asset given the large sample of solar-like stars \citep[About more than 20, stars for the core P1 sample,][]{Rauer2025}.  The code was then made fully functional and was tested for the PLATO mission on hundreds of solar-like stars.  The code can be found on Gitlab\footnote{git.ias.u-psud.fr/plato\_pdc/plato\_wp37/wp372\_msap3\\/msap3\_04/progfit}.  The specificity of \texttt{progFIT} is that it uses the \texttt{Minuit2} library for finding the highest likelihood, but it not only finds the highest likelihood (or $\log$(likelihood) minimum) it also check that the distance between the minimum found and the extrapolated minimum is small compared to a given criteria, i.e. providing the estimated distance to minimum (EDM).  The Hessian { (used for getting the parameter errors)} is computed on the fly as well as the EDM.  The EDM is a representation that in the found minimum is within a typical $\sigma$ distance in the hypersphere of the $N$ parameters modeling the power spectrum.  The EDM is a dimensionless quantity that can be written as
\begin{equation}
{\rm EDM}=\frac{1} {2}\sum_{i=1}^{i=N} \left(d^{r}_{i}\right)^{2}= \frac{N} {2} <d^{r}>^{2}
\end{equation}
where $d^{r}_{i}$ is the relative distance to minimum in units of $\sigma_i$, and $<d^{r}>$ is the mean relative distance over the $N$ parameters.  For 200 parameters and a mean $<d^{r}>$  of 3\%, the resulting EDM is 0.1. I used this latter value for the convergence of all fits.  For an EDM of 0.1, when the number of parameters is decreased to 100, the mean distance to the minimum increases by a factor $\sqrt{2}$, still sufficient for most applications.

For \texttt{progFIT}, the typical speed is 22 minutes on an Apple M3 Max, for a one-year duration time series, with 19 quadruplets $l$=0, 1, 2 and 3 for a total of 166 parameters: 10 parameters for the background (three Harvey-like profiles with exponent of 4 and one white noise), one common splitting for all degrees, one common inclination angle for all degrees, one common ratio { between the heights of the $l$=2 modes to $l$=0 modes},  one common ratio { between the heights of the $l$=3 modes to { $l$=1 modes} } and 8 parameters per quadruplet (four frequencies, two linewidths { (one for the $l=0-2$ pair and one for the $l=1-3$ pair)}, two mode heights) assuming a symmetrical Lorentzian profile; the ratio within a multiplet follows \citet{Toutain1993}.  For the 29-year duration time series, it takes 8 hours with 23 quadruplets.

Following the successful development of \texttt{progFIT}, another code was developed for fitting the NS and EW difference signals, and HV guiding signals.  Since these signals are by construction not sensitive to $l$=0 modes, the new code (\texttt{guiFIT}) took into account this lack of visibility, but also the different visibilities in a multiplet.  The code can also be found on Gitlab\footnote{git.ias.u-psud.fr/plato\_pdc/plato\_wp37/wp372\_msap3\\/msap3\_04/guifit}.

{ The \texttt{progFIT} code was used 62 times on the FD and guiding pixel power spectra, while \texttt{guiFIT} was used 116 times.  In total, the two codes were used less than about 180 times resulting in a bit less of 3 days of computing on a single processor, in comparison the former IDL code would have taken about 150 days of computing.  With several cores such as the M3 Max, the duration reduces even further to about five hours.}

\subsection{Fourier spectrum fitting}
This technique, known as {phase} fit, was introduced by \cite{JS92} and later refined by 
\cite{TALG98}.  The method relies on the knowledge of the 
leakage matrix for inferring the p-mode parameters.  This method can 
only be applied when the leakage matrix for a given degree can be inverted 
\citep{TALG98}.  

The p-mode parameters are usually locally extracted (around each p-mode) from the Fourier spectra 
using MLE \citep{JS92, TALG98} with an IDL code.  The splittings 
were derived from Clebsh-Gordan coefficients as in \cite{RITZ91}, the 
decomposition is given in Appendix~\ref{AppendixA}.  Three pixel noises were used 
for modeling the solar 
noise, while the leakage and noise covariance matrix 
were derived according to \citet{TALG98}.  Examples of Fourier spectra for various $l$ and $m$ can be found in \citet{TARS98}.

Here I did not study the odd $a$-coefficients that are related to the solar differential rotation.  Although results on these coefficients are available with the LOI \citep{TA1998} their use has to be combined with higher degree modes for getting a proper rotation inversion \citep[See][]{TA2002b}.  Since the LOI does not provide high degree modes, work using the GONG and HMI / MDI data provides odd $a$-coefficients covering all degrees from 0 to 200 with a single instrument \citep{Korzennik2023} allowing homogenous rotation inversion.

           \begin{figure*}[h]
   \centering
   \includegraphics[angle=180,width=9.cm]{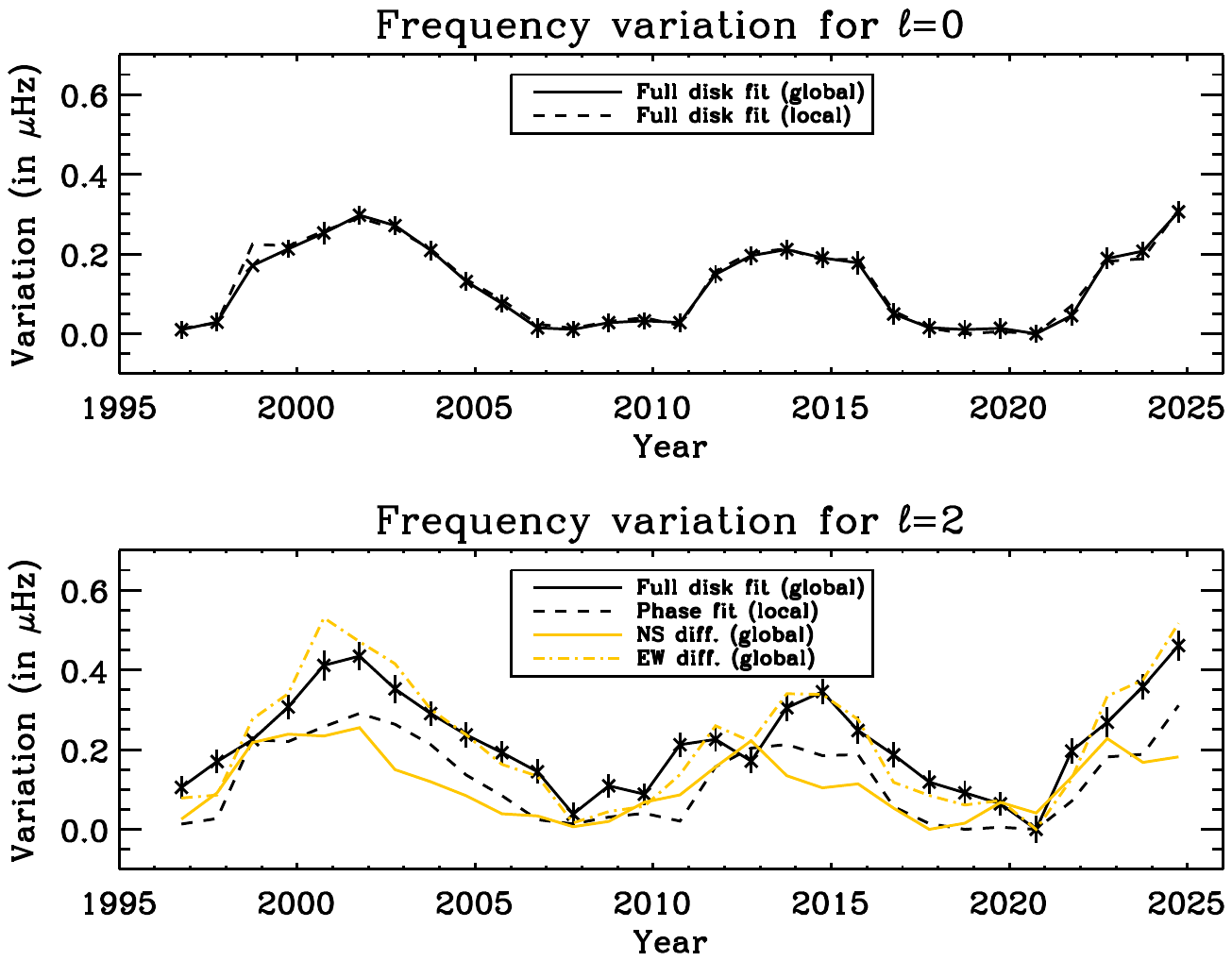}
            \includegraphics[angle=180,width=9.cm]{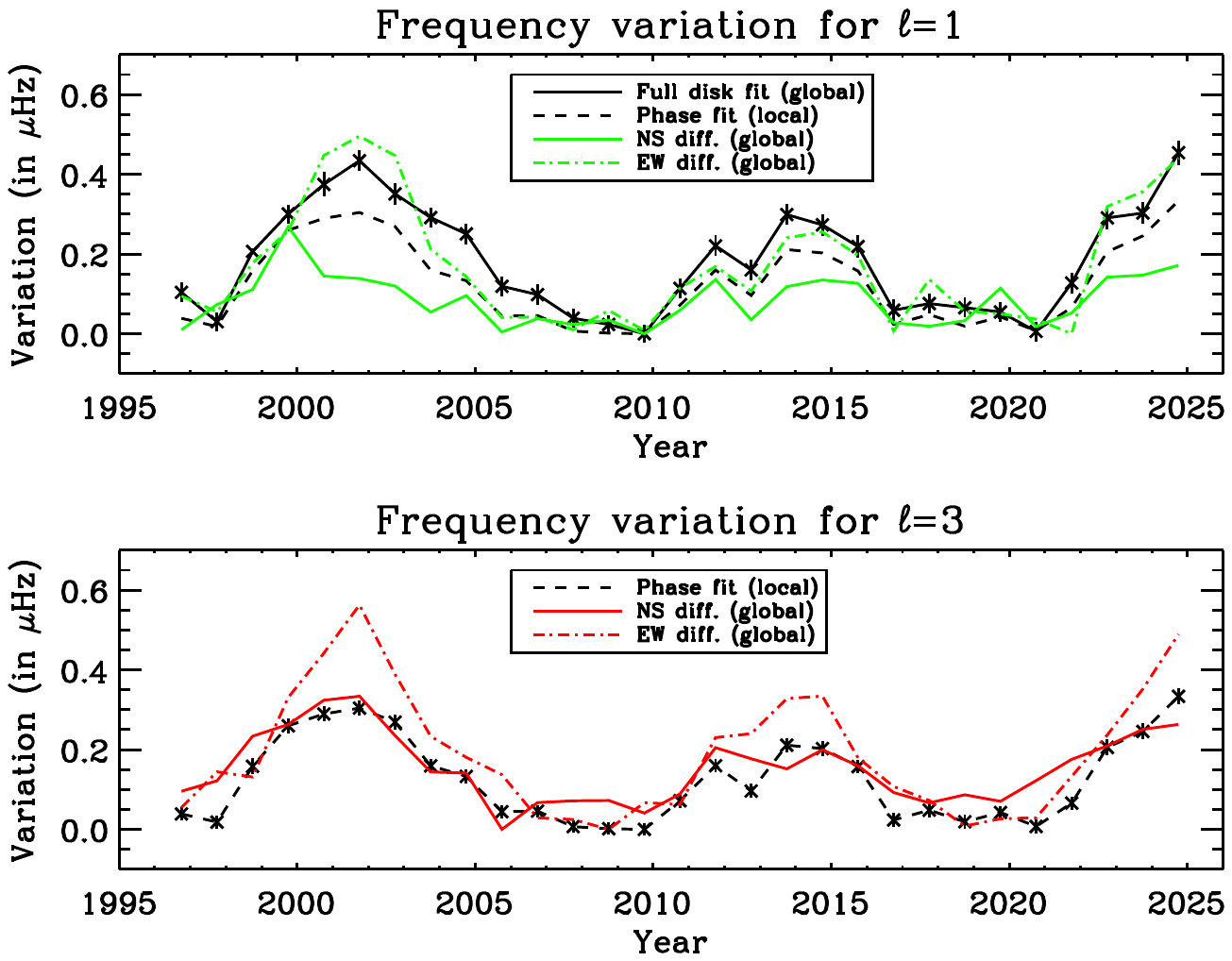}
      \caption{Mean mode frequency variation as a function of time. (Top, left) for the $l$=0 modes for the global fit (continuous line) and the local fit (dashed line). (Top, right) for the $l$=1 modes for the global fit (continuous line), the local phase fit (dashed line), the global fit of North-South difference (green line) and the global fit of East-West difference (green dashed line).  (Bottom, left) for the $l$=2 modes for the global fit (continuous line), the local phase fit (dashed line), the global fit of North-South difference (orange line) and the global fit of East-West difference (orange dashed line).  (Bottom, right) for the $l$=3 modes for the local phase fit (dashed line), the global fit of North-South difference (red line) and the global fit of East-West difference (red dashed line). { The $l=3$ modes are not shown because they are too faint to give a meaningful full-disk signals.}}
      \label{activity0123}
   \end{figure*}

              \begin{figure}[h]
   \centering
   \includegraphics[angle=180,width=9.cm]{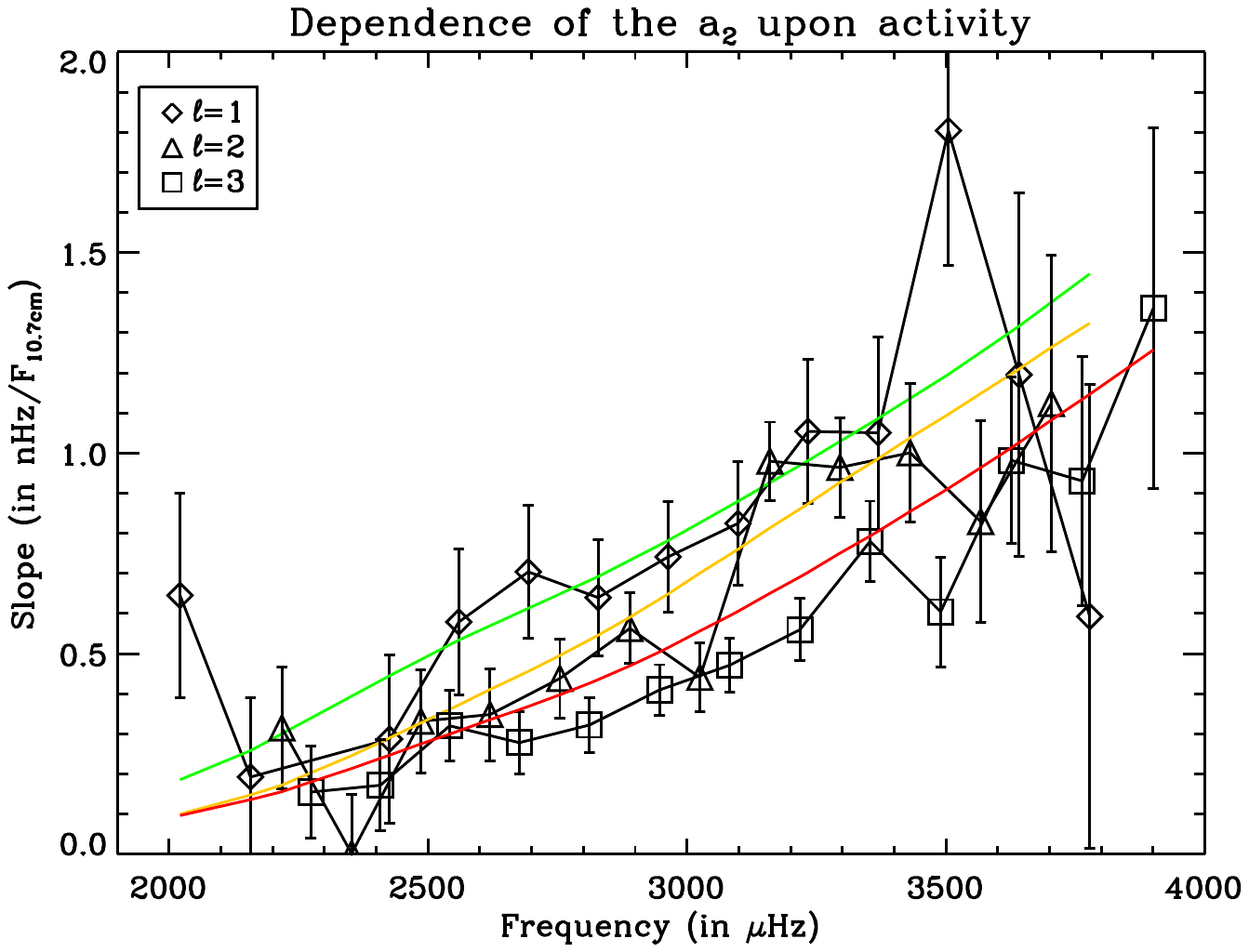}
      \caption{Slope of the dependence of the $a_2$-coefficient to the 10.7-cm flux as a function of frequency for degrees $l$=1, 2 and 3.  A second-order polynomial fit is shown for each degree: $l$=1 (green line), $l=2$ (orange line), $l=3$ (red line).  The data used were the resolved data locally fitted.}
      \label{activity_a2}
   \end{figure}

        \begin{figure*}[h]
   \centering
    \includegraphics[angle=180,width=9.cm]{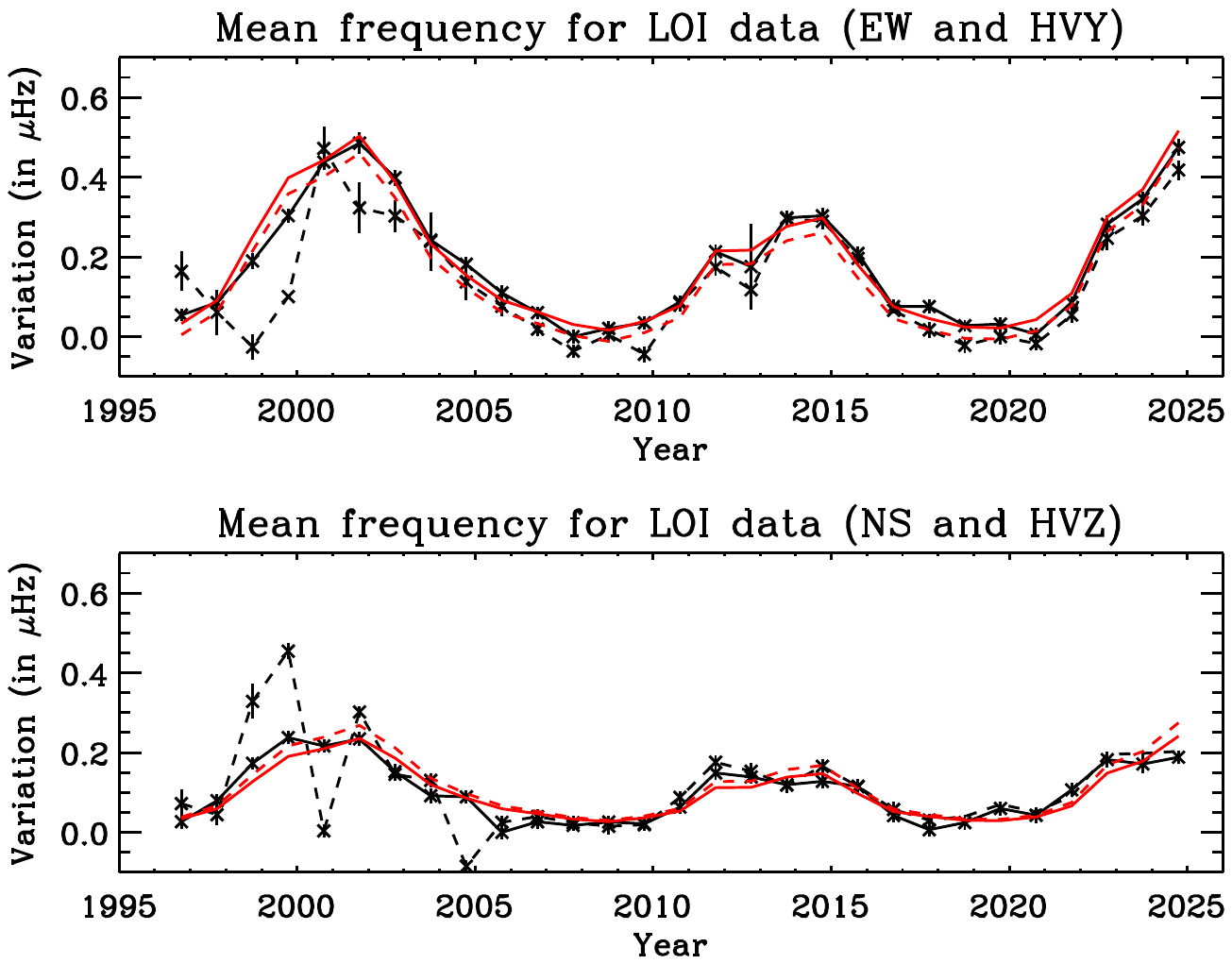}
   \includegraphics[angle=180,width=9.cm]{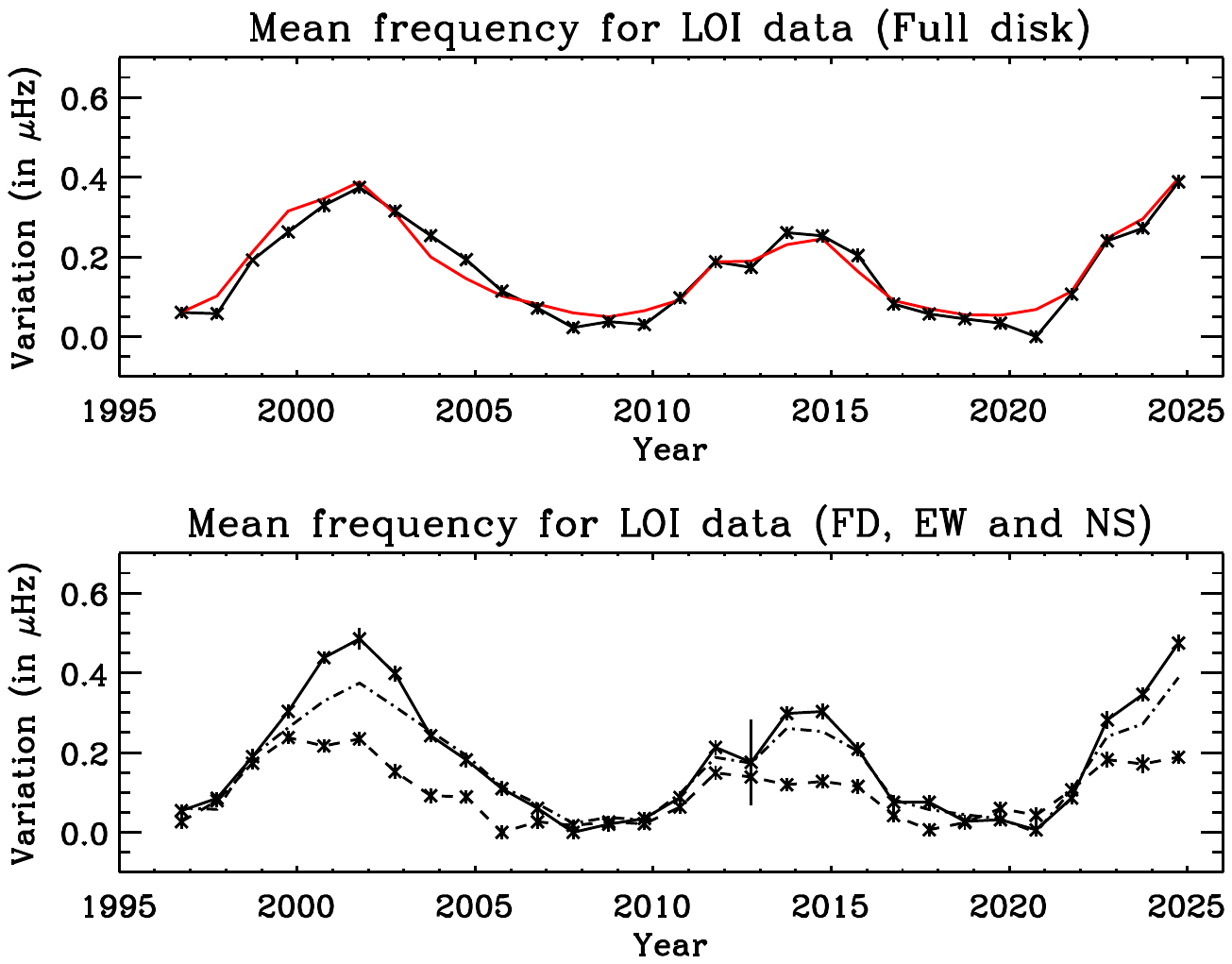}
           
      \caption{Mean mode frequency variation as a function of time.  (Top, left) averaged over the $l$=1 and 2 modes for the EW difference (continuous line), for the EW HV (dashed line), and the respective fits of the variations to the F10.7 cm flux (in red). (Top, right) averaged over the $l$=0,1 and 2 modes for the global fit of the FD data (continuous line) and the fit of the variations to the F10.7 cm flux (red line). (Bottom, left) averaged over the $l$=1 and 2 modes for the NS difference (continuous line), for the NS HV (dashed line), and the respective fits of the variations to the F10.7 cm flux (in red)  (Bottom, right) for comparison:  the averaged EW difference signal (continuous line),  the averaged NS difference signal (dashed line) and the averaged FD signal (dashed-dotted line).  }
      \label{activity_FDnsew}
   \end{figure*}

\section{p-mode parameters as a function of solar activity}
The dependence of the p-mode parameters upon solar activity was studied
against the flux at 10.7 cm.  This flux is the manifestation of the
existence of magnetic fields on the surface of the Sun.

The 10.7-cm flux could be downloaded at the SIDC (Solar Influence Data Analysis Center) in 2023 with a time series running from April 1st 1996 to March 31st, 2021; the data are at the time of writing no longer available.  The flux can now be found daily thanks to Space Weather Canada\footnote{spaceweather.gc.ca/forecast-prevision/solar-solaire/solarflux/sx-5-en.php}, and the two time series were merged into a single time series  from  April 1st 1996 to March 31st, 2025, for producing a yearly average.  Since I used the 29 yearly time series of the different signals for that study, I also averaged the flux over the same period of one year.  The 29 power spectra were fitted using a variable and a fixed noise background.  The two fits were used to reduce the bias on linewidths and mode heights that occurs when one uses a different background \citep[See for instance][]{TA2014}.

      \begin{figure}[!]
   \centering
   \includegraphics[angle=180,width=9.cm]{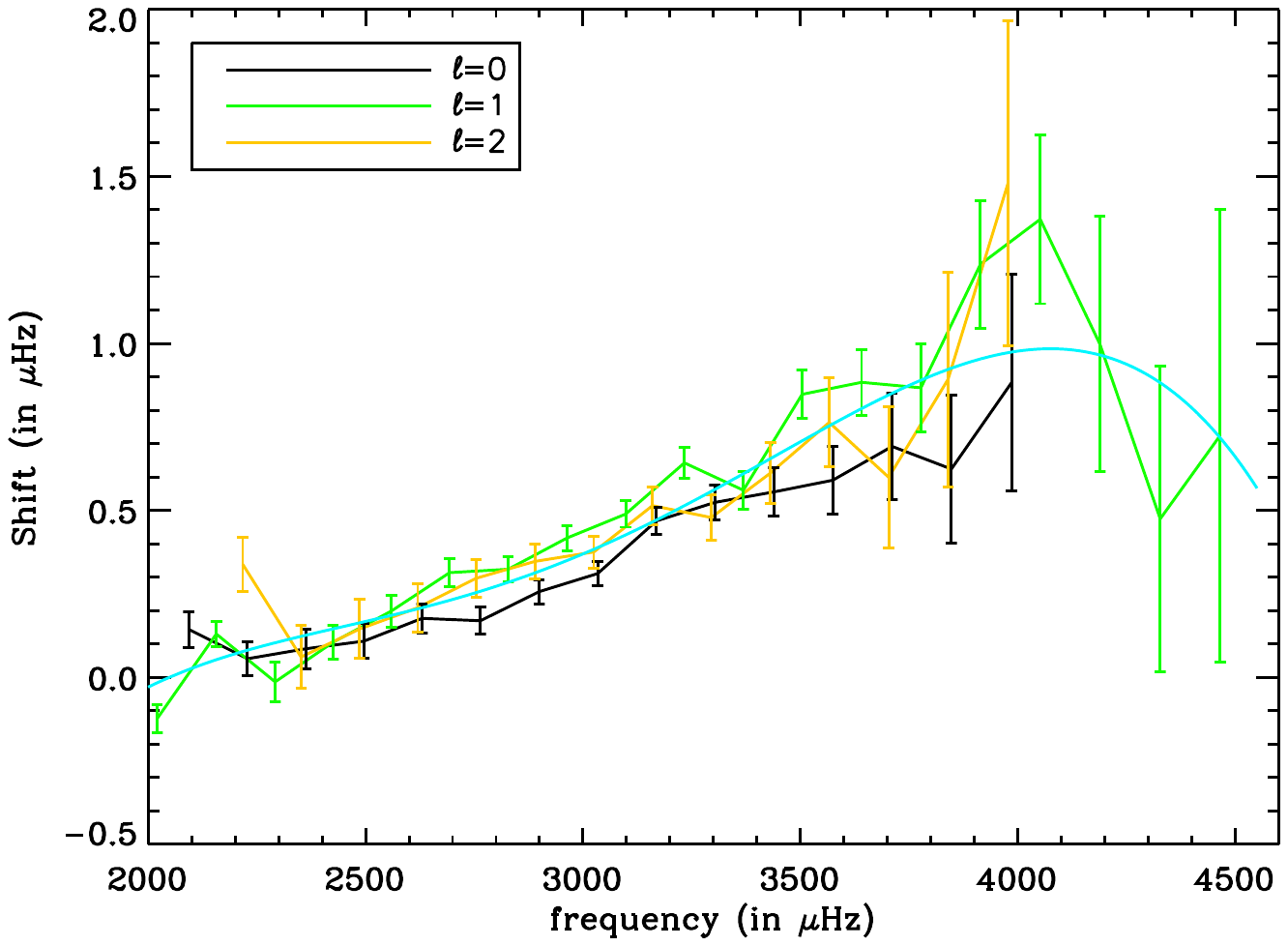}
         \caption{Frequency shifts as a function of frequency between maximum and minimum activity for various degree.  The cyan line is a fourth-order polynomial fit of the 3 shifts.}
                  \label{frequency_shift}
   \end{figure}

\subsection{Frequencies}

The impact of solar activity on the frequency and the various even $a$-coefficients of the splitting was first studied by \citet{KL90}.  The even coefficients are directly related to the decomposition of the magnetic field perturbation on even-order Legendre polynomials \citep{Antia2001}.  

{ For studying the frequency variation over the solar cycle, I used Eq. (5) of \citet{WJC2004} that includes a correction for the mode inertia and a function ${\cal F}(\nu)$ that measures the variation of the frequency shift over frequency, normalized to 1 at 3 $\mu$Hz.  This latter function was fitted from Figure 1 of \citet{WJC2004} as:
\begin{equation}
{\cal F}(\nu)=1.678-6.794(\nu_{3})+7.784(\nu_{3})^2-1.669(\nu_{3})^3
\end{equation}
where $(\nu_{3})$ is the frequency normalized to 3 $\mu$Hz.  The weighting as provided by their Eq. (5) is a way to relate all frequency variations to a single frequency of 3 $\mu$Hz, then ${\cal F}(\nu)$ can provide the frequency shift at any given frequency if needed.  The weighting scheme is used over the frequency range from 2400 $\mu$Hz to 3500 $\mu$Hz for obtaining the variations for each degree, and over all degrees up to $l=3$.}

     \begin{figure*}[h]
   \centering
   \includegraphics[angle=180,width=9.cm]{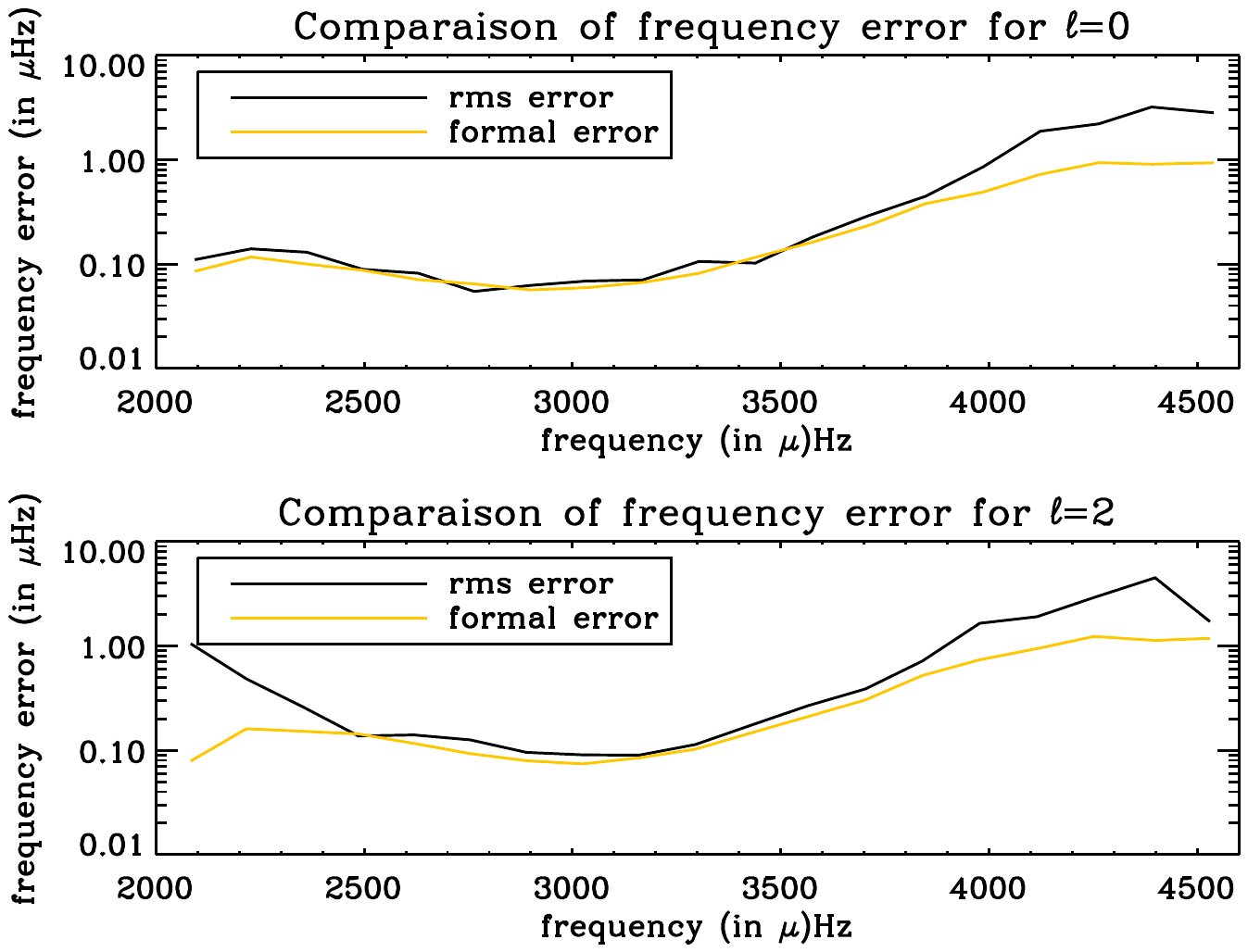}
   \includegraphics[angle=180,width=9cm]{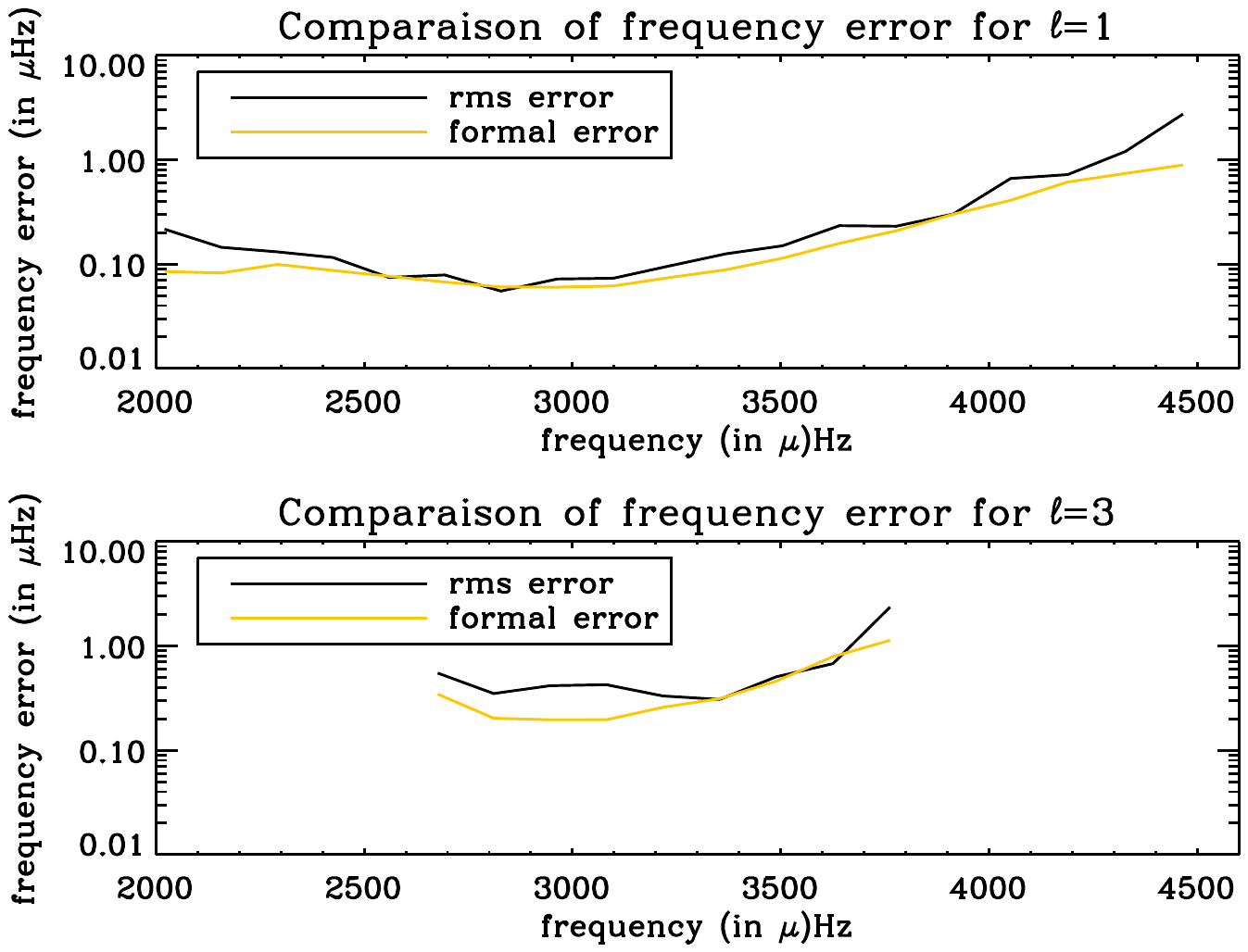}
      \caption{Frequency error as a function of frequency:  rms errors (black line), formal errors (orange line) for $l=0$ (Top, left), $l=1$ (Top, right), $l=2$ (Bottom, left) and $l=3$ (Bottom, right).}
         \label{comp_error}
   \end{figure*}

      \begin{figure}[h]
   \centering
   \includegraphics[angle=180,width=9.cm]{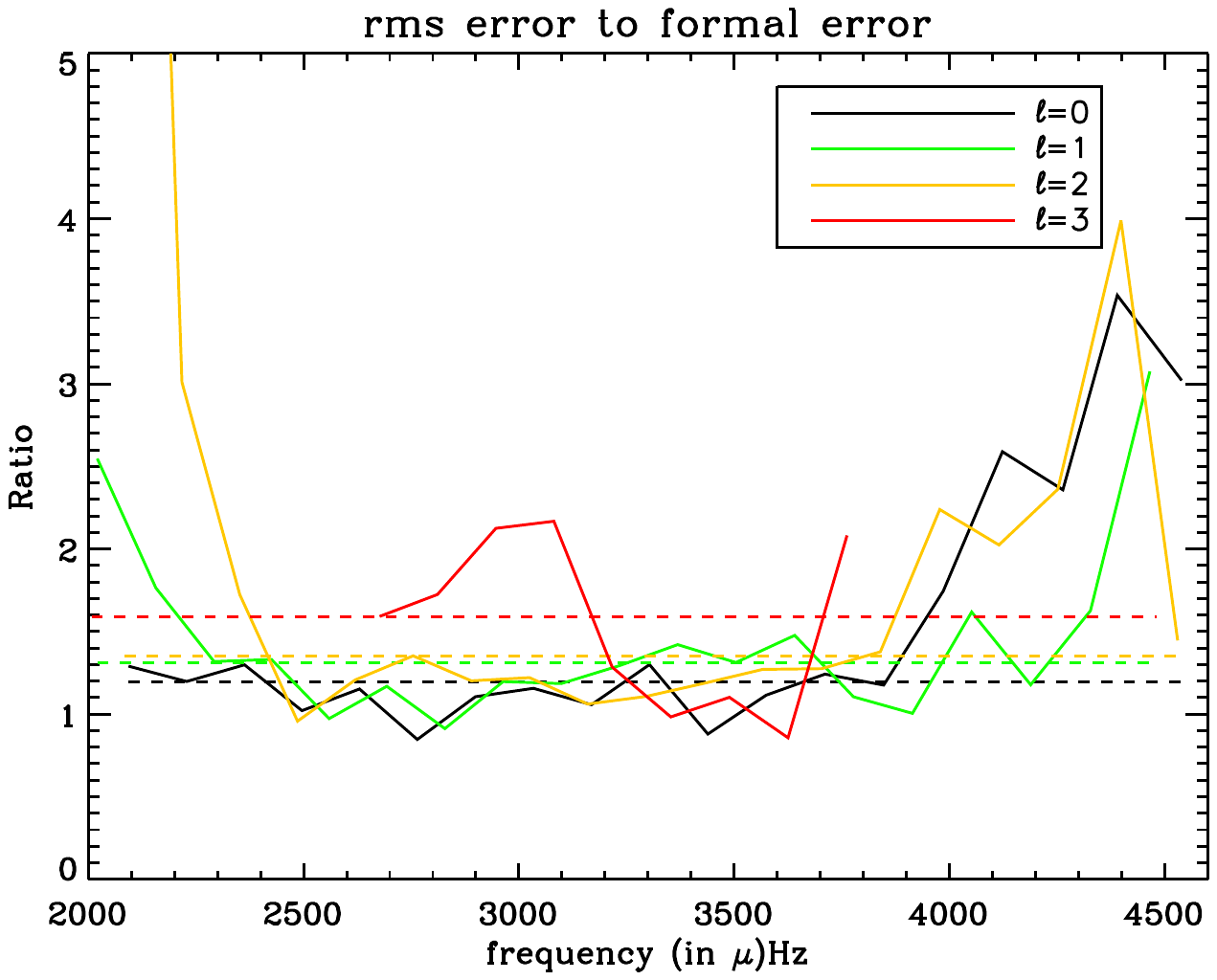}
         \caption{Ratio of rms error to the median formal error as a function of frequency for the $l$=0, 1, 2 and 3 modes observed in FD.}
          \label{comp_error1}
   \end{figure}
   
                       \begin{figure}[h]
   \centering
   \includegraphics[angle=180,width=9.cm]{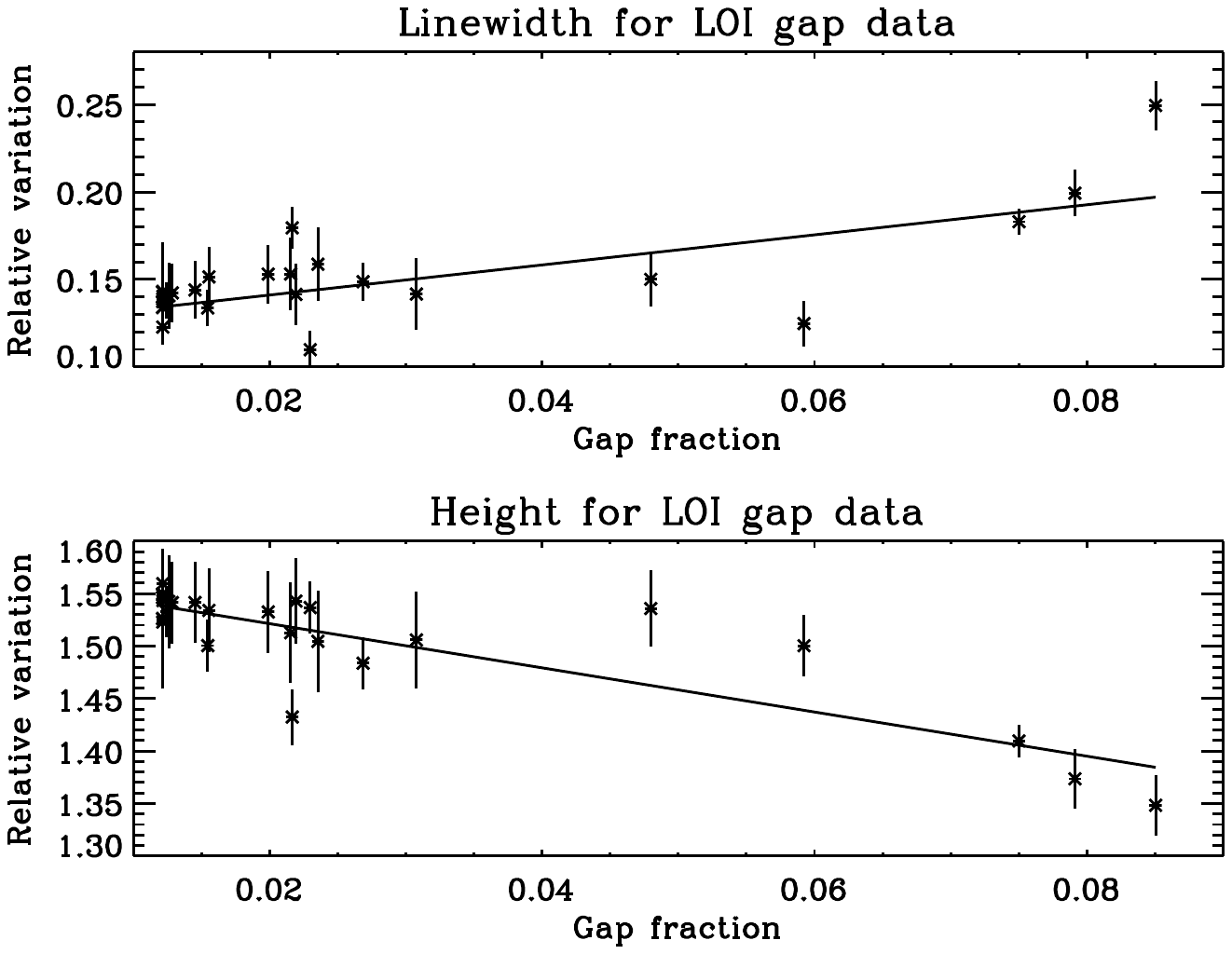}
      \caption{Log of mean mode linewidth (top) and mean mode height (bottom) as a function of gap fraction (crosses).  Linear fits are also shown (black line).}
      \label{linewidth_gap}
   \end{figure}
    
              \begin{figure}[h]
   \centering
   \includegraphics[angle=180,width=9.cm]{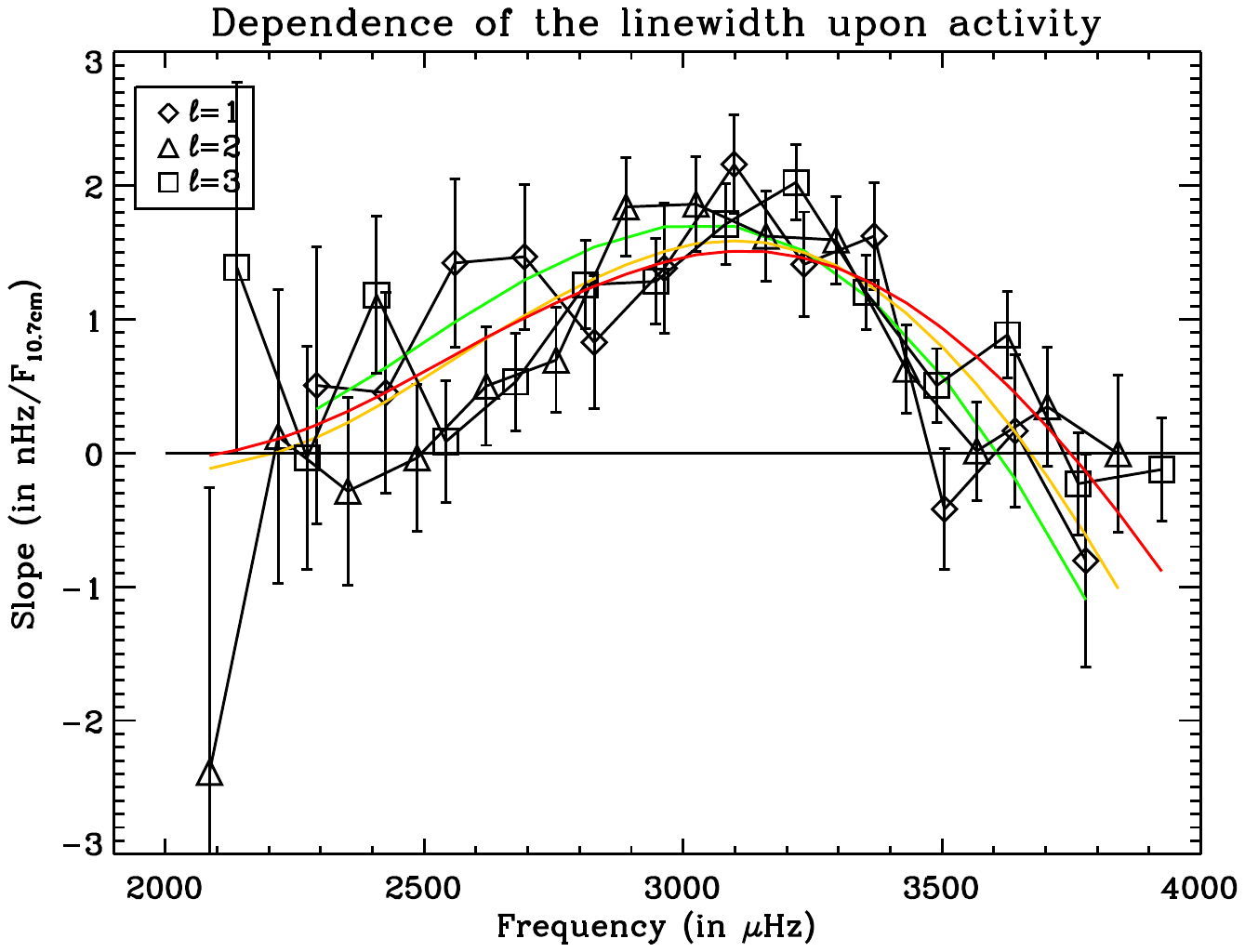}
      \caption{Slope of the dependence of the linewidth to the 10.7-cm flux as a function of frequency for degrees $l$=1, 2 and 3.  A 2nd-order polynomial fit is shown for each degree: $l$=1 (green line), $l=2$ (orange line), $l=3$ (red line).  The data used were the resolved data locally fitted.}
      \label{activity_linewidth}
   \end{figure}

  \begin{figure*}[h]
   \centering
   \includegraphics[angle=180,width=9.cm]{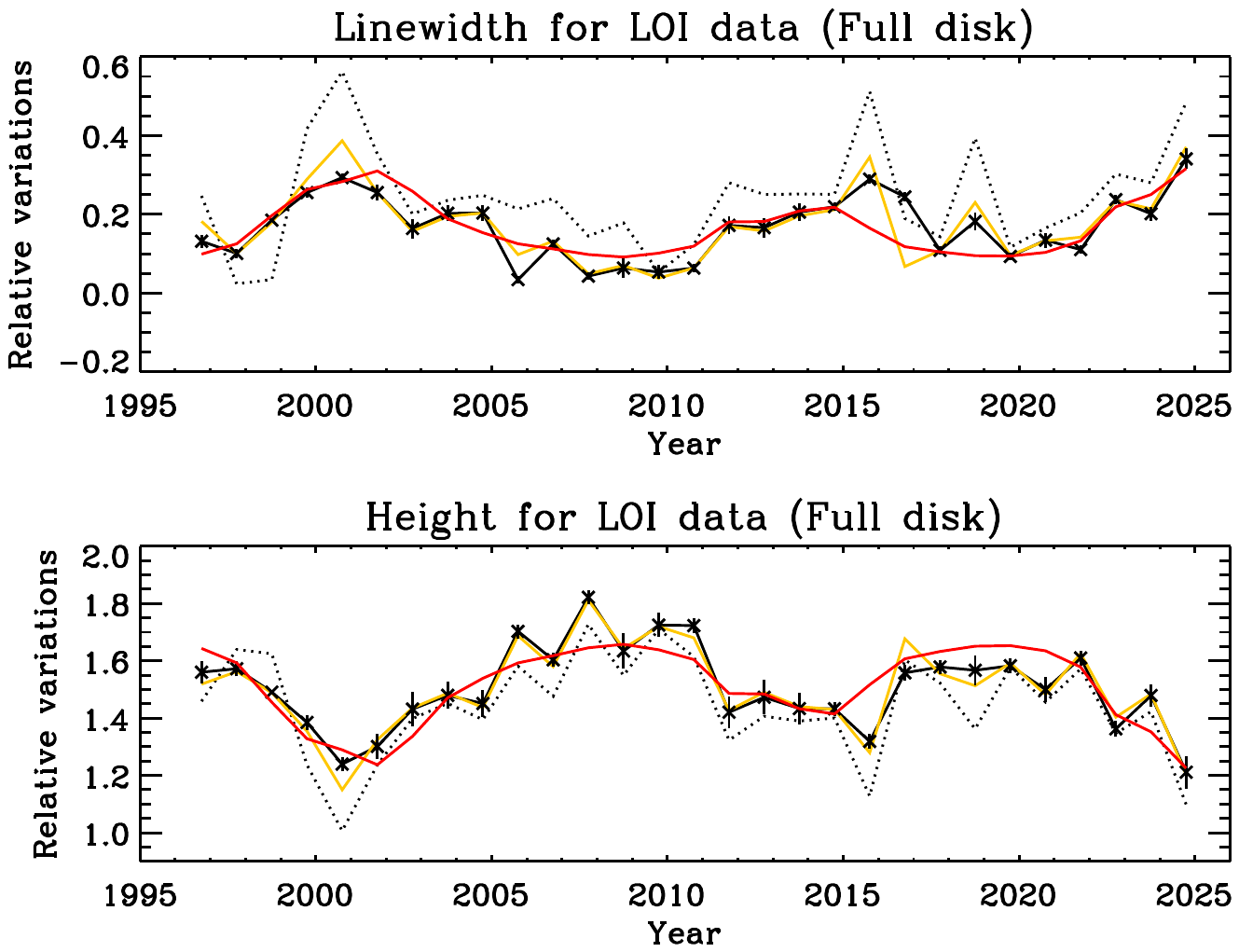}
                  \includegraphics[angle=180,width=9.cm]{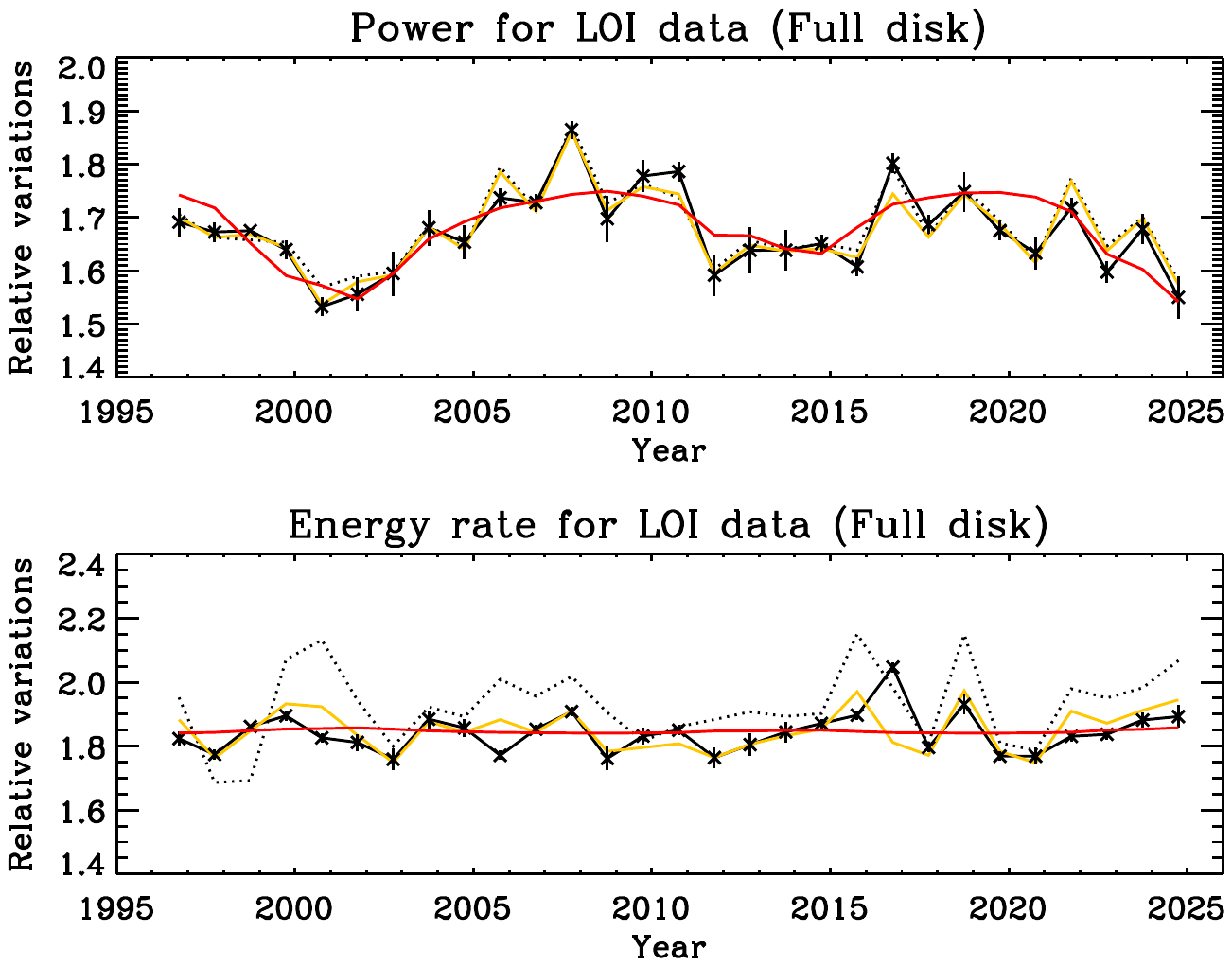}               
      \caption{Mode parameter variation as a function of time for linewidth (top, left), power (top, right), height (bottom, left) and energy rate (bottom, right).  The continuous { black} line is for the globally fitted data with a free background, while the dotted line is for a fixed background.  The correction of the parameters obtained for a fixed background is shown as an orange line; the correction is performed like in \citet{TA2014}.  { The red lines shows the linear fit of each parameter to the 10.7-cm flux.}}
      \label{linewidth_mode_height_activity}
   \end{figure*}

\subsubsection{Dependence on the degree and frequency}
Figure \ref{activity0123} shows the frequency shift for $l$=0, 1, 2 and 3, either fitted globally for the Sun-as-a-star signal and the difference signals of the resolved image (EW and NS) or locally with the resolved image.  The maximum frequency variation is typically 0.3 to 0.4 $\mu$Hz, depending on the degree.  For the $l=0$ mode, there is no difference between the local and the global fit.  For the other degrees, there are differences due to the impact of the $a_2$- and $a_4$-coefficients upon the frequency shift.  For $l$ different from 0, each signal has a different dependence upon the $a_2$- and $a_4$-coefficients.  The frequency shift for each signal depends upon the visibility of each $(l,m)$ component.  For the NS differences and the HV Z, the signals are primarily sensitive to the $m=0$ modes for $l \le 2$.  For the EW differences and the HV Y, the signals are primarily sensitive to the $m=\pm l$ modes for $l \le 2$.  For the FD signals, the sensitivity varies with the degree for which the dependence of the measured frequency was given by \citet{TA2007a}.  Table \ref{frequency_shift_a2} gives the frequency dependence for all these signals and degrees.  Figure \ref{activity0123} clearly shows the result of Table \ref{frequency_shift_a2}.  

Using the resolved data, I derived the dependence of the $a_2$-coefficient upon the 10.7-cm flux for the three different degrees.  Figure \ref{activity_a2} shows the slope of the linear fit of $a_2$ to the 10.7-cm flux.  Over the 29 years, the maximum variation of the $a_2$-coefficient is about 0.1 $\mu$Hz (with the impact of the $a_4$ being negligible) resulting in a difference between the EW- and NS-signal frequency shift of about 0.3 $\mu$Hz at the maximum of activity (See Fig.~\ref{activity0123} and Table \ref{frequency_shift_a2}).

\begin{table}[h]
  \caption{Frequency as a function of the degree and the signals. $ \nu^{res}_l$ is the frequency obtained using the phase fit of the resolved data.  For the FD signal and $l=2$, the dependence is derived from \citet{TA2007} for a linewidth of 1 $\mu$Hz.}
  \label{frequency_shift_a2}  

  \begin{center}
   \begin{tabular}{cccc}  
     \hline
Signal & $l=0$&  $l=1$ & $l=2$\\  
\hline    
EW / HV Y & - & $\nu^{res}_1+a_2$ & $\nu^{res}_2+2 a_2+2a_4$\\
Full disk &  $\nu^{res}_0$ & $\nu^{res}_1+a_2$ & $\nu^{res}_2+1.6 a_2+3a_4$\\
NS / HV Z & - & $\nu^{res}_1-2a_2$ & $\nu^{res}_2- a_2-8a_4$\\
      \hline
   \end{tabular}
  \end{center}
\end{table}

      \begin{figure}[h]
   \centering 
      \includegraphics[angle=180,width=9.cm]{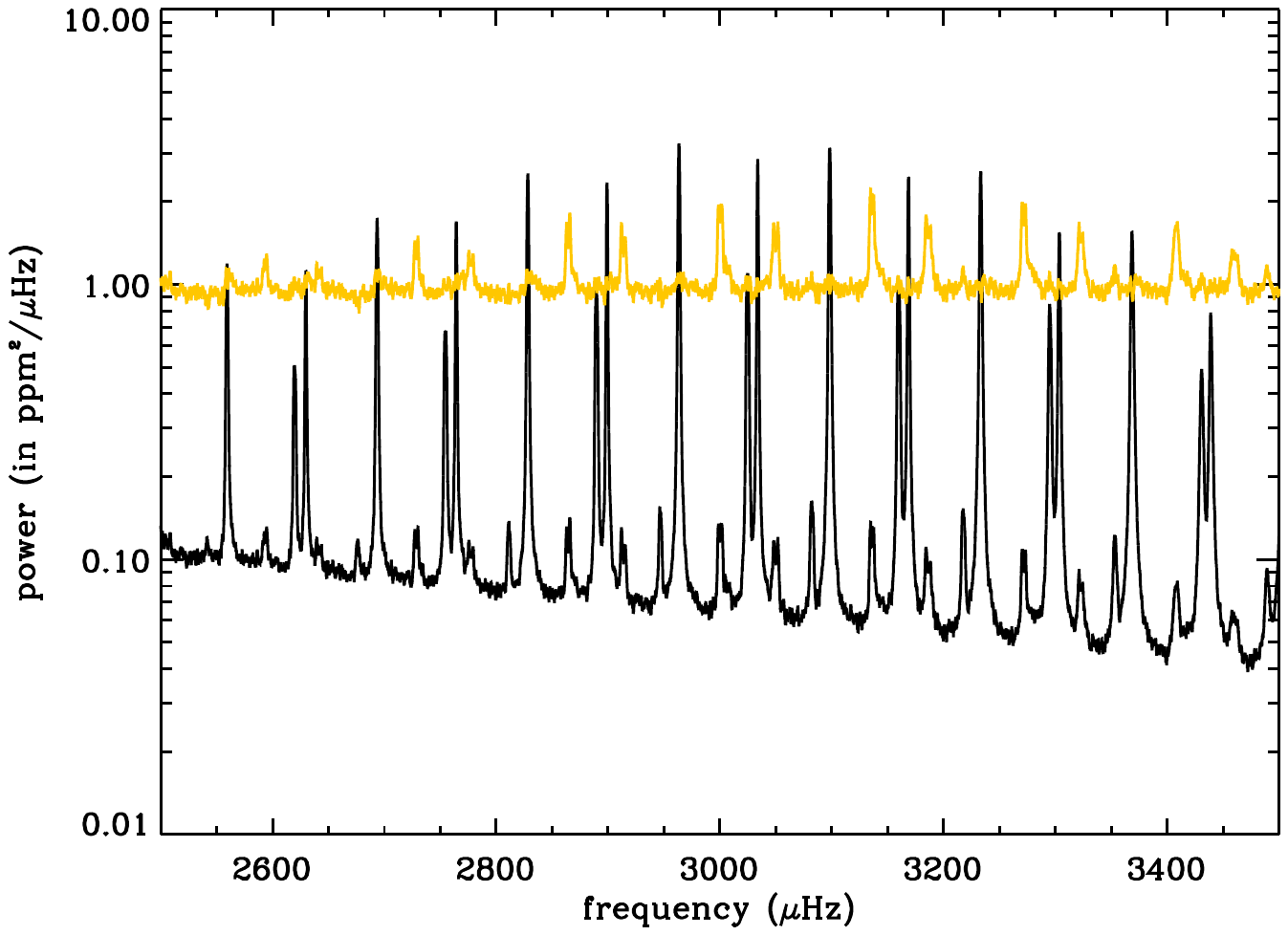}  
      \vspace{-1cm}    
      \caption{Power spectrum as a function of frequency for the 29 years (black line); ratio of the power spectrum to that of the fitted mode as a function of frequency (orange line) showing the remaining modes not included in the model.  { The remaining $l=4$ modes are located at the left hand side of the $l=0-2$ pair, while the remaining $l=5$ modes with a lower amplitude are located at the right hand side of the $l=0-2$ pair.  (See also Fig.~\ref{echelle})}}
      \label{power}
   \end{figure}

  \begin{figure}[h]
   \centering 
  \includegraphics[angle=180,width=9.cm]{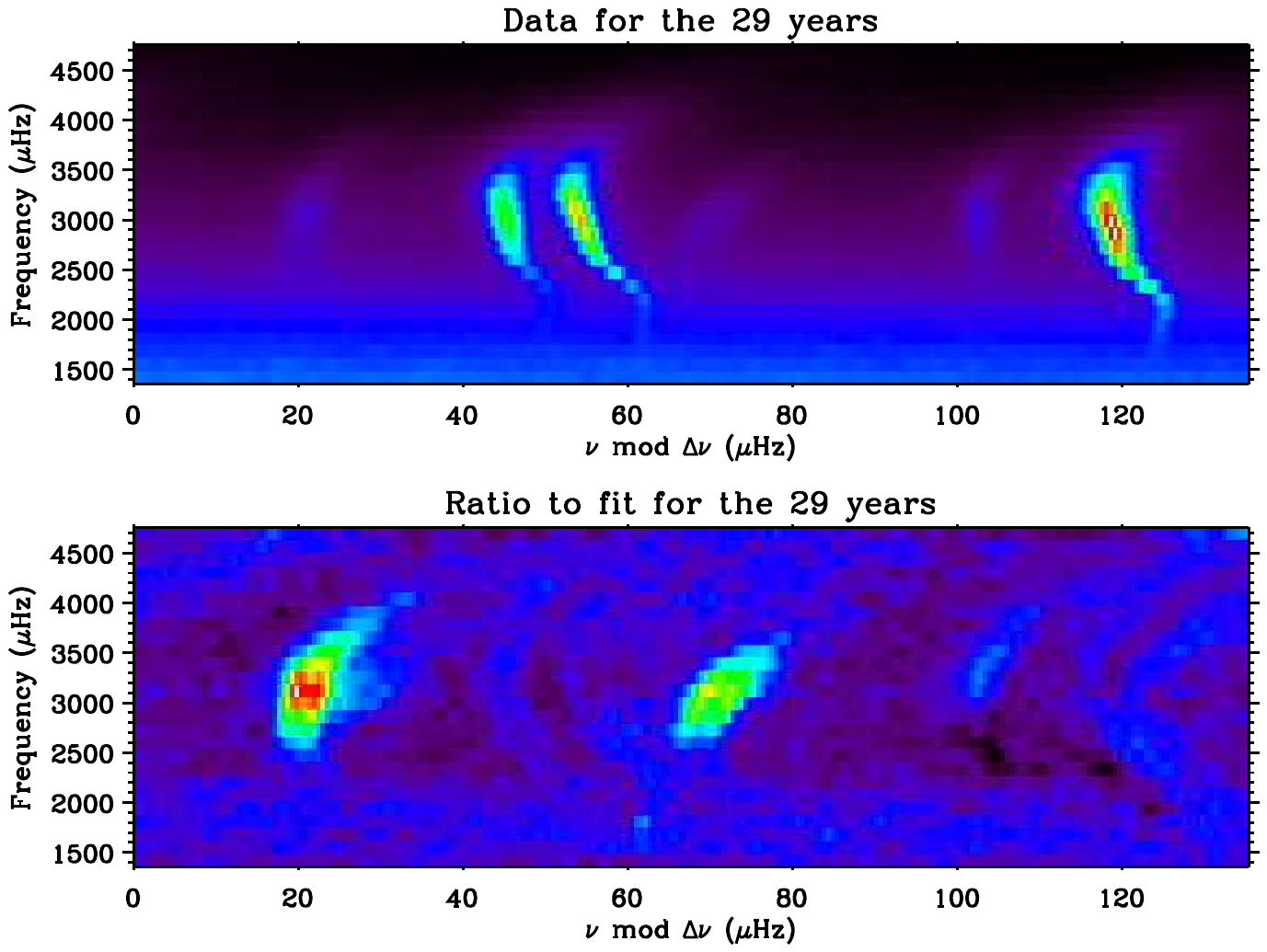}   
      \vspace{-1cm}    
      \caption{(Top) Echelle diagram of the power spectrum for the 29 years of LOI FD data.  The double ridges close to the centre are the $l=0-2$ modes, while the ridge at the right hand side are the $l=1$ modes, sided by the faint ridge of the $l=3$ modes. (Bottom) Echelle diagram of the ratio of the power spectrum to that of the fitted model for the 29 years.  The remaining ridge at the left hand side are the $l$=4 modes, with a faint ridge of the $l=7$ modes going through.  The remaining ridge at the center are the $l$=5 modes.}
      \label{echelle}
   \end{figure}

     \begin{figure}[h]
   \centering
   \includegraphics[angle=180,width=9.cm]{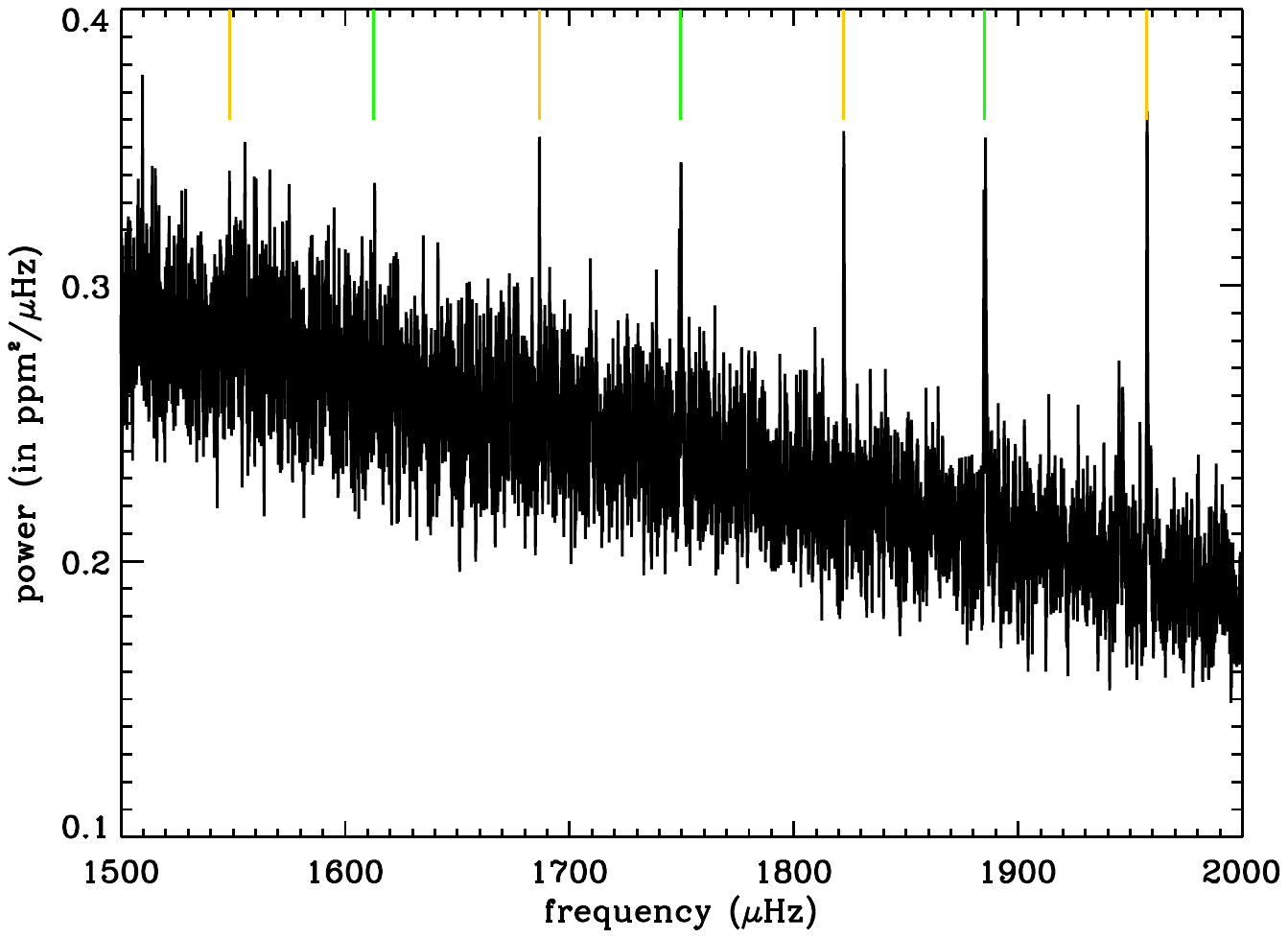}          
      \caption{power spectrum as a function of frequency for the 29 years at low frequency smoothed over 0.25 $\mu$Hz (black line); the $l=0$ and $l=1$ mode frequencies are indicated as vertical orange and green lines, respectively.  The $l=2$ and $l=3$ modes are fainter and not visible at low frequency.}
      \label{lowfreq}
   \end{figure}

           \begin{figure*}[h]
   \centering
   \includegraphics[angle=180,width=9.cm]{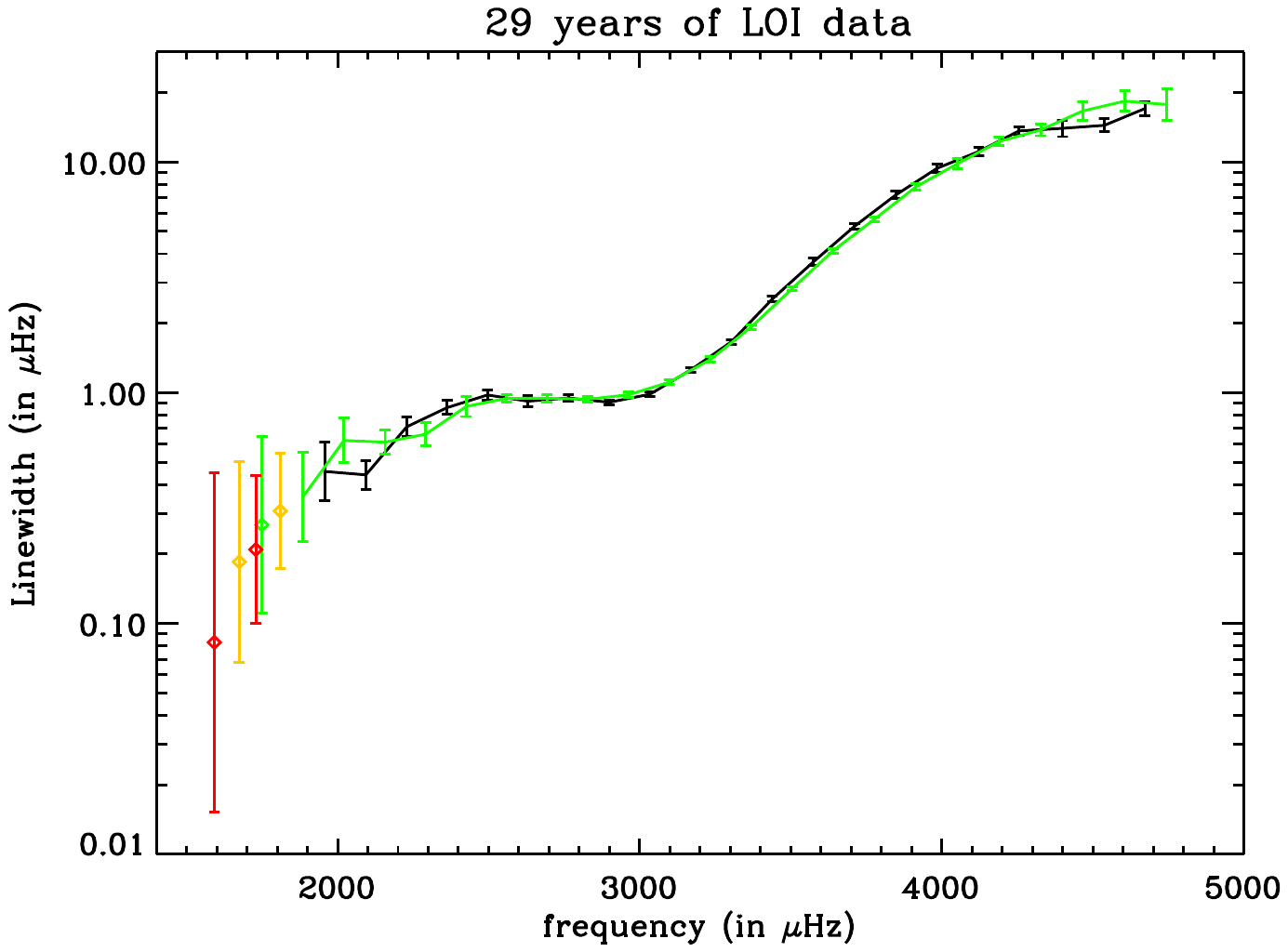}
            \includegraphics[angle=180,width=9.cm]{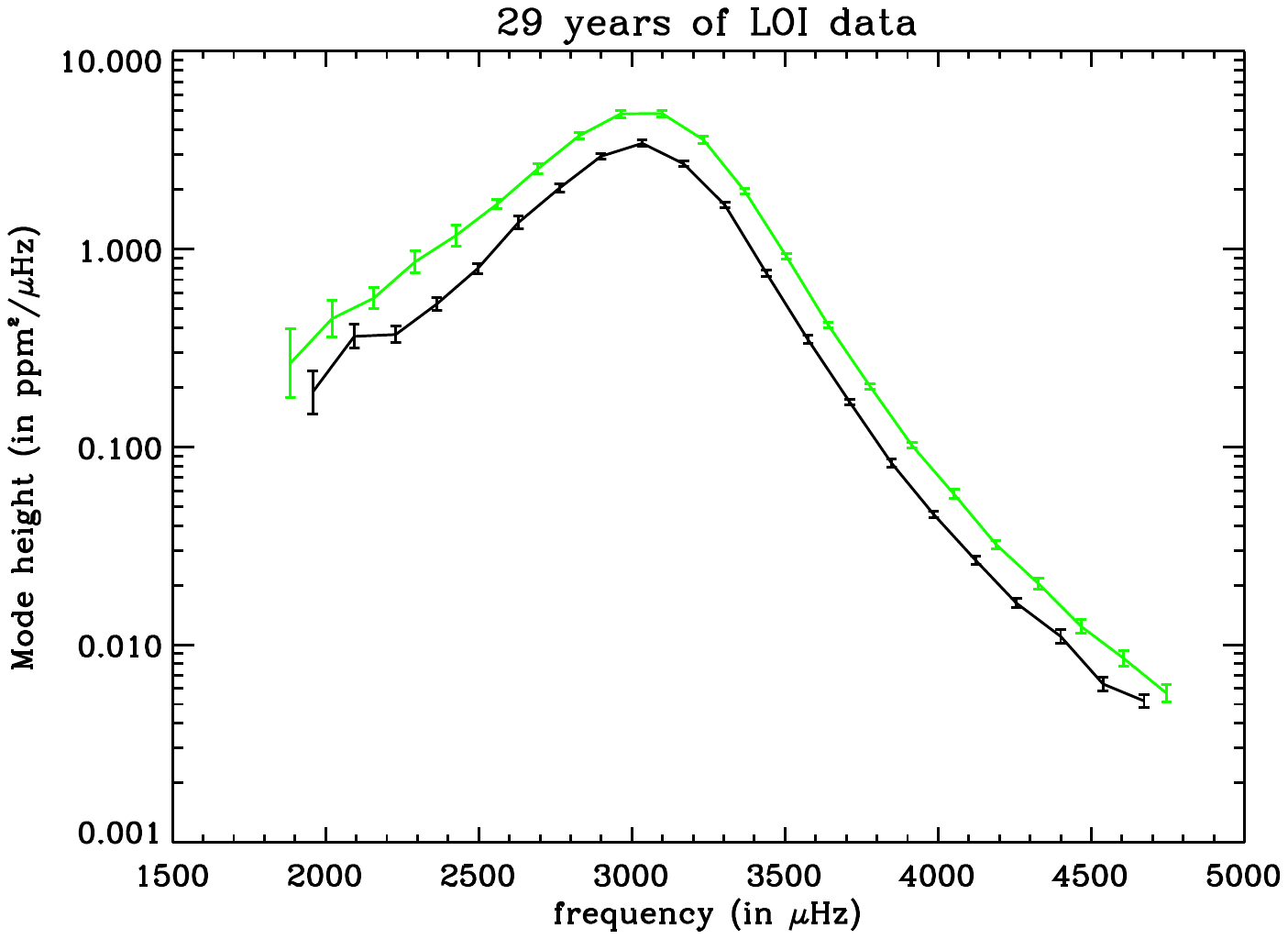}
      \caption{(Left) Mode linewidth as a function of frequency for the $l$=0-2 doublets (black line), for the $l=1$ modes (green line) obtained with the global fit of the power spectrum.  The linewidths at low frequencies for $l$=2 (orange diamond) and for $l$=1 (red diamond) were obtained using the collapsogram technique.(Right) Mode height as a function of frequency for the $l$=0-2 doublets (black line), for the $l=1$ modes (green line).  }
      \label{linewidth_mode_height}
   \end{figure*}   

\subsubsection{Degree and frequency averaging}

The averaging done over the frequency range can also be combined for averaging over the degrees.  Figure \ref{activity_FDnsew} shows the averaged frequency shift over various degrees for the FD signals, the NS signals (difference and HV Z) and the EW signals (difference and HVY).  It is clear that the difference in frequency shift is due to the different mode sensitivity whose variation with $m$ results in a different sensitivity to the solar activity belt.  In addition, the variations are the same for the NS / EW differences and the HV Z / HV Y signals (See Table \ref{frequency_shift_a2}).

The averaging over frequency and degree can also be done using a time-travel technique \citep{Vasilyev2024}.  Although their claims of being a {simpler} method than mode frequency fitting is matter of debate, they still find similar variations with solar activity for the mean mode time travel.

\subsubsection{Frequency error bars}
I also compared the formal error bars obtained with the global MLE estimation with that of the rms deviation of the individual frequencies.  Since the frequencies vary with the solar activity, I compensated the variations for each degree and the solar-activity level as given by the 10.7-cm flux.  After subtraction of the frequency shift which increases with frequency until 4 $\mu$Hz (See Figure \ref{frequency_shift}), I was able to infer the rms variations for each degree and frequency.  Figure \ref{comp_error} gives the rms and formal errors for each degree for the FD signal, while Figure \ref{comp_error1} gives the ratio of the rms errors to the formal errors for $l$=0 to 3.  The formal errors are typically smaller by 20\% to 30\% smaller than the rms errors.  Such underestimation was also measured for the 45-year of BiSON splittings \citep{Howe2023}.  This underestimation does not contradict the fact that the MLE errors are bounded on the low side by the Cramer-Rao theorem \citep{TALG98}.  The same bounding occurs with a Bayesian fit since in that case one must use the Van Trees \citep{van1968detection} inequality taking also into account the information prior.  The two inequalities are such that the Bayesian bound is lower than the classical MLE bound \citep[See also][ for a hands-on application]{Echeverria2016}.  When there is no information, the two bounds are identical.

\subsection{Linewidth and mode heights}
The variation of low-degree-mode linewidths and mode heights were reported by \citet{Chaplin2000} using BiSON data.  Similar variations were found by \citet{JR2003} for GOLF and VIRGO data.  \citet{Kiefer2018} reported
on the variation of these mode parameters for the GONG instruments over two solar cycles.  The increase in mode linewidth is related to the decreasing size of granulation that drives the excitation mechanism of the modes \citep{Kiefer2018}.  Such variations in granular size with solar activity were indeed measured by \citet{Ballot2021} but with a one-year delay.  Changes in linewidths using VIRGO/SPM data were also reported by \citet{TT2005a} during half a solar cycle.  Hereafter, I report on variations over two solar cycles taking into account the varying duty cycle in the data.

\subsubsection{Impact of gaps}
When measuring linewidths and mode heights, one has to take care about how the power is redistributed in the power spectrum.  When there are gaps in the data, the frequency bins become correlated \citep{MG94}.  The influence of such gaps in the power spectrum can be corrected by properly taking account the statistics as shown by \citet{Stahn2008}.  In practice, the power spectrum fitting is done assuming that the frequency bins are not correlated resulting in biasing the mode linewidth as shown by \citet{RK2000} for the GONG data, or by \citet{Chaplin2000} for the BiSON data.  For studying the impact of gaps on the mode linewidth and mode height, I created 29 times series using the first year as a reference, then applied the data gap structure of each time series to the reference time series, thereby creating 28 new time series.  I averaged the mode linewidth and mode height over the  frequency range from 2400 $\mu$Hz to 3500 $\mu$H (as for frequency), and over the degrees.  I did a straight average between over the frequency range, not using the weighting scheme of \citet{WJC2004}.  Figure \ref{linewidth_gap} provides the variation of the mean mode linewidth and mean mode height as a function of the gap fraction, together with a linear fit of the dependence.  On average for any 1\% additional gap, the mode linewidth is increased by 0.85 \%, while the mode height is reduced by 2.1 \%.  Values of the same order of magnitude were found by \citet{KH2019} for the GONG data.  The impact of gap on linewidth and height is the direct result of the dilution of power due the widening of the window function \citep{MG94}.  The dependence is used for compensating the effect of gaps on the mean mode linewidth and height (See Section 6.2.3).

\subsubsection{Linewidth dependence on the degree and frequency}
Using the resolved data, I inferred the dependence of the mode linewidth on the 10.7-cm flux as a function of frequency.  Then I computed a linear fit of the mode linewidth variation to the 10.7-cm flux.  Figure \ref{activity_linewidth} shows the slope as a function of frequency for $l$=1 to 3.  It confirms that the mode linewidth dependence to activity is different from that of the frequency dependence { (See Fig.~\ref{frequency_shift})}, confirming what \citet{Kiefer2018} found with the GONG data.

\subsubsection{Degree and frequency averaging}
I averaged the mode linewidth and mode height over { the frequency range [2400, 3500] $\mu$Hz}, and over the degrees.  I did a straight average between over the frequency range, not using the frequency weighting scheme of \citet{WJC2004}the var.  Figure \ref{linewidth_mode_height_activity} shows the variation obtained over the solar cycle for mode linewidth, height, power and energy rate.  In Figure \ref{linewidth_mode_height_activity}, I also show the variations of the linewidth measured with a fixed background but corrected for the effect of biases due to the different background and splitting; the procedure is the same as followed by \citet{TA2014}.  One can notice that the biases are indeed corrected by the procedure and that the variations follow the same trend with solar activity. The linewidth variations are correlated with the activity while the mode height and power are anti-correlated.  The energy rate is independent of solar activity.  The same results were observed by  \citet{AMB2015} and \citet{Kiefer2018} with the BiSON and GONG data, respectively.

\section{Results for the 29-year time series}
I used the single time series of 29 years of the FD data for getting robust estimate of mode frequencies, linewidths and heights.  Figure \ref{power} shows the power spectrum obtained with the 29 years of FD data together with the ratio of the power spectrum to the fitted model.  The regular repetition of the modes can be used to derive an {\'e}chelle diagram \citep{GG81}. Figure \ref{echelle} shows the {\'e}chelle diagram of power spectrum obtained with the 29 years of FD data together with the  {\'e}chelle diagram of the ratio of the power spectrum to the fitted model.  The fitted model was obtained with the (\texttt{progFIT} software.  Although that software is efficient at recovering nearly all the modes, modes with low signal-to-noise ratio and narrow linewidths are not fully recovered.  A Bayesian approach with the known location of these low frequency modes (derived from velocity instrument such as GOLF) is more efficient.  In addition, it has been demonstrated that there is an optimal way to smooth the power spectrum for increasing the detectability of low frequency modes \citep{Appourchaux2004}.  

Combining the collapsogram technique \citep{TA2024, Salabert2009} and the smoothing of the power spectrum, I could detect modes below 1800 $\mu$Hz (See Figure \ref{lowfreq}).  I used the collapsogram technique together with an optimization of the $a_1$-, $a_3$- and $a_5$-splitting coefficients.  The optimization is performed with a genetic algorithm already used by \cite{TA2020} taking into account the statistics of the maximum \citep[See][]{TA2024}.  After the optimization, the resulting collapsed power spectrum is fitted using MLE.  As shown by \citet{TA2003b}, the fitting of averaged power spectrum can be performed using the MLE technique for two degrees of freedom (d.o.f), but scaling of the error bars is required to take into account the number of d.o.f.   Table \ref{frequency_table_global} gives the frequencies obtained by globally fitting the power spectrum of the 29-year time series, while Table \ref{frequency_table_collapsogram} provides the frequencies obtained at low frequency using the collapsogram.  

Figure \ref{linewidth_mode_height} gives the mode linewidth and mode height obtained using the global fit and the collapsogram { (See Tables~\ref{linewidth_height_table_global0} and \ref{linewidth_height_table_global1})}.  Since the averaging is performed over nearly 2 solar cycles with varying linewidths and frequency shifting, the mode linewidths are slightly higher than at low activity.  The impact of the frequency shift upon the linewidth was studied by \citet{Chaplin2008} but not taking into account the variation of the linewidth; the typical overestimation they obtained over a solar cycle for the linewidth were less than 3\%.  For taking into account the change of all parameters over the solar cycle, I simulated the impact of the variations using the measured mode frequency, linewidth and height variations.  I created 29 yearly time series for a single p mode to obtain a time series spanning the two solar cycles.  { Each time series has { an injected} frequency shift according to Fig~\ref{frequency_shift} and a linewidth and height changes according to Fig~\ref{activity_linewidth}, all scaled to the 10.7-cm flux}.  I fitted the resulting power spectrum for four typical frequencies: 2500 $\mu$Hz, { 3000} $\mu$Hz, 3500 $\mu$Hz, {4000}  $\mu$Hz.  {While the resulting {fitted} frequency shift is the same as the {mean} injected frequency} shifts ({averaged over the 29 time series}), the mode linewidth variations are about 30\% larger, i.e. the linewidth are 5\% and 10\% larger, at 2500 / 3500 $\mu$Hz and {3000} $\mu$Hz; there is no impact at {4000} $\mu$Hz as there is no resulting mode linewidth variation (See Figure \ref{activity_linewidth}).  Therefore, the resulting simulated bias on linewidth is only related to the linewidth variation with a typical frequency dependence similar to that of Figure \ref{activity_linewidth}.  Despite this bias,  the mode linewidths are in good agreement with measurements obtained by BiSON \citep{Chaplin1997, Davies2014}, GOLF \citep{Gelly2002}, GONG \citep{Komm2000, Korzennik2023}, HMI \citep{Korzennik2023} and VIRGO/SPM \citep{Salabert2011a}. Note that the linewidths measured by \citet{Komm2000} with GONG are smaller than the other measurements above 3700 $\mu$Hz.

  \begin{figure*}
   \centering
   \includegraphics[angle=180,width=9.cm]{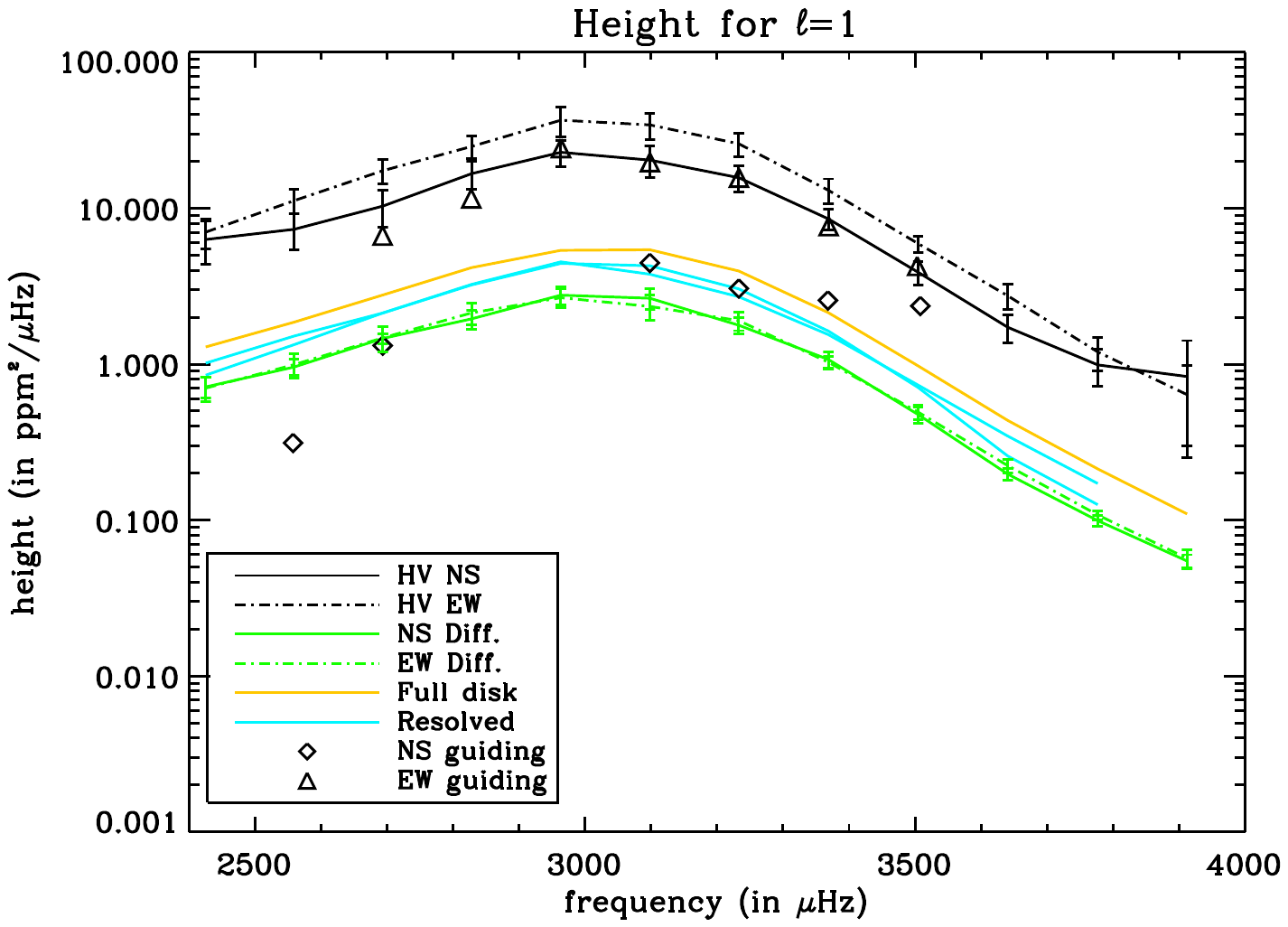}
   \includegraphics[angle=180,width=9.cm]{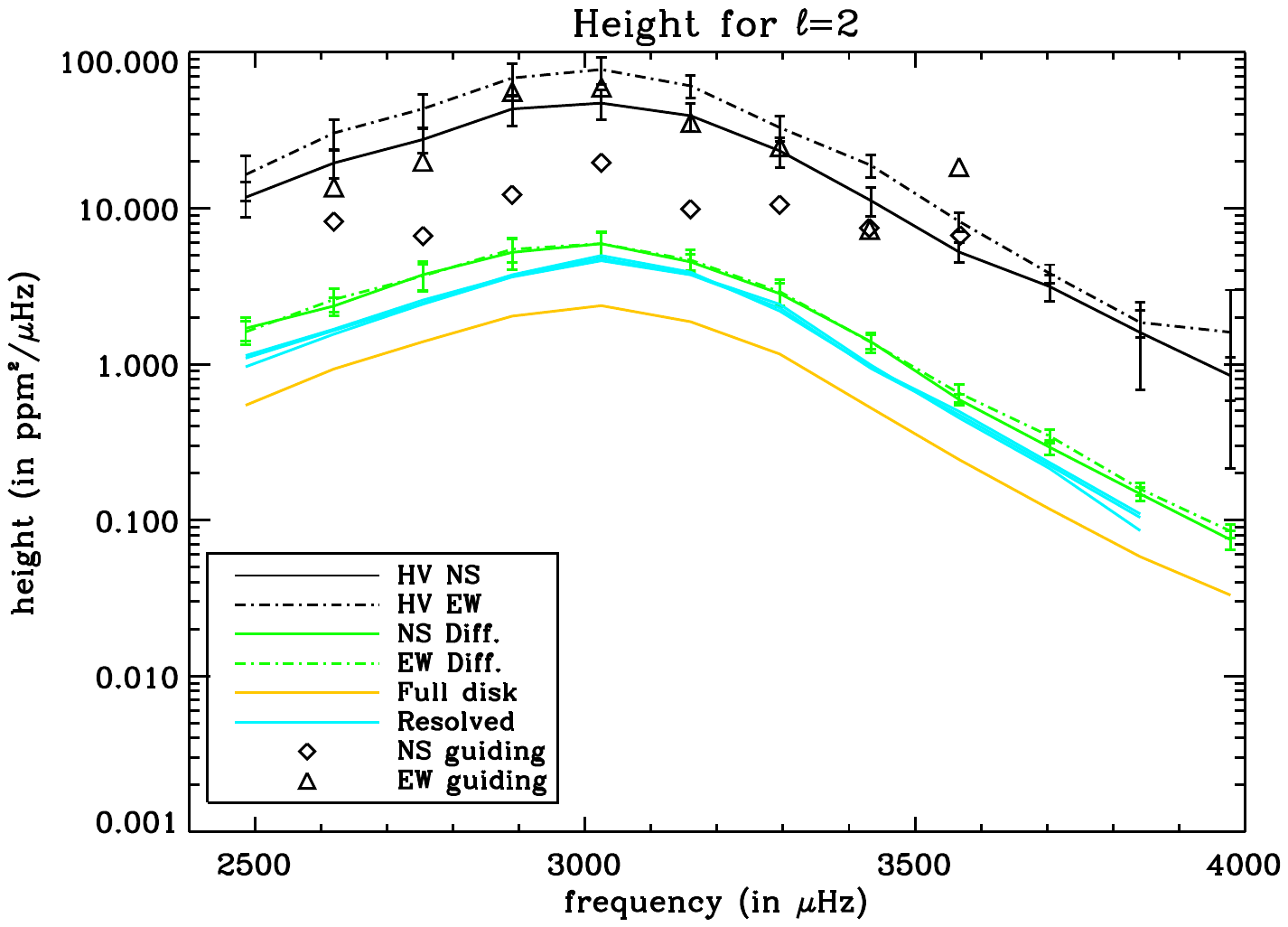}
      \caption{Mode height as a function of frequency for the $l$=1 modes (left hand side), and for the $l$=2 modes (right hand side) for different signals: North-South high voltage (continuous line), East-West high voltage (Dashed dotted line), North-South scientific pixels difference (green continuous line), East-West scientific pixels difference (green dash-dot line), FD (orange line), resolved (cyan line), North-South guiding pixels (diamond), East-West guiding pixels (triangle).  For the $l=1$ resolved data, the mode heights for $m=0$ are superimposed on those of the $m=\pm 1$ after correction of the visibilities.  For the $l=2$ resolved data, the mode heights for $m=0, \pm 1$ are superimposed on those of the $m=\pm 2$ after correction of the visibilities.}
      \label{amplitude_l12}
   \end{figure*}

  \begin{figure*}
   \centering
   \includegraphics[angle=180,width=9.cm]{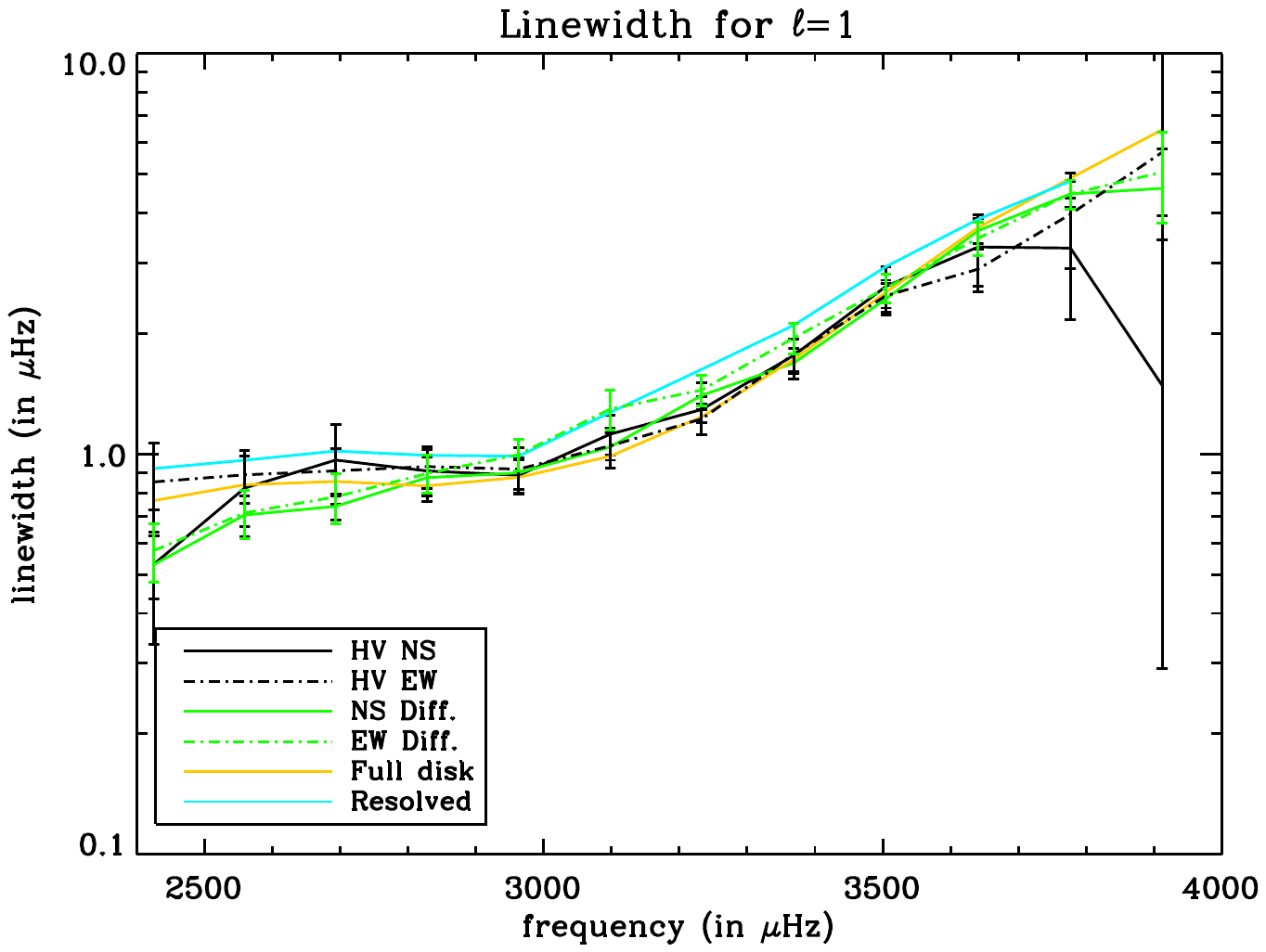}
   \includegraphics[angle=180,width=9.cm]{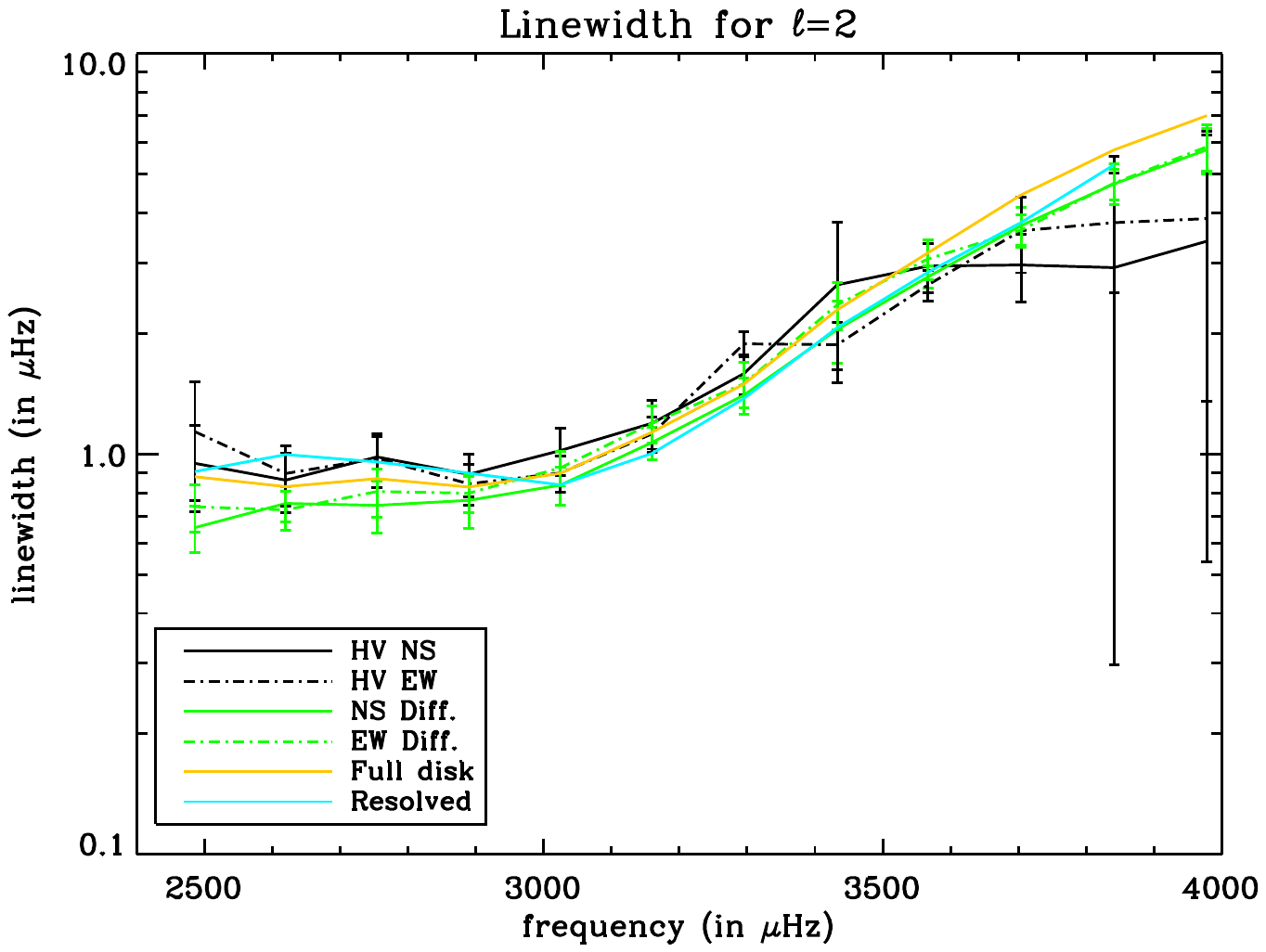}
      \caption{Mode linewidth as a function of frequency for the $l$=1 modes (left hand side), and for the $l$=2 modes (right hand side) for different signals: North-South high voltage (continuous line), East-West high voltage (Dashed dotted line), North-South scientific pixels difference (green continuous line), East-West scientific pixels difference (green dashed dotted line), FD (orange line), resolved (cyan line).}
      \label{linewidth_l12}
   \end{figure*}

\begin{table*}[h]
  \caption{Apparent radius of the outer dimension of the scientific pixels relative to the solar radius at 1 AU for two different limb darkening derived from fitting the ratio of pixel flux.}
  \label{LD_radius}  
  \begin{center}
   \begin{tabular}{cccccc}  
     \hline
Source & $u_2$ & $v_2$ &  radius (NS) & radius (EW) & radius (guiding)\\  
\hline    
 \citet{Allen1975} & 0.9700 & -0.2200 & 0.941 $\pm$ 0.014& 0.945 $\pm$ 0.014 &0.937 $\pm$ 0.006\\
 \citet{Neckel1994} & 0.9494 & -0.2080 & 0.943 $\pm$ 0.014& 0.947 $\pm$ 0.014 &0.938 $\pm$ 0.006\\
      \hline
   \end{tabular}
  \end{center}
\end{table*}

\section{Results for the guiding signals}
Figure \ref{amplitude_l12} shows the mode heights for $l=1$ and $l=2$ for the various guiding signals.  It is obvious that the mode heights for the HV signals are about 5 to 10 times larger than those of the FD and resolved signals.  On the other hand, the associated linewidths as shown in Figure \ref{linewidth_l12} are very similar, showing that the larger mode heights measured are not a result of a potential bias in the measured linewidths.  It was already obvious when comparing Figure~\ref{power-guiding} with respect to Figure~\ref{power-FD} that the mode height in the guiding signals was much than in the FD signals.  The higher mode heights shown on Figure \ref{amplitude_l12} were also observed in the MDI instrument by \citet{CT99}, the HMI instrument and the PICARD mission by \citet{Corbard2013}.  They all showed that the p-mode power increases towards the limb having power larger by a factor five,  for integration over an annulus with a width of 0.5\% R$_{\odot}$.  Their results agree with the ratio of mode heights reported.

\section{Visibilities and visibility ratios}
\subsection{Limb darkening and solar radius calibration}
The computation of the mode visibilities requires calibrating the limb darkening and the effective size of the solar radius on the LOI detector (See Appendix~ref{AppendixB).  The calibration was done by computing the ratio of the mean of the NS pixels (1, 2. 3, 4) to the Central pixels (9, 10, 11, 12), the ratio of the mean of the EW pixels (5, 6, 7, 8) to the Central pixels, and the ratio of the mean of the guiding pixels (12, 13, 14, 15)  as a function of the SoHO-Sun distance in AU.   The computation of the relative fluxes was performed using the shape of the pixels and the presence of the tracks as provided by the manufacturer\footnote{Advanced Micro Electronics, Horten, Norway}.  The tracks are non sensitive part of the detector required for getting the current resulting from the photoelectric effect.  The limb darkening is assumed to be expressed as:
\begin{equation}
I(\mu)=(1-u_2-v_2)+u_2 \mu + v_2 \mu^2
\end{equation}
where $\mu=\sqrt{1-r^2}$, and $r$ is the distance from the Sun disk center expressed in solar radius.

Figure \ref{plotratio} shows the ratios measured during the 29 years by plotting one point per month as a function of the Sun-SoHO distance, together with the fit of these ratios.  Unfortunately, this process does not allow to calibrate the limb darkening since there is a 100\% correlation between the apparent solar radius and the linear ($u_2$) and quadratic ($v_2$) terms of the limb darkening.  Instead, I used two sources for the limb darkening: \citet{Allen1975} and \citet{Neckel1994}.  For simplification, the limb darkening of \citet{Neckel1994} was reduced to that of the two terms above and interpolated at 500 nm.  Table \ref{LD_radius} gives the radius of the outer dimension of the scientific pixels with respectT to that of the Sun observed at 1 AU derived from the fit shown on Figure \ref{plotratio}.  This calibration is used for deriving the visibilities for the solar p modes observed either with the scientific pixels or with the guiding pixels.  I used the guiding pixel value of 0.937 $\pm$ 0.006 (Allen limb darkening).

\subsection{Visibilities for the different signals}
The theoretical visibilities of the p modes are directly related to the intensity perturbations which are related to the spherical harmonics and the limb darkening \citep[See][ and Appendix~\ref{AppendixB}]{TALG98}.
This is correct to a first approximation but \citet{Toutain99} showed that this approximation fails at the limb of the Sun because there the photons come from different heights in the solar atmosphere, thereby explaining the enhanced mode heights in the guiding pixels.  Further developments were carried by \citet{Kostogryz2021} that included other opacity contributions in addition to that of H$^{-}$. Developments that are used for understanding the difference between intensity and velocity measurements in HMI \citep{Fournier2025}.

Using the calibrated size of the guiding pixels with respect to the Sun radius at 1 AU, I derived the theoretical visibilities for the FD, the resolved, the EW and NS difference signals, the EW and NS HV signals and the EW and NS guiding pixels.  Then the theoretical ratios of visibilities are computed.

From the observations, I compute the average mode height for a given frequency over the 29 time series for all signals, as shown in Figure \ref{amplitude_l12}.  The ratio is then computed only for the modes spanning the frequency range [2500 $\mu$Hz, 3500 $\mu$Hz].  The ratios are then averaged over the modes for getting the mean ratio;  the error bars were deduced from the rms value divided by the square root of the number of frequencies.  Tables \ref{FD_visibilities}, \ref{resolved_visibilities},  \ref{EW_NS_visibilities}, \ref{gui_HV_EW_NS_visibilities} and \ref{gui_EW_NS_visibilities} give the comparison between the ratio averaged over the 29 time series and the theoretical ratio at the mean SoHO-Sun distance of 0.9906 AU.  

The radius of the outer boundary of the scientific pixels was also fitted in order to minimize the difference between the theoretical and observed FD ratios.  The adjusted value of 0.941~$\pm$~0.08 is very close to the value given in the previous section.  This coherence gives confidence in the value of the size and shape of the pixels, and in the understanding of the impact of the SoHO-Sun distance in the mode visibilities.

The FD visibility ratios shown in Table \ref{FD_visibilities} are in good agreement for all degrees but $l=3$.  The observed visibility ratio for $l=3$ is 3.7\% of the $l=1$ compared to a theoretical value of 2\%.  This is a large discrepancy that I did not study in detail but could be related to non-linearities due to the varying p angle (the inclination of the solar axis towards the observer).  The visibility ratios of Table \ref{FD_visibilities} are different from those of \citet{Ballot2011} mainly because of the presence of tracks in the FD signal of the LOI, not present in true Sun-as-a-star signals (the track size is about 1.6\% of the radius of the solar image).

The resolved visibility ratios shown in Table \ref{resolved_visibilities} are also in good agreement within a multiplet, while there is a systematic bias of about 19\% with respect to the FD visibilities.  The EW and NS difference signal visibility ratios shown in Table  \ref{EW_NS_visibilities} are in very good agreement showing no discrepancy.  

Finally, the visibility ratios for the EW and NS signals (guiding pixels and HV) are given in Table \ref{gui_HV_EW_NS_visibilities} showing very large discrepancies that are also clear in Figure \ref{amplitude_l12}.  The same discrepancy appears for the visibility ratios between the HV signals and the guiding pixels in Table \ref{gui_EW_NS_visibilities}.  One can notice that the theoretical visibilities for the EW and NS signals (guiding pixels and HV) for a given degree are identical, as demonstrated in Appendix~\ref{AppendixC}, then so are the ratios of visibilities.  I discuss the implication of these observed visibilities in the last section.

\section{Discussion and conclusion} 
In this article, I studied 29 years of observation of the LOI instrument aboard SoHO.  I corrected the impact of the so-called attractors that would lock the measured intensity flux to several digital values.  The flux intensities were also corrected to take into account the roll of SoHO due to the failed mechanism of the HGA.  Although the fluxes dropped by a factor 16, the power spectrum of the Sun-as-a-star time series did not drastically change between 1996 and 2004.  On the other hand, the power spectrum of the guiding pixels shows the solar p modes only in the 1996 time series.  After 1997, the noise is so high that the modes cannot be detected in the guiding pixels anymore.  Surprisingly, the HV signals for the EW and NS directions are much less noisy between 1996 and 2024, typically the noise signals became about four times lower making the detection of p modes much easier.   Using C++ codes developed for the PLATO mission, I globally fitted the power spectra of the various signals.

I studied the effect of solar activity upon mode frequency, mode linewidths and mode heights for various signals.   I confirm previous finding on the dependence of frequency with solar activity which is similar to the surface frequency effect.  These results are new, this is the first time that they are published for intensity measurements.  The impact of the $a_2$-coefficients upon the frequency shift of the different signals have an implication upon the measurement of the frequency shift in other stars \citep[Also studied by][]{Benomar2023}.  The NS difference or NS HV mimic the observations of star having a rotation axis inclined pole on (inclination angle of 0), while the EW difference or EW HV mimic that of a star observed with the rotation axis in the plane of the sky.  The difference between the frequency shift is larger for the EW signals compared to the NS signals.  For any other inclination angle the frequency shift will vary between these two extremes.  In other words, the level of activity as measured with the frequency shift depends upon the inclination angle.  Extreme care must be taken when drawing conclusions  on stellar activity with stars of similar age, radius and mass.  This may affect stellar activity studies performed with PLATO \citep{Rauer2025}.

I also studied the heights of the modes detected in the guiding pixels and HV signals.  The mode heights in the HV guiding signals are about 5-10 times larger than the modes observed in FD for $l=1$, while it is about 10-20s times larger for $l=2$ (after correction of the $l=2$ FD visibility).   Such higher mode heights are in agreement with other observations performed by \citet{CT99} and \citet{Corbard2013} with MDI, HMI and the PICARD mission.  On the other hand, the visibility ratios for the EW and NS signals (guiding pixels and HV) reported in this article are clearly not in agreement with the expectation of the simple theory provided in Appendix~\ref{AppendixC}.  There are several discrepancies with theory to be noted (See Table \ref{gui_HV_EW_NS_visibilities} and \ref{gui_EW_NS_visibilities}): 1) the mode visibilities of the EW signals are systematically higher than those of the NS signals; 2) the mode visibilities of the $l=2$ modes are significantly higher than those of the $l=1$ modes for the guiding pixels and the HV signals.  If one compares the results of the HV signals with the EW and NS difference (See Table~\ref{EW_NS_visibilities}), it is clear that for these latter signals 1) the mode visibilities in EW and NS are identical for $l=1$ and $l=2$; and 2) the ratio of mode visibilities of the $l=2$ to the $l=1$ modes are in agreement with theory.  It means that the intensity perturbation at the center of the disk ($\mu=1$) down to $\mu=\sqrt{1-0.941^2}=0.338$ follows a product of the limb darkening with a spherical harmonic (Eq.~C.1).  

When this is not the case, one can use the formalism given by \cite{Toutain99} for deriving the visibility ratio of $l=2$ with respect to $l=1$.  I show in Appendix~\ref{AppendixC} that the ratio of these visibilities for either the EW or NS signals is $(\Delta^{21}/\Delta^{11})^2$.  For spherical harmonics perturbations, this ratio is about 0.2 (See Table~\ref{gui_HV_EW_NS_visibilities}).  Using the relative intensity perturbation provided by \citet{Toutain99} in their Figures 5 and 6 for $l=2, m=1$ (used by $\Delta^{21}$) and for $l=1, m=1$ (used by $\Delta^{11}$), one can deduce that the ratio $(\Delta^{21}/\Delta^{11})^2$ should be much less than 1 as well.  The intensity perturbations provided by \citet{Toutain99} do not contradict the simple theory provided by Table~\ref{gui_HV_EW_NS_visibilities}.  The theory of \citet{Toutain99} was revised by \citet{Prokhorov2014} and \citet{Kostogryz2021}.  The latter used additional opacity in addition to the sole contribution of bound-free transitions of H$^{-}$ included in the model of \citet{Toutain99}.  Unfortunately,  \citet{Kostogryz2021} did not provide the intensity perturbations for $l=1$ and $l=2$.  The computation of such perturbation using the methodology of \citet{Kostogryz2021}  is beyond the scope of the current article.

This work may also serve future purposes for showing what can be achieved with a low resolution instrument detecting p modes in other stars.  Up to now, the observations of stellar p modes by CoROT \citep{Baglin2006}, Kepler \citep{Gilliland2010a}, TESS \citep{Ricker2014} and the coming PLATO \citep{Rauer2025} missions only allow the detection of low degree modes below $l=3$.  In the future, we hope to have large interferometers in space such as the concept of the Stellar Imager proposed by \citet{Carpenter2010}, which was aiming at providing about 30 resolution elements across a star ($l < 10$).  It would require the formation flying of 30 2-m diameter telescopes on a km baseline.  Such challenging developments are not impossible as a realistic road map for formation flying for space-based interferometry has been devised \citep{Monnier2019}.

Finally, this work will be used as the reference for the data reduction of the LOI instrument for which data will be archived at the European Space Astronomy Centre (ESAC) in Villafranca, Spain in the SOHO data archive.  At the time of writing, the  end of the SoHO mission is planned for September 2026 (D. M{\"u}ller, private communication).

\begin{acknowledgements}
      This article benefited from fruitful discussions I had with Bo Andersen, Antonio Jim\'enez and Torben Leifsen held during a memorable meeting in Svalbard.  This article represents the work performed during 3/4 of my career during which I owe a lot to several of my forefathers:  Jacques-Emile Blamont who introduced to the world of helioseismology and political maneuvering; Pierre Connes who introduced to the world of well-thought instrumentation, of spectroscopy and exoplanet search; David M.  Rust who had more faith and confidence in me that I will ever have which enabled me to be more aware of what I could do; Claus Fr{\"o}hlich who permitted me to be in charge and to develop the LOI instrument.  I owe a lot also to my career-long colleagues:  Bo N. Andersen who was at the start of the LOI concept, long standing friend whatever the circumstances; Antonio Jim\'enez still kicking the VIRGO Data Center and provider of the Level 0 data.  I owe also to the many ESTEC engineers that allowed me to design, to develop, to test the LOI especially Udo Telljohann.  Last but not least, I could not have done this career without the long time support of my wife, Maryse, there since the university years; my two sons, K\'evin and Thibault, and our many gone or still-living cats. A special mention to Roger-Maurice Bonnet, Alan Gabriel and Jean-Claude Vial for permitting me to go back to instrumentation development in France.  SoHO is a mission of international collaboration between ESA and 
      NASA.  The development of the (\texttt{progFIT} code was initiated by Bernard Leroy and Coralie Neiner at my prodding, code which was transformed into a regularly usable code with the major contribution of Claude Mercier, and the timely contribution of Vincent Buttice.  
\end{acknowledgements}

\bibliographystyle{aa}

\bibliography{thierrya}

\begin{appendix}

\section{Clebsh-Gordan expansion of the mode splitting}
\label{AppendixA}
In the early days of helioseismology (before the 1990's), mode 
splittings were usually expressed in term of Legendre polynomials.  
Unfortunately, these polynomials are orthogonal only on a continuous 
space (between [-1,1]) not on a discrete set such as (-1, 0,+1) for 
$l$=1, for example. Therefore, other expansion are required that can 
either be computed by hand or derived from quantum mechanics.  
\cite{RITZ91} derived these polynomials from quantum mechanics and 
used a normalization such that even
for high degree modes their $a_1$ are asymptotically equal to the 
usual Legendre polynomials.  It is
now common practice to have a normalization such that this is the 
case.  
The splittings are expressed as follows:

\begin{equation}
\nu_{(l,m)}-\nu_{(l,0)}=\sum_{i=1}^{i=n}a_{i} l {\cal{P}}_{l}^{i}(m)
\end{equation}
where $l$ is the degree, $m$ the order, and selected ${\cal{P}}_{l}^{i}(m)$ are given as follows:
\begin{equation}
{\cal{P}}_{l}^{1}(m)=\frac{m}{l}
\end{equation}
\begin{equation}
{\cal{P}}_{l}^{2}(m)=\frac{6 m^{2}-2L^2}{6 l^{2}-2L^2}
\end{equation}

\begin{equation}
{\cal{P}}_{l}^{3}(m)=\frac{20 m^{3}-4m(3L^2)-1)}{20 l^{3}-4l(3L^2-1)}
\end{equation}
\begin{equation}
{\cal{P}}_{l}^{4}(m)=\frac{70m^{4}-10m^{2}(6L^2-5)+6L^2(L^2-2)}{70l^{4}-10l^{2}(6L^2-5)+6L^2(L^2-2)}
\end{equation}
\begin{equation}
{\cal{P}}_{l}^{5}(m)=\frac{252m^5-140(2L^2-3)m^3+(20L^2(3L^2-10)+48)}{252l^5-140(2L^2-3)l^3+(20L^2(3L^2-10)+48)}
\end{equation}
{ with $L^2=l(l+1)$}.  Please note that for all $i$ we have ${\cal{P}}_{l}^{i}(l)=1$.  The 
polynomials are derived from Eqs (39) to (44) of \cite{RITZ91}. 

\section{LOI visibilities}
\label{AppendixB}
The LOI visibilities are
written as:
  \begin{eqnarray}                      
S_{nlm}=\sqrt{\frac{1}{N}\sum_{i=1}^{i=N}\tilde{y}_{i}^{(l,m)*}\tilde{y}_{i}^{(l,m)}}
\label{d13}
 \end{eqnarray}
where $N$ is the number of pixels, and $\tilde{y}_{i}^{(l,m)}$ are 
given by
 \begin{equation}               
\tilde{y}_{i}^{(l,m)}=\frac{\int_{{\cal{D}}_{i}}\sqrt{4 
\pi}Y_{l}^{m}(\theta,\phi)I(\sin \theta \cos \phi) \sin^2 
\theta \cos \phi {\rm d}\theta {\rm d}\phi}{ \int_{{\cal{D}}_{i}}I(\sin \theta 
\cos \phi) \sin^2 
\theta \cos \phi {\rm d}\theta d\phi}
    \label{ddsmall}
 \end{equation}
where ${\cal{D}}_{i}$ is the domain representing pixel $i$, 
$Y_{l}^{m}(\theta,\phi)$ 
is the spherical harmonic and $(\theta, \phi)$ are the spherical 
angular coordinates of a point on the surface of the Sun and $I$ is the limb darkening.

Please note that with the definition of Eq. (\ref{ddsmall}), 
$\tilde{y}_{i}^{l,m}=1$, and therefore the visibility of the LOI data 
for 
$l=0, m=0$ is simply $\sqrt{12/12}=1$.

\begin{table}[h]
  \caption{Visibilities of the LOI instrument}
  \label{diff}  

  \begin{center}
   \begin{tabular}{ccc}  
     \hline
$l$ & $m$ &  $S_{nlm}$\\  

\hline    
        0&0&1.000\\
        1&0&0.749\\
        1&1&1.069\\
        2&0&0.744\\
        2&1&0.877\\
        2&2&1.045\\
        3&0&0.695\\
        3&1&0.783\\
        3&2&0.857\\
        3&3&0.983\\
      \hline
   \end{tabular}
  \end{center}
\end{table}

\section{HV guiding signals and guiding pixel visibilities}
\label{AppendixC}
The HV guiding signals are proportional to the difference in the guiding pixel fluxes (EW or NS).  Therefore, for computing the ratio of visibilities for the
HV guiding signals, one has to compute the intensity perturbation obtained from the guiding pixels.

The reference axis for spherical harmonics is linked to the solar rotation axis, $x$ being in the direction of the observer, $z$ being the solar rotation axis
and $y$ making the $(x,y, z)$ axis system right-handed.  The corresponding spherical coordinates are $(r, \theta, \phi)$

Since we are trying to integrate the flux over the guiding detectors having a circular symmetry, we use the $(x', y, z')$ system with $z'$ pointing toward the observer.
A rotation by 90 degrees transforms the system $(x, y, z$) to $(x', y, z')$.  The associated spherical coordinates are then $(r', \theta', \phi')$.  In that framework,
the East and West guiding pixels are defined by $\phi' \in$ [-135,-45] and $\phi' \in$ [+45,+135], respectively; while the North and South guiding pixels are defined by 
by $\phi' \in$ [-45, +45] and $\phi' \in$ [+135,+225].

\subsection{Intensity perturbation}
Assuming that the intensity perturbation is proportional to the spherical harmonics, the perturbation integrated over the guiding pixels { for a given $(l,m)$}, can be written as:
\begin{equation}
\Delta I_l^m=\int_{\theta'} \int_{\phi'} Y_l^m(\theta,\phi) I_{LD}(\sin \theta \cos \theta) {\rm d}\Sigma
\end{equation}
where ${\rm d}\Sigma$ is the projected surface element, { $(\theta, \phi)$ are the spherical angular coordinates}, and $I_{LD}$ is the limb darkening.  The spherical harmonics expressed in the $(x, y, z)$  system can be decomposed using Edmonds rotation matrices { $D^{(l)}({\cal R})$} as a function of spherical harmonics expressed in the
$(x', y, z')$ system as:
\begin{equation}
Y_l^m(\theta,\phi)=\sum_{m'=-l}^{m'=+l} D^{(l)}_{mm'}({\cal R}) Y_l^{m'}(\theta',\phi')
\end{equation}
Since we also have { the following expression} :
\begin{equation}
Y_l^{m'}(\theta',\phi')=c_l^{m'} P_l^{m'} (\cos \theta') {\rm e}^{{\rm i} m' \phi'}
\end{equation}
We can rewrite Eq. (C.2) as:
\begin{equation}
Y_l^m(\theta,\phi)=\sum_{m'=-l}^{m'=+l} D^{(l)}_{mm'}({\cal R}) c_l^{m'}  P_l^{m'} (\cos \theta') {\rm e}^{{\rm i} m' \phi'}
\end{equation}

Using Eq. (C.3) we can then rewrite Eq. (C.1) as:

\begin{equation}
\Delta I_l^m=\int_{\theta'} \int_{\phi'} \left(\sum_{m'=-l}^{m'=+l} D^{(l)}_{mm'}({\cal R}) c_l^{m'}  P_l^{m'} (\cos \theta') {\rm e}^{i m' \phi'}\right)I_{LD}(\mu){\rm d}\Sigma
\end{equation}

The surface element can be written as ${\rm d}\Sigma=\mu {\rm d}\mu {\rm d}\phi'$, with $\mu=\cos \theta'$.  Equation (C.5) can then be separated in $\theta'$ and $\phi'$ as:
\begin{equation}
\Delta I_l^m=\sum_{m'=-l}^{m'=+l} D^{(l)}_{mm'}({\cal R})  \left(   \int_{\theta'}c_l^{m'} P_l^{m'} (\mu)I_{LD}(\mu) \mu {\rm d}\mu  \right) \left( \int_{\phi'} {\rm e}^{{\rm i} m' \phi'} {\rm d}\phi' \right)
\end{equation}

We can rewrite Eq. (C.6) as:
\begin{equation}
\Delta I_l^m=\sum_{m'=-l}^{m'=+l} [D^{(l)}_{mm'}({\cal R})]  \Delta^{lm'} \Delta J^{m'}
\end{equation}
with
\begin{equation}
\Delta J^{m'}=\int_{\phi'} {\rm e}^{{\rm i} m' \phi'} {\rm d}\phi'
\end{equation}
and:
\begin{equation}
 \Delta^{lm'}= \int_{\theta'}c_l^{m'} P_l^{m'} (\mu) I_{LD}(\mu) \mu {\rm d}\mu
\end{equation}
Since we have $P_{l'}^{-m'}(\mu)=(-1)^{m'} P_{l'}^{m'}(\mu)$, we have also $\Delta^{l'-m'}=(-1)^{m'} \Delta^{l'm'}$

Here I note that the perturbation used in Eq. (C.1) is supposed to be a spherical harmonic.  \citet{Toutain99} demonstrated that any perturbation can be decomposed using Edmonds rotations matrices.  As a result, Eq. (C.7) does not only apply to spherical harmonics but to any perturbation using spherical harmonics as a base, involving its derivatives, or any other combination.

\subsection{$\Delta J^{m'}$ for the E and W guiding pixels}
The guiding pixel integrals can be calculated for the E and W pixels (E: $\phi' \in$ [-135,-45] and W:$\phi' \in$ [+45,+135]):
\begin{equation}
\Delta J^{m'}_{E}=({\rm i} ^{m'}) \left(\frac{2\sin m' \pi/4} {m'}\right) 
\end{equation}
and 
\begin{equation}
\Delta J^{m'}_{W}=(-1)^{m'}) ({\rm i} ^{m'})   \left(\frac{2\sin m' \pi/4} {m'}\right) 
\end{equation}

\subsection{$\Delta J^{m'}$ for the N and S guiding pixel}
The guiding pixel integrals can be calculated for the N and S pixels (N: $\phi' \in$ [-45, +45] and S: $\phi' \in$ [+135,+225]):
\begin{equation}
\Delta J^{m'}_{N}=\left(\frac{2\sin m' \pi/4} {m'}\right) 
\end{equation}
and 
\begin{equation}
\Delta J^{m'}_{S}=(-1)^{m'}  \left(\frac{2\sin m' \pi/4} {m'}\right) 
\end{equation}

\subsection{Rotation matrices $D^{(l)}({\cal R})$}

\begin{equation}
D^{(1)}({\cal R})=\begin{pmatrix}
1/2 &\sqrt{2}/2& 1/2 \\
-\sqrt{2}/2&0&\sqrt{2}/2\\
1/2&-\sqrt{2}/2& 1/2
\end{pmatrix}
\end{equation}

\begin{equation}
D^{(2)}({\cal R})=\begin{pmatrix}
1/4 & 1/2 &\sqrt{3/8}& 1/2&  1/4\\
-1/2 &-1/2 &0&1/2 &1/2 \\
\sqrt{3/8}&0&-1/2 &0&\sqrt{3/8}\\
-1/2 &1/2 &0&-1/2&1/2 \\
1/4& -1/2 &\sqrt{3/8}&-1/2&  1/4
\end{pmatrix}
\end{equation}
\begin{equation}
\tiny
D^{(3)}({\cal R})=\begin{pmatrix}
\frac{1}{8} & \sqrt{\frac{3}{32}} &\frac{\sqrt{15}}{8}& \frac{\sqrt{5}}{4}&\frac{\sqrt{15}}{8}&   \sqrt{\frac{3}{32}}&\frac{1}{8}\\
\sqrt{\frac{3}{32}} &\frac{1}{2}&\sqrt{\frac{5}{32}}& 0&\sqrt{\frac{5}{32}}&  \frac{1}{2}& \sqrt{\frac{3}{32}}\\
\frac{\sqrt{15}}{8} &\sqrt{\frac{5}{32}}&-\frac{1}{8} & \frac{\sqrt{3}}{4} &-\frac{1}{8}&\sqrt{\frac{5}{32}}&\frac{\sqrt{15}}{8} \\
\frac{\sqrt{5}}{4}&0& \frac{\sqrt{3}}{4} &0&\frac{\sqrt{3}}{4} & 0& \frac{\sqrt{5}}{4}\\
\frac{\sqrt{15}}{8} &-\sqrt{\frac{5}{32}}&-\frac{1}{8}& \frac{\sqrt{3}}{4} &-\frac{1}{8}&\sqrt{\frac{5}{32}}&\frac{\sqrt{15}}{8}\\
\sqrt{\frac{3}{32}}&\frac{1}{2}&-\sqrt{\frac{5}{32}}& 0&\sqrt{\frac{5}{32}}&-\frac{1}{2}& \sqrt{\frac{3}{32}}\\
\frac{1}{8} & -\sqrt{\frac{3}{32}} &\frac{\sqrt{15}}{8}& -\frac{\sqrt{5}}{4}&\frac{\sqrt{15}}{8}&  - \sqrt{\frac{3}{32}}&\frac{1}{8}\\
\end{pmatrix}
\end{equation}

\subsection{$\Delta I_l^m$ for the HV voltage}
The HV signals are proportional to the difference in the intensity between
the E (N) and W (S) pixels.  Then we can write:
\begin{equation}
\Delta I_l^m({\rm HV}_{{\rm EW}})=\Delta I_l^m({\rm E})-\Delta I_l^m({\rm W})
\end{equation}
and
\begin{equation}
\Delta I_l^m({\rm HV}_{{\rm NS}})=\Delta I_l^m({\rm N})-\Delta I_l^m({\rm S})
\end{equation}

Using the rotation matrices given above, Eq. (C.10) to (C.13) and Eq. (C.7), Tables (C.1) to (C.3)
provides the results for Eq. (C.17) and (C.18).  Table (C.4) gives the resulting visibilities in power after summing over 
all components in a multiplet.

\begin{table}[h]
  \caption{Calculated $l=1$ $\Delta I_l^m$ response for the HV guiding voltage.}
  \label{gui_EW_NS_visibilities_calculated_1}  

  \begin{center}
   \begin{tabular}{ccc}  
     \hline
Order& EW & NS\\  
\hline    
$\pm 1$ & $2 \sqrt{2} i \Delta^{11} $& 0\\
$0$&0 & $4 \Delta^{11}$\\
      \hline
   \end{tabular}
  \end{center}
\end{table}

\begin{table}[h]
  \caption{Calculated $l=2$ $\Delta I_l^m$ response for the HV guiding voltage.}
  \label{gui_EW_NS_visibilities_calculated_2}  

  \begin{center}
   \begin{tabular}{ccc}  
     \hline
Order& EW & NS\\  
\hline    
$\pm 2$ & $2 \sqrt{2} i \Delta^{21} $& 0 \\
$\pm 1$& 0& $2 \sqrt{2} i \Delta^{21}$\\
$0$& 0& 0\\
      \hline
   \end{tabular}
  \end{center}
\end{table}

\begin{table}[h]
  \caption{Calculated $l=3$ $\Delta I_l^m$ response for the HV guiding voltage.}
  \label{gui_EW_NS_visibilities_calculated}  
   
  \begin{center}
   \begin{tabular}{ccc}  
     \hline
Order & EW & NS\\  
\hline    
$\pm 3$ & $-\frac{\sqrt{2}}{6} i \Delta^{33}+\frac{\sqrt{30}}{2} i \Delta^{31} $& 0 \\
$\pm 2$ & 0 & $-\frac{\sqrt{3}}{3}  \Delta^{33} - \sqrt{5}  \Delta^{31}$\\
$\pm 1$& $- \frac{\sqrt{30}}{6}  i \Delta^{33} - \frac{\sqrt{2}}{2} i \Delta^{31} $ & 0\\
$0$ & 0 &$-\frac{\sqrt{10}}{3}   \Delta^{33} +\sqrt{6}  \Delta^{31}$\\
      \hline
   \end{tabular}
  \end{center}
\end{table}

\begin{table}[h]
  \caption{Calculated visibilities for the HV guiding voltage.}
  \label{gui_EW_NS_visibilities_calculated}  
  
  \begin{center}
   \begin{tabular}{ccc}  
     \hline
Degree & EW & NS\\  
\hline    
1 & 16$(\Delta^{11})^2$ & 16$(\Delta^{11})^2$\\
2& 16$(\Delta^{21})^2$  & 16$(\Delta^{21})^2$\\
3 & 16$(\Delta^{31})^2$ + 16 $(\Delta^{33})^2$/9&16$(\Delta^{31})^2$ + 16 $(\Delta^{33})^2$/9\\
      \hline
   \end{tabular}
  \end{center}
\end{table}

\subsection{$\Delta I_l^m$ for the EW and NS guiding pixels}
The EW guiding pixel signals are proportional to the mean intensity of
the E (N) and W (S) pixels.  Then we can write:
\begin{equation}
\Delta I_l^m(EW)=\frac{1}{2}(\Delta I_l^m(E)+\Delta I_l^m(W))
\end{equation}
and
\begin{equation}
\Delta I_l^m(NS)=\frac{1}{2}(\Delta I_l^m(N)+\Delta I_l^m(S))
\end{equation}

\begin{table}[h]
  \caption{Calculated $l=1$ $\Delta I_l^m$ response for the EW and NS guiding pixels.}
  \label{gui_EW_NS_visibilities_calculated_gui_1}  
   
  \begin{center}
   \begin{tabular}{ccc}  
     \hline
Order & EW & NS\\  
\hline    
$\pm 1$ & $ \sqrt{2} \Delta^{10}\pi/2$& $ \sqrt{2} \Delta^{10}\pi/2$\\
$0$& 0& 0\\
      \hline
   \end{tabular}
  \end{center}
\end{table}

\begin{table}[h]
  \caption{Calculated $l=2$ $\Delta I_l^m$ response for EW and NS guiding pixels.}
  \label{gui_EW_NS_visibilities_calculated_2}  
   
  \begin{center}
   \begin{tabular}{ccc}  
     \hline
Order & EW & NS\\  
\hline    
$\pm 2$ & $-\Delta^{22}/2 + \sqrt{3}  \Delta^{20}\pi/4 $& $\Delta^{22}/2 + \sqrt{3}  \Delta^{20}\pi/4 $ \\
$\pm 1$& 0& 0\\
$0$& $-\sqrt{3}\Delta^{22}/2 - \Delta^{20}\pi/4 $ & $\sqrt{3}\Delta^{22}/2 - \Delta^{20}\pi/4$\\

      \hline
   \end{tabular}
  \end{center}
\end{table}

\begin{table}[h]
  \caption{Calculated visibilities for the EW and NS guiding pixels.}
  \label{gui_EW_NS_visibilities_calculated}  
   
  \begin{center}
   \begin{tabular}{ccc}  
     \hline
Degree & EW & NS\\  
\hline    
1 & $(\Delta^{10})^2(\pi)^2$ &  $(\Delta^{10})^2(\pi)^2$ \\
2& 5$(\Delta^{22}/2)^2+7(\Delta^{20})^2(\frac{\pi}{4})^2$  & 5$(\Delta^{22}/2)^2+7(\Delta^{20})^2(\frac{\pi}{4})^2$\\
      \hline
   \end{tabular}
  \end{center}
\end{table}

\subsection{The multiplet}
\subsubsection{$l$=1 modes}
In this case, the $m=\pm 1$ responses for the EW / NS pixels and the HV EW signals are different from zero, while nothing is transmitted for $m=0$.  For the NS HV signal, only $m=0$ is transmitted.  The visibilities\footnote{sum over the transmitted modes} in the power spectrum for the EW signals and for the NS signals are identical.

\subsubsection{$l$=2 modes}
In this case, the $m=\pm 2$ responses for the EW guiding signals are different from zero, while nothing is transmitted for other $m$.  For the NS guiding signals, only $m=\pm 1$ are transmitted.  For the EW and NS pixels, the responses are different from 0 for $m=\pm 2$ and $m=0$.  The visibilities in the power spectrum for the EW signals and for the NS signals are identical.

\subsubsection{$l$=3 modes}
In this case, the $m=\pm 1, \pm 3$ responses for the EW guiding signals are different from zero, while nothing is transmitted for other $m$.  For the NS guiding signals, only $m=\pm 2, 0$  are transmitted.  The visibilities in the power spectrum for EW and NS guiding signals are identical.

\clearpage
\section{Observed visibility ratios}

\begin{table}[h]
  \caption{Observed and theoretical log of visibility ratios for FD signals, obtained by minimizing the difference with the observed ratios; adjusted outer science  pixel radius is 0.941~$\pm$~0.08 derived from minimizing visibility differences}
  \label{FD_visibilities}  
   
  \begin{center}
   \begin{tabular}{ccccc}  
     \hline
Ratio & theor. &  obs. & diff.\\  
\hline    
$l=1 / 0$& +0.488  & +0.498 $\pm$ 0.030&+0.010 $\pm$ 0.020\\
$l=2 / 0$& -0.446 & -0.444 $\pm$ 0.009&+0.002 $\pm$ 0.009\\
$l=3 / 1$ & -3.912 & -3.276 $\pm$ 0.028&+0.636 $\pm$ 0.020\\
      \hline
   \end{tabular}
  \end{center}
\end{table}

\begin{table}[h]
  \caption{Observed and theoretical log of visibility ratios for resolved signals, obtained using the adjusted outer science  pixel radius of 0.941~$\pm$~0.08 derived from minimizing visibility differences of Table \ref{FD_visibilities}}
  \label{resolved_visibilities}  
   
  \begin{center}
  \begin{small}
   \begin{tabular}{cccccc}  
     \hline
Degree&Ratio & theor. &  obs. & diff.\\  
\hline    
1&$m=0 / 1$& -0.701  & -0.684 $\pm$ 0.078&+0.017 $\pm$ 0.078\\
2&$m=0 / 2$& -0.674 & -0.464 $\pm$ 0.046&+0.210 $\pm$ 0.046\\
3&$m=1 / 2$ & -0.348 & -0.286 $\pm$ 0.038&+0.062 $\pm$ 0.038\\
3&$m=0 / 3$ & -0.698 & -0.677 $\pm$ 0.066&+0.021 $\pm$ 0.066\\
3&$m=1 / 3$ & -0.448 & -0.489 $\pm$ 0.052&--0.041 $\pm$ 0.052\\
3&$m=2 / 3$ & -0.274 & -0.341 $\pm$ 0.034&--0.067 $\pm$ 0.034\\
1&$m=1$ / FD& -0.068& -0.266 $\pm$ 0.020& --0.198 $\pm$ 0.020\\
2&$m=2$ / FD& +0.821& 0.653 $\pm$  0.019& --0.180 $\pm$ 0.019\\
      \hline
   \end{tabular}
   \end{small}
  \end{center}
\end{table}

\begin{table}[h]
  \caption{Observed and theoretical log of visibility ratios for the EW and NS difference signals, obtained using the adjusted outer science  pixel radius of 0.941~$\pm$~0.08 derived from minimizing visibility differences of Table \ref{FD_visibilities}.}
  \label{EW_NS_visibilities}  

  \begin{center}
   \begin{tabular}{ccccc}  
     \hline
Ratio & theor.&  obs.& diff.\\  
\hline    
$l=2 / 1$ (EW) & +0.762 & +0.717 $\pm$ 0.030&-0.045 $\pm$ 0.020\\
$l=3 / 1$ (EW) & +0.216 & +0.340 $\pm$ 0.008&+0.124 $\pm$ 0.008\\
$l=2 / 1$ (NS) & +0.730  & +0.716 $\pm$ 0.020&-0.014 $\pm$ 0.020\\
$l=3 / 1$ (NS) & +0.318 & +0.234 $\pm$ 0.008&-0.084 $\pm$ 0.007\\	
$l=1$ EW / FD & -0.652 & -0.685 $\pm$ 0.061& -0.023 $\pm$ 0.060\\
$l=2$ EW / FD & +1.059 & +0.982 $\pm$ 0.017& -0.077 $\pm$ 0.060\\
$l=1$ NS / FD & -0.770 & -0.707 $\pm$ 0.063& +0.063 $\pm$ 0.060\\
$l=2$ NS / FD & +0.909 & +0.934 $\pm$ 0.022& +0.025 $\pm$ 0.060\\
      \hline
   \end{tabular}
  \end{center}
\end{table}

  \begin{figure}[!]
   \centering
   \includegraphics[angle=180,width=9.cm]{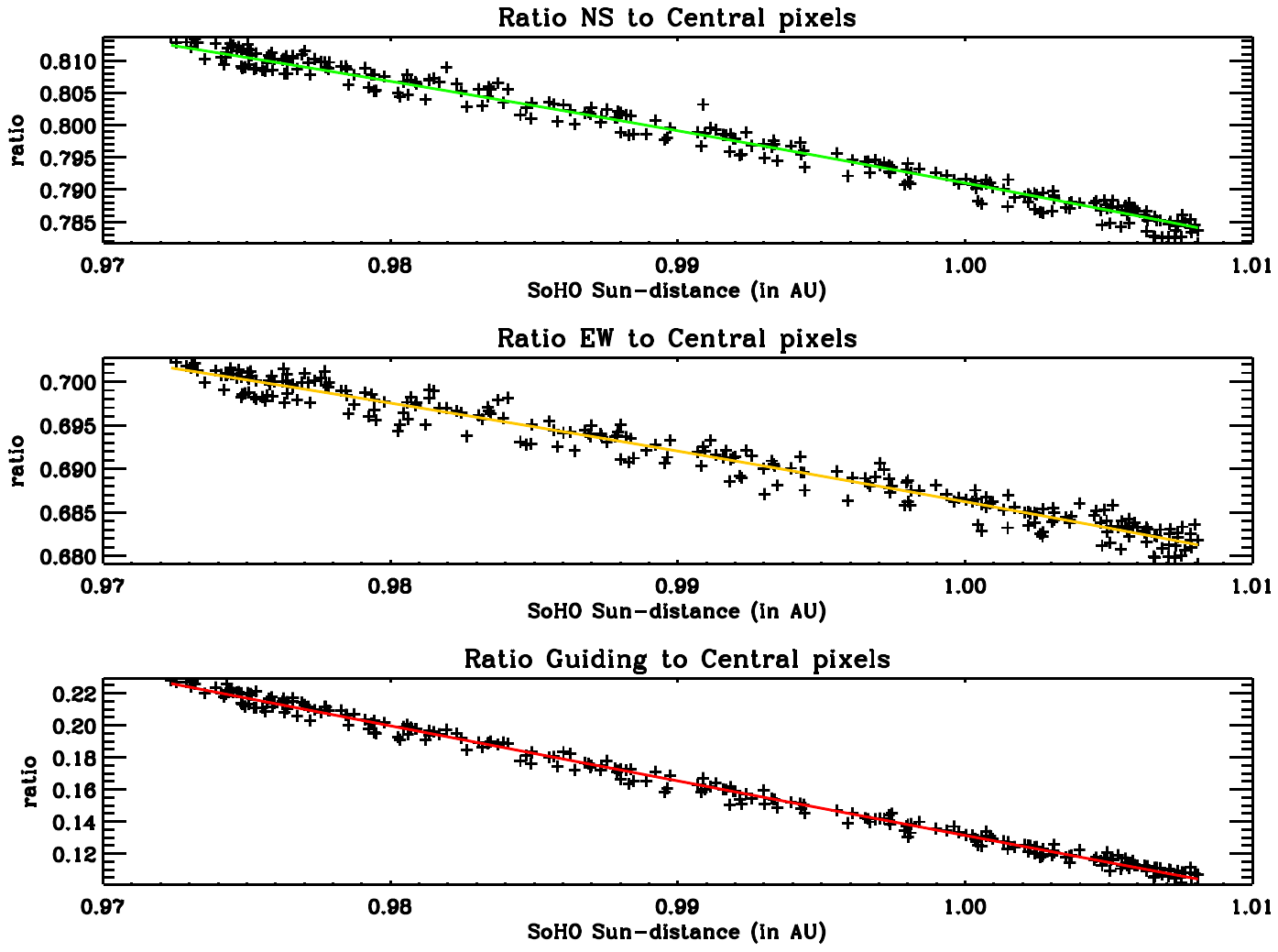}
      \caption{Ratio of pixel signals as a function of the SoHO Sun-distance and the associated fit: (top) ratio of the North-South pixels to the central pixels (crosses), fit (green line); (middle) ratio of the East-West pixels to the central pixels (crosses), fit (orange line); (bottom) ratio of the Guiding pixels to the central pixels (crosses), fit (red line).}
      \label{plotratio}
   \end{figure}

\begin{table}[h]
  \caption{Observed and theoretical log of visibility ratios for the EW and NS HV signals and the EW and NS guiding pixels, obtained using the adjusted outer science  pixel radius of 0.941~$\pm$~0.08 derived from minimizing visibility differences of Table \ref{FD_visibilities}}
  \label{gui_HV_EW_NS_visibilities}  
   
  \begin{center}
  \begin{small}
   \begin{tabular}{ccccc} 
     \hline
Ratio & theor. &  obs. & diff.\\  
\hline    
$l=2 / 1$ (EW$_{HV}$) & -1.570 & 0.848 $\pm$ 0.042&+2.418 $\pm$ 0.042\\
$l=3 / 1$ (EW$_{HV}$) & -0.414  & 0.455 $\pm$ 0.028&+0.869 $\pm$ 0.028\\
$l=2 / 1$ (NS$_{HV}$) & -1.570  & 0.762 $\pm$ 0.047&+2.332 $\pm$ 0.047\\
$l=3 / 1$ (NS$_{HV}$) & -0.414  & 0.539 $\pm$ 0.036&+0.953 $\pm$ 0.036\\
$l=1$ (EW / NS)$_{HV}$& 0.   &0.349 $\pm$ 0.067& +0.349 $\pm$ 0.067\\
$l=2$ (EW / NS)$_{HV}$& 0.    &0.411 $\pm$ 0.039&+0.411 $\pm$ 0.039 \\
      \hline
   \end{tabular}
   \end{small}
  \end{center}
\end{table}

\begin{table}[h]
  \caption{Observed and theoretical log of visibility ratios for the EW and NS HV signals and the EW and NS guiding pixels, obtained using the adjusted outer science  pixel radius of 0.941~$\pm$~0.08 derived from minimizing visibility differences of Table \ref{FD_visibilities}.  The visibility ratios are only an estimate since there is only a one-year time series for the guiding pixels.}
  \label{gui_EW_NS_visibilities}  
   
  \begin{center}
   \begin{tabular}{ccccc}  
     \hline
Ratio & theor. &  obs. & diff.\\  
\hline    
$l=1$ EW HV/gui.&  +2.91 & $\approx$ 0.22 & $\approx$ -2.70\\
$l=1$ NS HV/gui.&   +2.91& $\approx$ 1.61 & $\approx$ -1.30\\
$l=2$ EW HV/gui.&  -1.56 & $\approx$ 0.07 &$\approx$ +1.63 \\
$l=2$ NS HV/gui.&  -1.56 & $\approx$ 0.91 & $\approx$ +2.57\\
      \hline
   \end{tabular}
  \end{center}
\end{table}

\clearpage

\onecolumn
\section{Frequencies}

\begin{table}[h]
   \caption{Mode frequencies in $\mu$Hz from global fitting of the 29 years of FD data. $^\dagger${These frequencies were obtained with a local fit of the FD data.}}

  \begin{center}
   \begin{tabular}{ccccc}  
     \hline
$n$ & $l=0$ &  $l=1$& $l=2$& $l=3$\\  

\hline    
11	&	{1686.599 $\pm$  0.011}$^\dagger$			&	-	&		-	&		-	\\
12	&	{1822.300 $\pm$ 0.044}$^\dagger$			&	1885.134	$\pm$	0.023	&	1945.860	$\pm$	0.057	&			\\
13	&	1957.538	$\pm$	0.033	&	2020.900	$\pm$	0.030	&	2082.156	$\pm$	0.036	&			\\
14	&	2093.543	$\pm$	0.019	&	2156.918	$\pm$	0.025	&	2217.730	$\pm$	0.048	&			\\
15	&	2228.846	$\pm$	0.021	&	2292.082	$\pm$	0.015	&	2352.384	$\pm$	0.042	&			\\
16	&	2362.889	$\pm$	0.020	&	2425.705	$\pm$	0.013	&	2486.010	$\pm$	0.030	&			\\
17	&	2496.286	$\pm$	0.017	&	2559.291	$\pm$	0.015	&	2619.766	$\pm$	0.020	&			\\
18	&	2629.754	$\pm$	0.010	&	2693.494	$\pm$	0.009	&	2754.594	$\pm$	0.015	&	2811.334	$\pm$	0.079  \\
19	&	2764.271	$\pm$	0.011	&	2828.281	$\pm$	0.012	&	2889.676	$\pm$	0.015	&	2946.893	$\pm$	0.048  \\
20	&	2899.098	$\pm$	0.012	&	2963.441	$\pm$	0.012	&	3024.819	$\pm$	0.013	&	3082.315	$\pm$	0.032  \\
21	&	3033.864	$\pm$	0.012	&	3098.310	$\pm$	0.011	&	3159.989	$\pm$	0.016	&	3217.731	$\pm$	0.038  \\
22	&	3168.738	$\pm$	0.013	&	3233.317	$\pm$	0.013	&	3295.155	$\pm$	0.020	&	3353.318	$\pm$	0.067   \\
23	&	3303.663	$\pm$	0.016	&	3368.741	$\pm$	0.016	&	3430.861	$\pm$	0.028	&	3489.452	$\pm$	0.104 \\
24	&	3439.152	$\pm$	0.023	&	3504.460	$\pm$	0.023	&	3566.921	$\pm$	0.043	&	3625.931	$\pm$	0.179 \\
25	&	3575.007	$\pm$	0.028	&	3640.773	$\pm$	0.032	&	3703.428	$\pm$	0.067	&	3762.616	$\pm$	0.360 \\
26	&	3711.196	$\pm$	0.051	&	3777.291	$\pm$	0.044	&	3840.297	$\pm$	0.096	&	3900.347	$\pm$	0.595 \\
27	&	3847.985	$\pm$	0.057	&	3914.350	$\pm$	0.064	&	3977.802	$\pm$	0.189	&	4043.565	$\pm$	0.016 \\
28	&	3985.001	$\pm$	0.122	&	4051.804	$\pm$	0.084	&	4115.178	$\pm$	0.199	&	4197.790	$\pm$	0.893 \\
29	&	4122.931	$\pm$	0.147	&	4189.064	$\pm$	0.107	&	4262.748	$\pm$	0.087	&	4339.006	$\pm$	0.900 \\
30	&	4255.467	$\pm$	0.075	&	4327.551	$\pm$	0.118	&	4391.455	$\pm$	0.155	&	4473.272	$\pm$	1.319 \\
31	&	4399.915	$\pm$	0.192	&	4466.338	$\pm$	0.175	&	4530.130	$\pm$	0.178	&			\\
32	&	4538.289	$\pm$	0.140	&	4605.468	$\pm$	0.237	&	4683.535	$\pm$	0.417	&			\\
33	&	4672.324	$\pm$	0.168	&	4745.060	$\pm$	0.261	&	-	&			\\
      \hline
   \end{tabular}
  \end{center}
  \label{frequency_table_global}
\end{table}

\begin{table}[h]
  \caption{Mode frequencies from local fitting of the 29 years of resolved data with the collapsogram technique.  $^\dagger$These frequencies were also obtained with the global fit.}
  \label{diff}  
   
  \begin{center}
   \begin{tabular}{ccccc}  
     \hline
$n$ & $l=0$ &  $l=1$& $l=2$& $l=3$\\  

\hline    
9	&					-				&						-			&					-			       &1591.559 $\pm$     0.012\\	
10    &				-					&				-					& 1674.546 $\pm$      0.015                   & 1729.144 $\pm$      0.021\\
11	&	-			&	1749.334 $\pm$      0.028			&	1810.301 $\pm$      0.016					&		-	\\	
12    &	-			&	{1885.145 $\pm$ 0.020}$^\dagger$								& {1945.886 $\pm$ 0.016}$^\dagger$   & - \\
      \hline
   \end{tabular}
  \end{center}
  \label{frequency_table_collapsogram}
\end{table}

\FloatBarrier
\clearpage
\section{Linewidths and mode heights}
\begin{table}[h]
\centering
\begin{minipage}{0.48\textwidth}
  \caption{Log of mode linewidths (in $\mu$Hz) and mode height (in ppm$^2/\mu$Hz) from global fitting of the 29 years of FD data for the $l=0$ modes.  $^\dagger$These linewidths are derived from the collapsogram analysis of the $l$=2 modes.}
      \centering
    \scriptsize
  \vspace{1ex} 
  \resizebox{\linewidth}{!}{
   \begin{tabular}{cccc}  
     \hline
$n$ &frequency ($\mu$Hz)& log(linewidth) &  log(height)\\  

\hline    
   {11}$^\dagger$&       {1674.55} &   {-1.688  $\pm$  0.333 } & - \\
  {12$^\dagger$}&      {1810.30} &   {-1.181  $\pm$    0.191 } &        - \\
      13 &       1957.54 &      -1.666 $\pm$    0.084 &     -0.785 $\pm$    0.097\\
      14 &       2093.54 &      -1.017 $\pm$    0.046 &     -0.820 $\pm$    0.048\\
      15 &       2228.85 &     -0.995 $\pm$    0.032 &     -0.340 $\pm$    0.032\\
      16 &       2362.89 &     -0.641 $\pm$    0.025 &     -0.149 $\pm$    0.023\\
      17 &       2496.29 &     -0.228 $\pm$    0.020 &    -0.022 $\pm$    0.017\\
      18 &       2629.75 &      +0.309 $\pm$    0.026 &    -0.085 $\pm$    0.019\\
      19 &       2764.27 &      +0.711 $\pm$    0.015 &    -0.052 $\pm$    0.010\\
      20 &       2899.10 &       +1.079 $\pm$    0.012 &    -0.096 $\pm$   0.008\\
      21 &       3033.86 &       +1.230 $\pm$    0.013 &    -0.015 $\pm$   0.008\\
      22 &       3168.74 &      +0.993 $\pm$    0.010 &      +0.227 $\pm$   0.007\\
      23 &       3303.66 &      +0.512 $\pm$    0.011 &      +0.508 $\pm$   0.007\\
      24 &       3439.15 &     -0.281 $\pm$    0.011 &      +0.936 $\pm$   0.009\\
      25 &       3575.01 &      -1.055 $\pm$    0.016 &       +1.306 $\pm$    0.012\\
      26 &       3711.20 &      -1.782 $\pm$   0.009 &       +1.661 $\pm$   0.009\\
      27 &       3847.98 &      -2.493 $\pm$    0.015 &       +1.974 $\pm$    0.012\\
      28 &       3985.00 &      -3.087 $\pm$    0.013 &       +2.241 $\pm$    0.013\\
      29 &       4122.93 &      -3.625 $\pm$    0.016 &       +2.408 $\pm$    0.014\\
      30 &       4255.47 &      -4.121 $\pm$    0.017 &       +2.613 $\pm$    0.014\\
      31 &       4399.91 &      -4.509 $\pm$    0.026 &       +2.639 $\pm$    0.028\\
      32 &       4538.29 &      -5.063 $\pm$    0.025 &       +2.668 $\pm$    0.022\\
      33 &       4672.32 &      -5.260 $\pm$    0.024 &       +2.836 $\pm$    0.023\\   
        \hline
   \end{tabular}
}
    \label{linewidth_height_table_global0}
\end{minipage}\hfill
\begin{minipage}{0.48\textwidth}
      \caption{Log of mode linewidths (in $\mu$Hz) and mode height (in ppm$^2/\mu$Hz) from global fitting of the 29 years of FD data for the $l=1$ modes.   $^\dagger$These linewidths are derived from the collapsogram analysis of the $l$=3 modes.}
      \centering
  \scriptsize
  \vspace{1ex}
  \resizebox{\linewidth}{!}{
   \begin{tabular}{cccc}  
  \hline
$n$ &frequency ($\mu$Hz)& log(linewidth) &  log(height)\\  

\hline    
   {10}$^\dagger$&       {1591.56} &   {\-2.492  $\pm$  0.566} & - \\
  {11}$^\dagger$&      {1729.14} &   {-1.565  $\pm$   0.246 } &        - \\
      12 &       1885.13 &      -1.328 $\pm$     0.132 &      -1.041 $\pm$     0.148\\
      13 &       2020.90 &     -0.811 $\pm$    0.070 &     -0.477 $\pm$    0.074\\
      14 &       2156.92 &     -0.572 $\pm$    0.040 &     -0.494 $\pm$    0.041\\
      15 &       2292.08 &     -0.149 $\pm$    0.044 &     -0.413 $\pm$    0.038\\
      16 &       2425.70 &      +0.157 $\pm$    0.041 &     -0.137 $\pm$    0.033\\
      17 &       2559.29 &      +0.521 $\pm$    0.018 &    -0.056 $\pm$    0.013\\
      18 &       2693.49 &      +0.935 $\pm$    0.020 &    -0.056 $\pm$    0.013\\
      19 &       2828.28 &       +1.319 $\pm$    0.014 &    -0.065 $\pm$   0.009\\
      20 &       2963.44 &       +1.573 $\pm$    0.014 &    -0.020 $\pm$   0.009\\
      21 &       3098.31 &       +1.579 $\pm$    0.013 &     +0.105 $\pm$   0.008\\
      22 &       3233.32 &       +1.272 $\pm$    0.015 &      +0.331 $\pm$   0.009\\
      23 &       3368.74 &      +0.674 $\pm$    0.012 &      +0.654 $\pm$   0.008\\
      24 &       3504.46 &    -0.086 $\pm$    0.010 &       +1.038 $\pm$   0.007\\
      25 &       3640.77 &     -0.882 $\pm$    0.010 &       +1.418 $\pm$   0.007\\
      26 &       3777.29 &      -1.597 $\pm$    0.010 &       +1.732 $\pm$   0.008\\
      27 &       3914.35 &      -2.285 $\pm$    0.011 &       +2.054 $\pm$    0.010\\
      28 &       4051.80 &      -2.854 $\pm$    0.018 &       +2.286 $\pm$    0.016\\
      29 &       4189.06 &      -3.439 $\pm$    0.017 &       +2.506 $\pm$    0.014\\
      30 &       4327.55 &      -3.895 $\pm$    0.020 &       +2.623 $\pm$    0.019\\
      31 &       4466.34 &      -4.391 $\pm$    0.028 &       +2.812 $\pm$    0.030\\
      32 &       4605.47 &      -4.764 $\pm$    0.029 &       +2.910 $\pm$    0.034\\
      33 &       4745.06 &      -5.169 $\pm$    0.035 &       +2.876 $\pm$    0.052\\
        \hline
   \end{tabular}
   \label{linewidth_height_table_global1}
}
\end{minipage}
\end{table}

\end{appendix}

\end{document}